\documentclass{aa}
\usepackage[varg]{txfonts}
\usepackage{graphicx}
\usepackage{lscape}

\begin{document}
\title{Characterisation of red supergiants in the Gaia spectral range}
\author{Ricardo Dorda\inst{\ref{inst1}}
\and Carlos Gonz\'alez-Fern\'andez\inst{\ref{inst2}}
\and Ignacio Negueruela\inst{\ref{inst1}}}
\institute{Departamento de F\'{\i}sica, Ingenier\'{\i}a de Sistemas y Teor\'{\i}a de la Se\~nal, Universidad de Alicante, Carretera de San Vicente s/n, San Vicente del Raspeig E03690, Alicante, Spain\label{inst1}
\and Institute of Astronomy, University of Cambridge, Madingley Road, Cambridge CB3 0HA, United Kingdom\label{inst2}}

\abstract{The infrared Calcium Triplet and its nearby spectral region have been used for spectral and luminosity classification of late-type stars, but the samples of cool supergiants (CSGs) used have been very limited (in size, metallicity range, and spectral types covered). The spectral range of the \textit{Gaia} Radial Velocity Spectrograph (RVS) covers most of this region but does not reach the main TiO bands in this region, whose depths define the M sequence.}{We study the behaviour of spectral features around the Calcium Triplet and develop effective criteria to identify and classify CSGs, comparing their efficiency with other methods previously proposed.}{We measure the main spectral features in a large sample (almost 600) of CSGs from three different galaxies, and we analyse their behaviour through a principal component analysis. Using the principal components, we develop an automatised method to differentiate CSGs from other bright late-type stars, and to classify them.}{The proposed method identifies a high fraction ($0.98\pm0.04$) of the supergiants in our test sample, which cover a wide metallicity range (supergiants from the SMC, the LMC, and the Milky Way) and with spectral types from G0 up to late-M. In addition, it is capable to separate most of the non-supergiants in the sample, identifying as supergiants only a very small fraction of them ($0.02\pm0.04$). A comparison of this method with other previously proposed shows that it is more efficient and selects less interlopers. A way to automatically assign a spectral type to the supergiants is also developed. We apply this study to spectra at the resolution and spectral range of the \textit{Gaia} RVS, with a similar success rate.}{The method developed identifies and classifies CSGs in large samples, with high efficiency and low contamination, even in conditions of wide metallicity and spectral-type ranges. As this method uses the infrared Calcium Triplet spectral region, it is specially useful for surveys looking for CSGs in high-extinction regions. In addition the method is directly applicable to the \textit{Gaia} spectra.}


\maketitle
\section{Introduction}

Red supergiants (RSGs) are evolved high-mass stars, characterised by very high luminosities $\log(L/L_{\sun})\sim4.5$\,--\,5.8
\citep{hum1979} and late spectral types (K and M). They are the result of the evolution of moderately high-mass stars with masses from $\sim8$ to $\sim40\:$M$_{\sun}$, which represent the overwhelming majority of high-mass stars \citep{eks2013}. Since this phase is short ($\lesssim10$\% of their lifetime), evolutionary models for high-mass stars find a strict test-bed in the RSG phase.

The interest of RSGs goes beyond their role as evolutionary model constraints. Due to their high luminosity and low temperature, RSGs appear very bright in the infrared, and thus are easily observable at very large distances, even if they are affected by high extinction. Thanks to this, in the past few years, several massive and highly reddened clusters have been discovered in the inner Galaxy \citep[for example][]{dav2007,neg2012}, in the region where the tip of the Galactic bar is believed to touch the base of the Scutum arm, revealing recent widespread massive star formation in this part of the Milky Way.

Stars massive enough to pass through the RSG phase are expected to end their lives as core-collapse supernovae (SNe). In fact, RSGs are the progenitors of type-IIP SNe  \citep{gro2013, sma2015}, which is the most frequent SN type. Thus, the characterisation of RSGs (individually and as population) has an obvious interest for SN studies.

From a theoretical point of view, high- and intermediate-mass stars are easy to tell apart because of their very different evolutionary paths. High-mass stars are those with enough mass ($\gtrsim8\:M_{\sun}$) to end their lives as SNe after a few million years of life, while intermediate-mass stars will go through the asymptotic giant branch (AGB) phase, lose their envelopes, and finally become white dwarfs (except in a few cases, close to the limit with high-mass stars, where an electron-capture SN is possible).

Despite their different natures, RSGs are hard to distinguish from other late type stars, such as AGB or red giant branch (RGB) stars, by using only photometry. The intrinsic colours of all these stars are the same, as their temperatures are similar. Of course, the bolometric magnitude, $M_{\mathrm{bol}}$, of RSGs is much higher than that of RGB and most AGB stars, but this is not really helpful when the distances and extinctions are unknown, and there are many less luminous but closer foreground stars, as is the case of the Galactic plane. To break this degeneracy, spectroscopic studies are necessary.

Classical spectral classification criteria were originally defined for the optical range
\citep[e.g.][and references therein]{tur1985,kee1987}. However, as has been mentioned before, late stars are more easily accessible in the near infrared (NIR) than in the optical. The Calcium Triplet (CaT) spectral region, from $\sim8400\:$\AA{} to $\sim8900\:$\AA{}, has many advantages for a spectral study of RSGs. Firstly, this region is close to the emission peak of stars with temperatures typical of RSGs, and it is less affected by extinction than the optical range. Secondly, it is not affected by strong telluric absorption, as it is inside an atmospheric window. Thirdly, it is rich in spectral features that can be used for spectral classification, and many works have already studied them \citep{kir1991,gin1994,car1997,mun1999}. In fact, the CaT itself is a well-known luminosity discriminator \citep[e.g.][]{dia1989}.

There are still some unresolved issues related to the spectral type and luminosity classification of RSGs. The number of standard stars of the MK system classified as RSGs is very limited, and some of them are not very reliable standards because they present spectral variations of a few subtypes. Even more dramatic is the situation among the M-type RSGs: only a handful of them have been sufficiently well-characterised. Thus, the number of RSGs studied in works dealing with cool stars in general \citep[e.g.][]{kir1991,gin1994,car1997} is really low, and they cover only the K sequence. In their spectral atlas of the CaT region at moderately-high resolution, \cite{mun1999} reach later spectral types, but only for dwarf and giant stars, never supergiants (SGs). In fact, the number of SGs considered by Munari~\&~Tomasella is extremely low.

For M1 and later types, there are TiO molecular bands growing deeper in this wavelength range. Classical criteria define the M sequence by the presence and depth of the TiO bands in the optical spectral range. In the CaT range there are two main TiO bands (with bandheads at $8432+8442+8452\:$\AA{} and $8859\:$\AA{}), plus a weak VO band (at $8624\:$\AA{}),  which are used for the same purpose (see Fig.~\ref{sec_spt}). However, these bands represent a major complication for luminosity classification. At low and mid-resolutions these bands erode the continuum, weakening other spectral features, and even erasing them \citep{dor2013}. Therefore, the bands affect the line ratios and other measurements used as luminosity class (LC) criteria, rendering most of them useless except for the earliest M subtypes (those earlier than M3). Being so, extrapolation of the classification scheme from earlier types cannot be used. Even the CaT becomes unable to separate clearly LC~I from LC~II and LC~III \citep{neg2011}. However, the classification is still roughly possible if the spectral type (SpT) is known, as this will predict the TiO bandhead depths, and therefore warn about the erosion suffered by other spectral features \citep{neg2011,neg2012}.

Interest in the CaT range has grown in recent years because this is the spectral range that is being observed  by the \textit{Gaia} space telescope. \textit{Gaia} uses its Radial-Velocity Spectrometer (or RVS) to observe the centre of the CaT range (from $8470\:$\AA{} to $8740\:$\AA{}) at medium resolution \citep[R$=\lambda/\Delta\lambda=11\,500$;][]{kat2004}. Unfortunately, the spectral range observed does not cover any of the two main TiO bandheads in the region. Thus, for all M stars observed, all the spectral features in the RVS spectral range are affected by molecular band erosion, while the corresponding bandheads are not seen (and therefore their depth cannot be measured). In consequence, a strong degeneracy between SpT and luminosity appears for all bright late stars. Given the high extinction towards the inner Galaxy, many of the stars observed by the RVS in this area of the sky will be luminous cool stars. Because of their brightness in the CaT region, they will be excellent tracers of structure in these obscured regions. A good luminosity classification, however, will be necessary to make use of this information.

In this paper, we set out to derive spectral and luminosity classification criteria making use of the spectral features available in the spectra of cool luminous stars. These criteria will be useful for the analysis not only of \textit{Gaia} spectra, but also of the products of forthcoming spectroscopic surveys, such as those that will be conducted with William herschell telescope Enhanced Area Velocity Explorer (WEAVE).

\section{Measurements}
\label{measurements}
\begin{figure}[ht!]
   \centering
   \includegraphics[trim=1cm 0.5cm 2cm 1.2cm,clip,width=9cm]{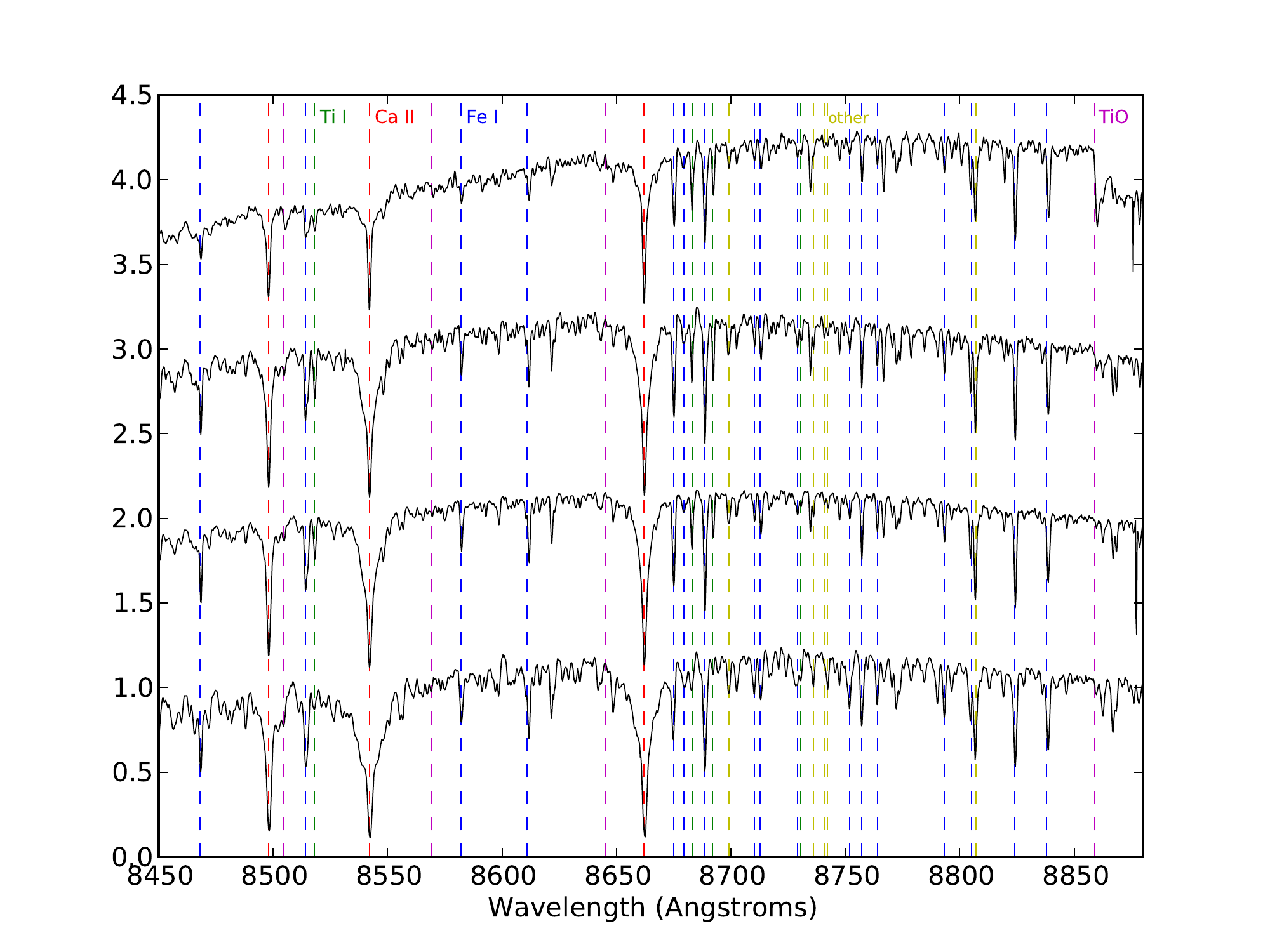}
   \caption{Example of the spectra used. This is a SpT sequence displaying all the features measured. The stars shown have about the same luminosity class and similar metallicity (they are all from the LMC). From bottom to top: [M2002]169754 (G5\:Ia), [M2002]168047 (K4\:Iab), [M2002]130426 (M2\:Iab), and [M2002]130426 (M5\:Iab). The dashed lines indicate the spectral features measured (shortened list, see text). Their colour represents the dominant chemical species in each feature: red for Ca\,{\sc{ii}}, blue for Fe\,{\sc{i}}, green for Ti\,{\sc{i}}, yellow for other atomic lines (Mn\,{\sc{i}}, Si\,{\sc{i}} and Mg\,{\sc{i}}), and magenta for the TiO bands. For more details, see Section~\ref{measurements}.}
   \label{sec_spt}
\end{figure}

\begin{figure}[ht!]
   \centering
   \includegraphics[trim=1cm 0.5cm 2cm 1.2cm,clip,width=9cm]{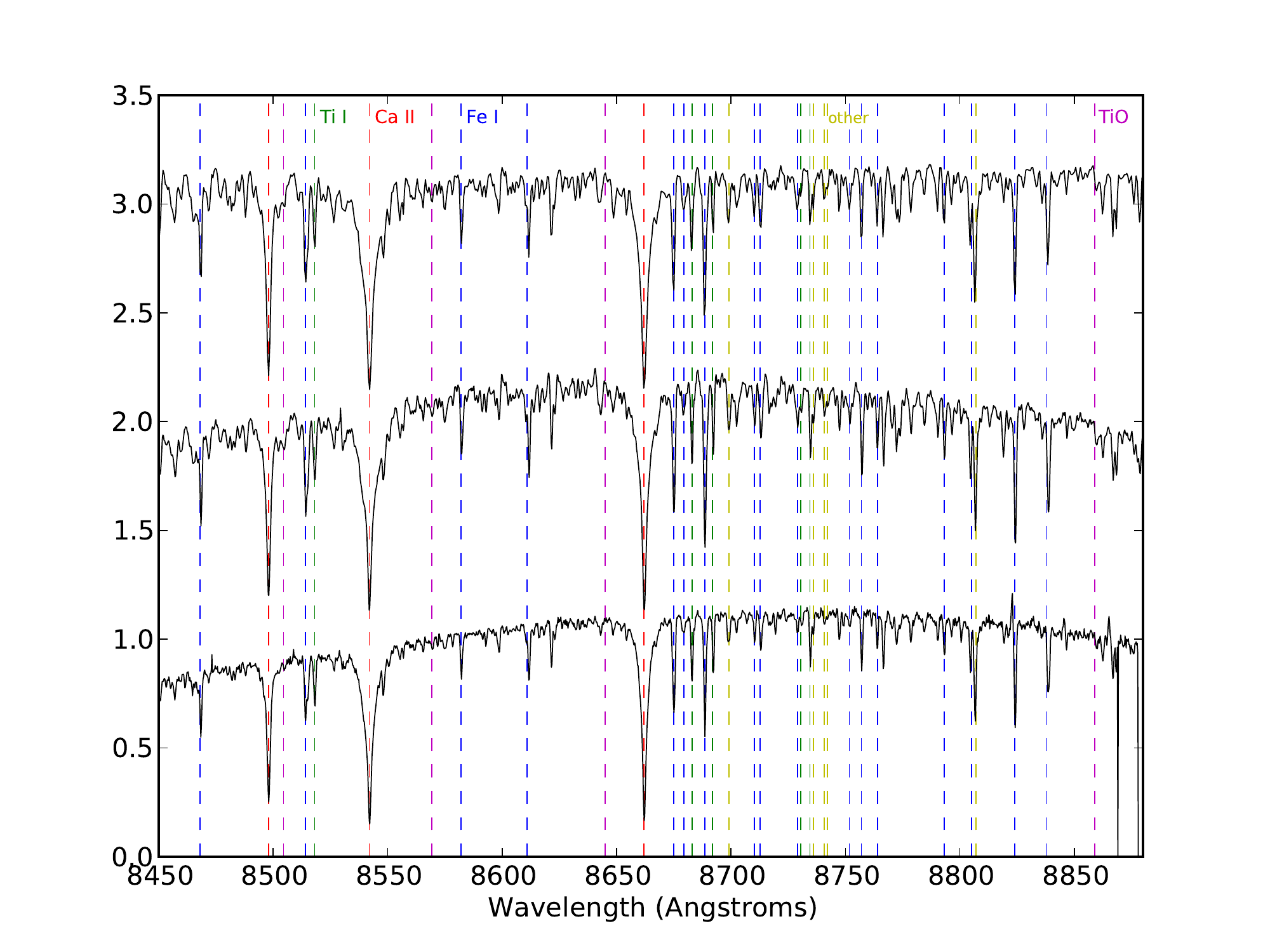}
   \caption{Example of the spectra used. This is a metallicity sequence, showing stars with the same SpT and LC (M1\:Iab) in different galaxies. From bottom to top: SMC381 (SMC), [M2002]135754 (LMC), and HD~13658 (Milky Way). The dashed lines are as in Fig.~\ref{sec_spt}. We note that the curved shape of the spectra are because of an instrumental effect.}
   \label{sec_gal}
\end{figure}

We have made use of a very large spectroscopic database, comprising stars from the Milky Way and both Magellanic Clouds (MCs). We show examples of the spectra used, indicating the features measured, in Figs.~\ref{sec_spt} and~\ref{sec_gal}.

\subsection{Data from the Magellanic Clouds}

\cite{gon2015} have recently published one of the largest spectroscopic catalogues of cool supergiants (CSGs) to date, including samples from both MCs. From now on, we will refer to this work as GDN2015. The present work makes use of this catalogue to bring some light to the identification and classification of RSGs, because it is large enough to obtain statistically significant results about RSGs from both galaxies.

The sample from GDN2015 also has another critical feature for our work: all objects were simultaneously observed in both the optical and CaT spectral ranges. As the spectral and luminosity classifications were done using classical criteria in the optical range, the assigned SpT and LC are independent of the features in the CaT spectral range. A general description of the observation, reduction, and classification of the data from the MCs is presented in GDN2015. Here, we offer a brief summary of the data that have been used for the present work.

The data from GDN2015 were obtained with the fibre-fed dual-beam AAOmega spectrograph on the Anglo Australian Telescope. This instrument is capable of taking a simultaneous observation of two different spectral ranges. For the "blue" range, in the 2012 and 2013 campaigns, the 580V grating was used. For the "red" range the 1700D grating was used, providing  a resolution $R\sim11\,000$ around the CaT. The 1700D grating has a $500\:$\AA{} wide spectral range centred on $8700\:$\AA{}, but the projection of the spectrum from each fibre on the CCD depends on its position on the plate, displacing the range limits up to $\sim$20\AA{}. Therefore, it is not possible to define a precise common range.

GDN2015 observed the Small Magellanic Cloud (SMC) on three epochs (2010, 2011, and 2012) and the Large Magellanic Cloud (LMC) on two (2010 and 2012). However, here we are only using the data from SMC-2012 and LMC-2013. There are two reasons for this. Firstly, the spectral range observed in 2010 did not cover the TiO bandhead at $8859\:$\AA{}. Secondly, in 2010 and 2011 only previously known RSGs were observed. The surveys of 2012 and 2013, however, were exploratory, using a candidate selection plus a significant number of previously known RSGs -- in fact, almost all the known RSGs observed in 2010 and 2011. Therefore mixing these epochs would lead to redundancy for the hundred previously known RSGs, while leaving out the 2012 or 2013 campaigns would exclude a large number of new CSGs.

The spectral and luminosity classifications were done using the optical spectra of the observed targets. There are a number of targets observed more than once on the same epoch, but all the spectra were classified independently without knowing their identities. GDN2015 used these redundant targets to estimate the internal coherence of the classification, finding that the mean differences between the spectral classifications of the redundant targets are about one subtype for the SpT and half a subclass for the LC. We use these values as uncertainties for the SpT and LC in this work. The results of this coherence test are detailed in Table~2 of GDN2015.

For this work, we have used all the stars from the 2012 and 2013 campaigns except the Carbon stars and those with SpTs earlier than~G. We removed these stars because their NIR spectra are very different to our target late stars and thus they are easily identifiable. Nevertheless, we kept late-type, non-supergiant objects, because these stars passed the cut of the photometric criteria, and therefore represent the kind of interlopers to handle in a survey looking for RSGs. Therefore any useful spectral criteria should be able to separate these stars from the CSGs. These interlopers are mainly foreground stars (G, K and early-M~dwarfs and giants), with a smaller number of AGB stars (high-luminosity and late-SpT giants) from the MCs themselves (but mainly from the LMC). Table~\ref{observations_MC} contains a summary of the data used for the present work.

\subsection{Data from the Galaxy}

We decided to complement the GDN2015 data with stars from the Milky Way, for two main reasons. Firstly, although GDN2015 provides a statistically significant number of CSGs, the number of mid- and late-M stars in this sample was small. These subtypes are not as frequent in the MCs as they are in the Milky Way, because CSGs tend to have earlier SpTs at lower metallicities \citep{eli1985}. Secondly, we wanted to include objects covering a broad range of metallicities, so that we could understand the effect of chemical composition on our criteria. In addition, a sample at approximately Solar metallicity is needed to check the validity of our criteria for the range of metallicities that we may expect to find among the stars that \textit{Gaia} will observe towards the Inner Galaxy. Thus, we have observed a significant number of MK standards \citep[or at least stars with spectral classification determined by][presenting LCs from I to III]{kee1989}, and other well known CSGs in the Perseus arm. All these objects have spectral \textbf{and luminosity} classifications that have been repeatedly confirmed in the literature. We therefore considered them well characterised and did not perform our own optical classification. Table~\ref{observations_MC} has a summary of the data used for the present work, but the details about the observed Galactic stars are shown in Table~\ref{stdr}.

The stars from the Milky Way were observed along four different campaigns, one in 2011, two in 2012, and one in 2015, using the Intermediate Dispersion Spectrograph (IDS) attached to the 2.5~m Isaac Newton Telescope (INT) in La Palma (Spain). We used grating R1200R with the RED+2~CCD mounted with its 4096 pixel axis along the wavelength direction. This configuration covers a $572\:$\AA{} wide spectral range centred at $8500\:$\AA{} (i.e. the spectral range around the CaT), with a resolution  $R\sim10\,500$, very similar to the resolution of the data from GDN2015 ($R\sim11\,000$).

IDS is a classical long-slit spectrograph, and so reduction of these observations was carried out in the standard manner, using the {\sc iraf} facility.\footnote{IRAF is distributed by the National Optical Astronomy Observatories, which are operated by the Association of Universities for Research in Astronomy, Inc., under cooperative agreement with the National Science Foundation} As a last step, we obtained velocities along the line of sight for these objects. This calculation follows that outlined in \citet{neg2012}, with only one particularity: to correct the wavelength calibration from instrumental flexures, we use the subtracted sky spectrum to cross-correlate the sky emission lines between objects. By doing this, we guarantee that all our measurements will have the same instrumental signature that can then be corrected by comparing with velocity standards.

\begin{table}
\caption{Summary of the data used in this work, split by origin, LC and SpT.}
\label{observations_MC}
\centering
\begin{tabular}{c | c | c c c c}
\hline\hline
\noalign{\smallskip}
From&SpT&\multicolumn{4}{c}{Number of}\\
&&LC I&LC II\,--\,III&LC IV\,--\,V&Total\\
\noalign{\smallskip}
\hline
\noalign{\smallskip}
&G&116&25&66&207\\
SMC&K&151&54&36&241\\
&M&36&7&7&50\\
\noalign{\smallskip}
\hline
\noalign{\smallskip}
&G&6&2&0&8\\     
LMC&K&94&16&4&114\\
&M&124&37&1&162\\
\noalign{\smallskip}
\hline
\noalign{\smallskip}
&G&2&0&0&2\\        
Galaxy&K&11&7&0&18\\
&M&49&47&0&96\\
\noalign{\smallskip}
\hline
\noalign{\smallskip}
&G&124&27&66&217\\
Total&K&256&78&40&374\\
&M&209&91&32&332\\
&All&589&195&114&898\\
\noalign{\smallskip}
\hline
\end{tabular}
\tablefoot{Non-CSGs from the SMC and the LMC are mostly foreground objects (see text) from those surveys, and therefore they do not belong to the MCs.}
\end{table}

\subsection{Spectral features measured}

To characterise the CaT region we selected the main atomic and molecular features present along the common spectral range that we have for all stars (from $8450\:$\AA{} to $8870\:$\AA{}), including known SpT and LC indicators, but also those not used previously. In total, we measured 47 atomic features and four molecular bandheads. See Tables~\ref{atomic_ranges} and~\ref{molecular_ranges} for the complete and detailed list.

The first step to measure these spectral features was to re-sample the spectra from each instrument to the same uniform resolution: $R\sim10\,000$, by convolving the spectra with a gaussian kernel. Later, we compared the measurements done with and without the convolution, and we found the mean difference for each measurement to be smaller than the corresponding typical uncertainty. We conclude that this convolution from $R\sim11\,000$ and $R\sim10\,500$ to $R\sim10\,000$ does not affect significantly our results and, in consequence, future works using a similar resolution (as is the case of the \textit{Gaia} RVS) may utilise our results without the need to re-sample their spectra to $R\sim10\,000$.

The second step was to shift each spectrum to their rest wavelength, using the previously calculated radial velocities. The third step was to measure the spectral features. This was done differently for the two types of features present in our spectra, atomic lines and molecular bandheads.

\subsubsection{Atomic lines}
\label{atomic_lines}

For the atomic spectral lines, we decided to measure their equivalent widths (EWs). Previous works \citep[e.g.][and references therein]{kir1991,gin1994,cen2001} proposed different methods for this calculation, through an estimation of the true continuum. However, because of the effect of the TiO bands over the continuum, these methods become useless for mid- and late-M stars. As we want uniform EW measurements for the whole SpT range, from early-G to late-M, we use local pseudo-continua to calculate the EWs. Table~\ref{atomic_ranges} shows the ranges used for each line.

To define the local pseudocontinuum for each line, we studied the behaviour with SpT of its neighbouring spectral region. For each line we selected two spectral ranges, one to the red and one to the blue of the line, not affected by atomic lines at any of the SpT in our range (from G to M). Then for each line we calculated the linear regression with wavelength of its two pseudo-continuum ranges, and from it we obtained its EW. This method using local pseudo-continua does not measure a true EW, because we are not necessarily using the real continuum. With it, we only measure the apparent strength, at this resolution, of the line at a given SpT. Uncertainties on the EWs were calculated using the method proposed by \cite{vol2006}.

The atomic feature centred at $8468\:$\AA{} (which spans from $\sim8462\:$\AA{} to $\sim8474\:$\AA{}) is a blend of many lines (mainly Ti\,{\sc{i}} and Fe\,{\sc{i}}) and multiple molecular bandheads of CN \citep{gin1994,car1997}. It is considered a powerful luminosity indicator for stars with SpT earlier than M3 \citep{neg2011}, but it is also sensitive to chemical anomalies \citep{car1997}. The measurement of this feature was not easy, because the continuum on its blue side changes strongly with SpT due to the presence of molecular bandheads and atomic lines at different SpTs. This feature, however, is too important to be discarded. Thus, we selected two continuum ranges on its blue side, one useful for early subtypes and the other for the M subtypes. We calculated two EWs for this feature in each case but then we kept only that with the slope of its regression closest to 0, as this is the measurement less affected by changes in the continuum.

\subsubsection{Molecular bandheads}
\label{bands}

The rise of a molecular band changes drastically the shape of the nearby spectral range, creating a shoulder at the wavelength of the bandhead. For wavelengths redder than this, the apparent continuum is not flat, but has a positive slope, tending to the original continuum level, giving the whole molecular feature the shape of a saw tooth.

We defined two criteria to detect the presence of a bandhead. The first criterion is to test the shape of the bandhead. For this, we take a small spectral range of a certain width, centred on the bandhead wavelength. To accept the presence of a band, the maximum intensity in this range has to be to the blue of the minimum intensity. The second criterion requires that the difference between the maximum and minimum intensities within the range defined for the previous step is larger than the standard deviation of the intensity in the continuum. With this criterion we make sure that the bandhead has a significant depth.

If a given bandhead passes both criteria, we measure its depth, if not, we assign a value of 0 to it. For this step, we defined for each molecular band a bandhead central wavelength at the point where the bandhead starts (i.e. the point where the flux suddenly decreases), and a pseudo-continuum range bluewards of it. We also defined a small range to the red of the bandhead centre, where the minimum flux of the bandhead should be. The values defined for each bandhead are given in Table~\ref{molecular_ranges}. For the measurement, we first find the point where the flux is the lowest in the range defined to the red of the centre of the bandhead. This is the bandhead bottom (\textit{BB}). After this, we perform a linear regression fit to the pseudo-continuum range, because we want to take into account the slope of the pseudocontinuum, and we extrapolate the resultant fit line over the defined centre of the bandhead. This value is the bandhead top (\textit{BT}). From these values, we obtain the bandhead depth (\textit{BD}):

\begin{displaymath}
BD = \frac{BT-BB}{BT}.
\end{displaymath}

To measure the uncertainty in a band's depth, we use a method analogous to that of \cite{vol2006}, but with two differences. We start from the expression for our $BD$, and we measure the signal-to-noise (S/N) of the pseudo-continuum using the residua from the fit to the pseudo-continuum instead of using the differences to the mean flux value of the pseudo-continuum. We do this because the pseudo-continuum is slopped by the effect of other molecular bands to its blue side, making the difference to mean flux meaningless. The resultant expression is:

\begin{displaymath}
\sigma(BD)=\frac{1}{\mathrm{S/N}}\sqrt[]{\frac{BB}{BT}+\left(\frac{BB}{BT}\right)^{2}}.
\end{displaymath}

\section{Analysis}
\label{Analy}

\subsection{Principal Component Analysis: calculation and applications}
\label{pca_cal}

Principal Component Analysis (PCA) is a statistical method that finds the directions of maximum variation in the multidimensional space of the input data. These directions are the Principal Components (PCs), determined by a list of coefficients, that is, the coordinates of these directions in the input data space.

The vast majority of our stars have no emission lines. In the few RSGs and AGBs which present this effect, they are weak, with the line profiles dominated by the absorption components. Therefore, there are no stars in our sample with negative EW values because of emission. Such values come exclusively from lines that disappear almost completely because of the presence of molecular bands, but also when the atomic line is too weak to be measured (in most cases because the star has a spectral type too early to display it). So, we have assigned a value of 0 to all the negative EWs for the PCA calculation.

We performed the PCA through a bootstrapping process, as a way to insure that the results obtained would not be driven by the stochasticity of a particular sample. For this, we first took a random subsample of 500 stars from our whole sample of almost 900 stars. Then we performed the PCA for this subsample. This process was repeated $10\,000$ times. In this way, for each PC coefficient we obtained a distribution of values, taking its median as the final value, and having its standard deviation as a measure of its reliability.

In a first study we used all the EWs and bandheads detailed in Tables~\ref{atomic_ranges} and \ref{molecular_ranges}. We tried then a shortened line list, including only those lines whose identification is certain, finding that it leads to less noisy results. This reduced list contains all the lines of Ca\,{\sc{ii}}, Fe\,{\sc{i}}, Ti\,{\sc{i}}, Mn\,{\sc{i}}, Si\,{\sc{i}} and Mg\,{\sc{i}}, except those marked with "?" because we are not sure about their chemical species. We also removed  Fe\,{\sc{i}}~8621.5\:\AA{} and Ti\,{\sc{i}}~8623\:\AA{} and the VO bandhead at 8624.5\:\AA{}, because they may be affected by the Diffuse Interstellar Band (DIB) at $\sim8621$\:\AA{}, in an unpredictable way. The exact position of the DIB in the spectra depends on the relative radial velocities between the interstellar medium that generates the DIB and the observed target, and thus it may appear at different positions around $\sim8621$\:\AA{}. The depth of this DIB depends on the amount of extinction to the target, which is not a problem for our sample, but will be for other samples affected by moderate or high extinction. In total, our shortened list contains 29 atomic lines and three bandheads.

\begin{figure}[ht!]
   \centering
   \includegraphics[trim=1cm 0.5cm 0.5cm 1.2cm,clip,width=9cm]{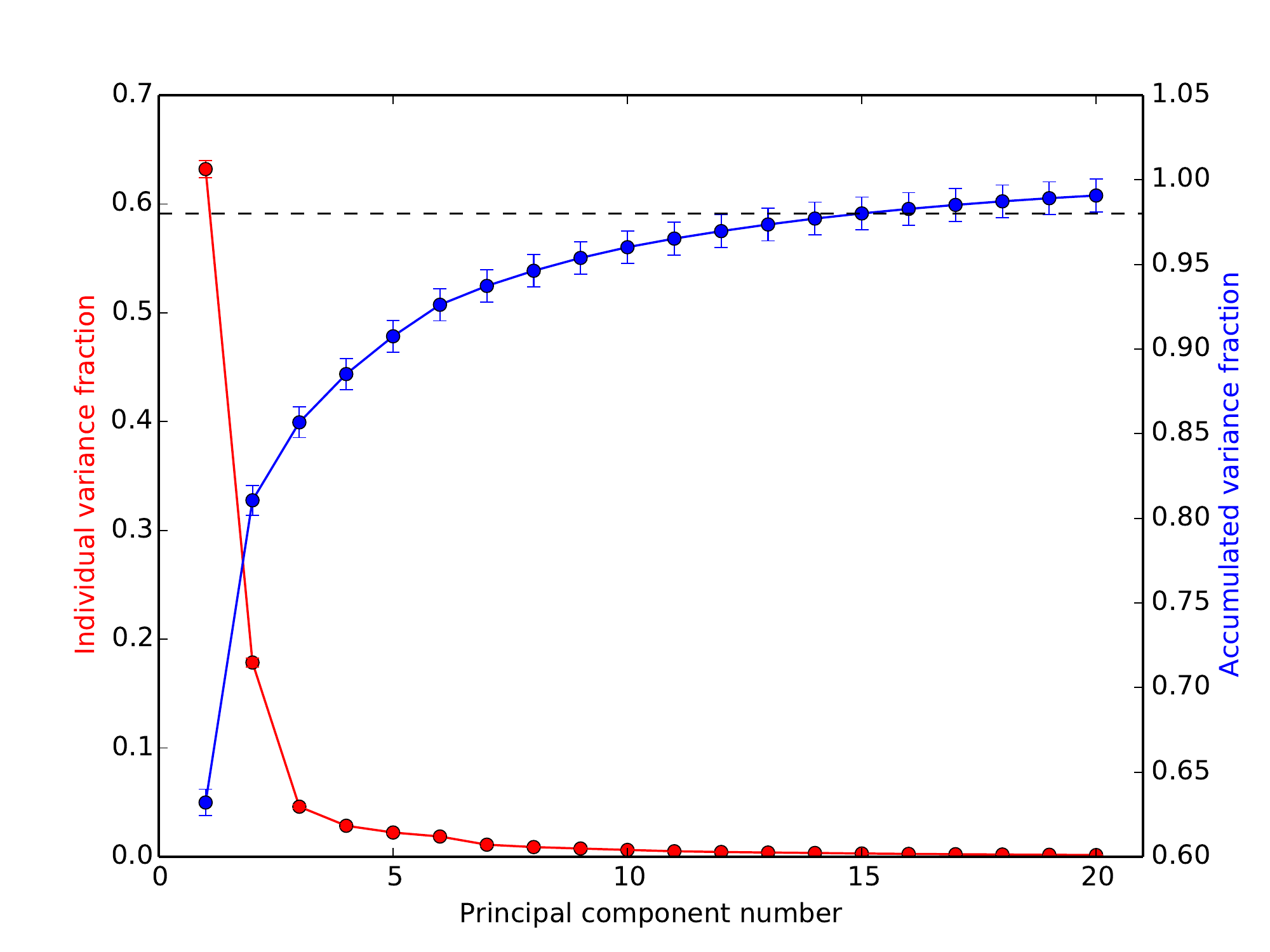}
   \caption{Summary of variance fractions of the principal components calculated from the shortened input list. The red circles are the individual variance (left vertical axis). The blue circles are the accumulated variance (right vertical axis). As the PCA calculations were done for $10\,000$ random samples, each circle is the median of the $10\,000$ variances obtained for each PC. The error bar in each point is its correspondent standard. The circles without error bars have errors smaller than the circle itself. Only the first 20 PCs are displayed here. The horizontal dashed line marks 98\% of the accumulated variance.}
   \label{variance_short}
\end{figure}

\begin{figure}[ht!]
   \centering
   \includegraphics[trim=1cm 0.5cm 2cm 1.2cm,clip,width=9cm]{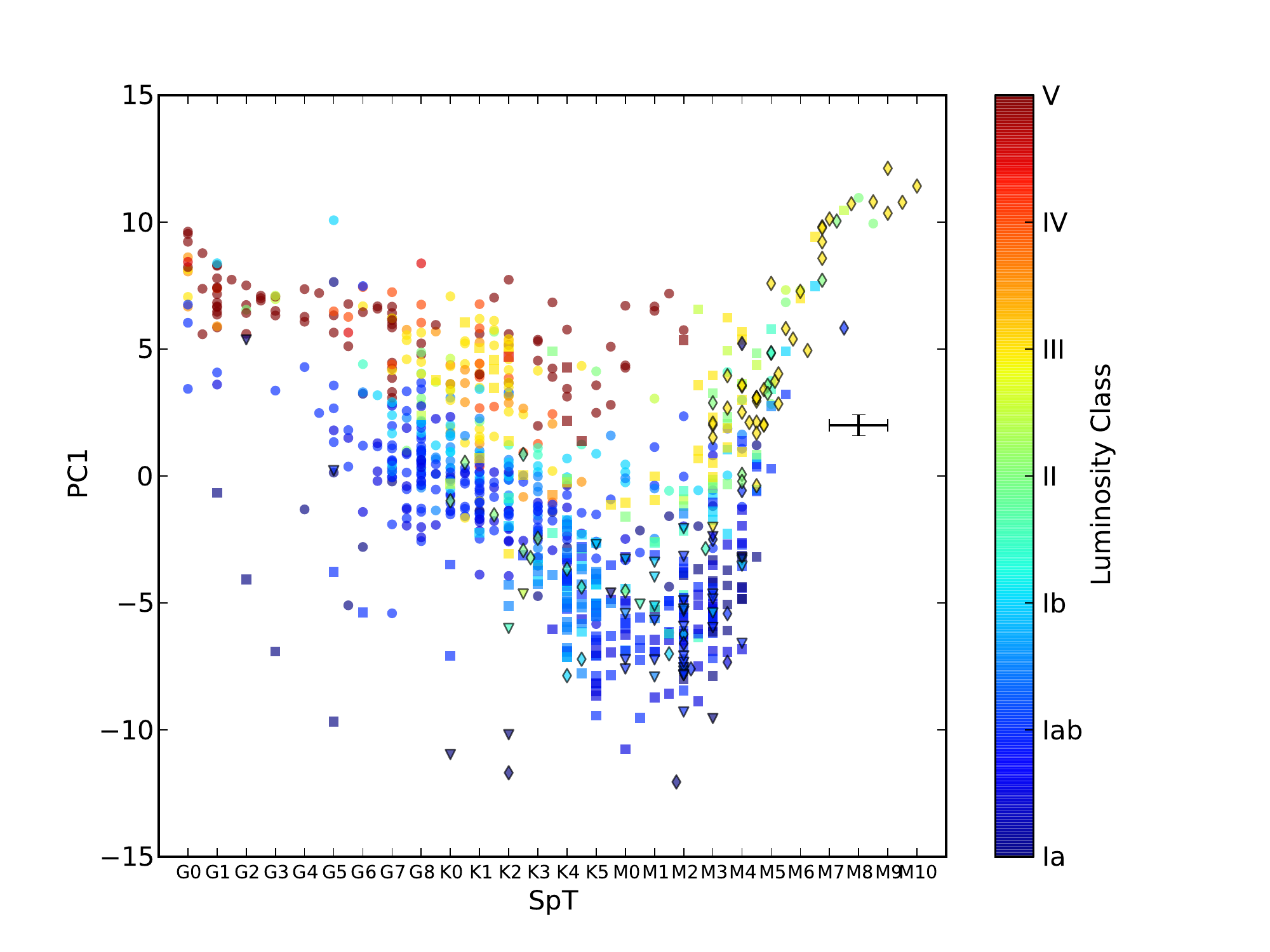}
   \caption{Spectral type against PC1 for all the stars in our sample. The colour indicates the luminosity class. The shape indicates their origin, circles are from the SMC survey, squares are from the LMC survey, diamonds are galactic standards, inverted triangles are from the Perseus arm survey. The error bars indicate the estimated uncertainty for SpT from GDN2015, and the median uncertainty for PC1, which has been calculated by propagating the uncertainties through the lineal combination of the input data (EWs and bandheads) to the PCA coefficients.}
   \label{PC1_spt}
\end{figure}

\begin{figure}[ht!]
   \centering
   \includegraphics[trim=1cm 0.5cm 2cm 1.2cm,clip,width=9cm]{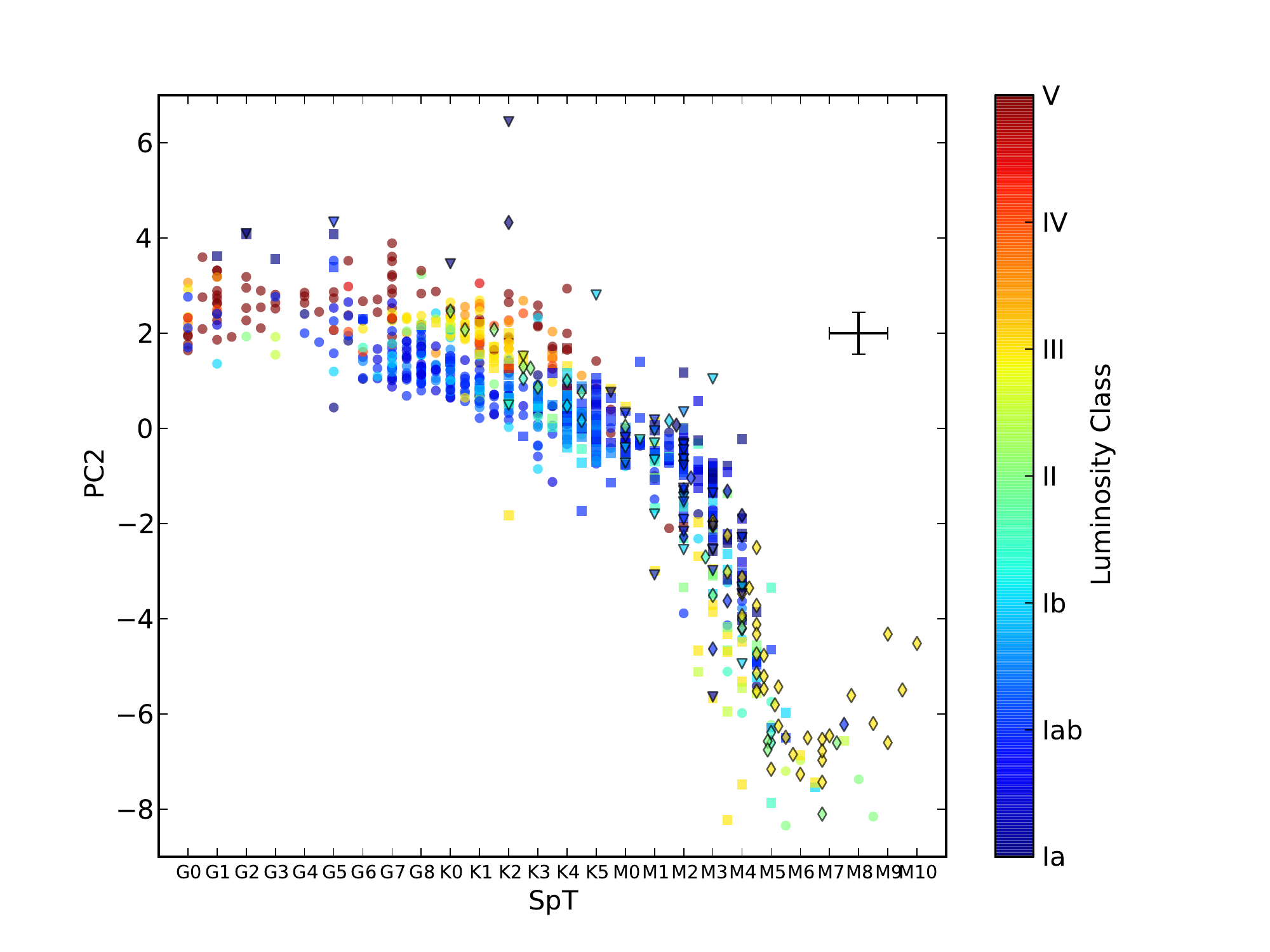}
   \caption{Spectral type against PC2. Symbols used are the same as in Fig.~\ref{PC1_spt}.}
   \label{PC2_spt}
\end{figure}

Because of the nature of the PCA, the first few PCs contain most of the variance, while later PCs contain progressively less, as shown in Fig.~\ref{variance_short}. That is why the standard deviation values for the coefficients of PC6 and later are significantly higher. In our PCA, the principal components PC1, PC2, and PC3 together contain more than 80\% of the accumulated variance. To reach 98\%, the first 15~PCs are necessary. Therefore we only show here the first three PCs, which display the clearest correlations with SpT and LC (see Figs.~\ref{PC1_spt} and~\ref{PC2_spt}). Tables~\ref{PC_reduc_a}, ~\ref{PC_reduc_b}, and ~\ref{PC_reduc_c} show the linear combinations through which PCs were calculated.

PC1 has a clear relation with both LC and SpT (see Fig.~\ref{PC1_spt}). For early SpTs, the SpT distribution of the SGs has a negative correlation with PC1 down to M2, where the minimum value of PC1 is reached. Non-SGs have a similar behaviour, forming a strip parallel to the SGs. From M2 toward later subtypes, both SGs and non-SGs are mixed in the same strip, which has a positive correlation between PC1 and SpT. The change in the behaviour of PC1 around M2 may be caused by the appearance of molecular bands from approximately M2 onward (see Sect.~\ref{tio_bandhead}). As these bands grow in depth, the atomic lines become progressively weaker due to the erasure of the continuum.

PC2 has an easier interpretation (see Fig.~\ref{PC2_spt}). It shows a very clear correlation with SpT for all SGs, and it seems not to be affected by metallicity, as the stars from different galaxies present the same behaviour. The separation between early SGs and non-SGs is less clear than in PC1, but still significant for SpT earlier than late K subtypes.

We can divide PC2 into three regions regarding its relation with SpT. The first is for subtypes earlier than approximately M2, with a negative slope. The second range goes from M2 to M7, with a more markedly negative slope. Finally, for stars M7 or later, which are poorly sampled, the slope seems to become positive. The two changes are caused by the behaviour of TiO bands. As with PC1, the change at M2 is because this is when the TiO bands first appear. The break at M7 is because the TiO band at $8432\:$\AA{} has become so strong that it is affecting the depth of the other bands, which probably are saturated: instead of growing, the bands become weaker for later SpTs (see Sect.~\ref{tio_bandhead}), while atomic lines have almost disappeared.

\subsection{Identifying supergiants}

\subsubsection{Principal components}
\label{sect_PCA}

In order to generate diagnostics, we have combined the three first PCs in diagrams, because each one contains significant information about the LC and SpT.

\begin{figure*}[ht!]
   \centering
   \includegraphics[trim=1cm 0.5cm 2cm 1.2cm,clip,width=9cm]{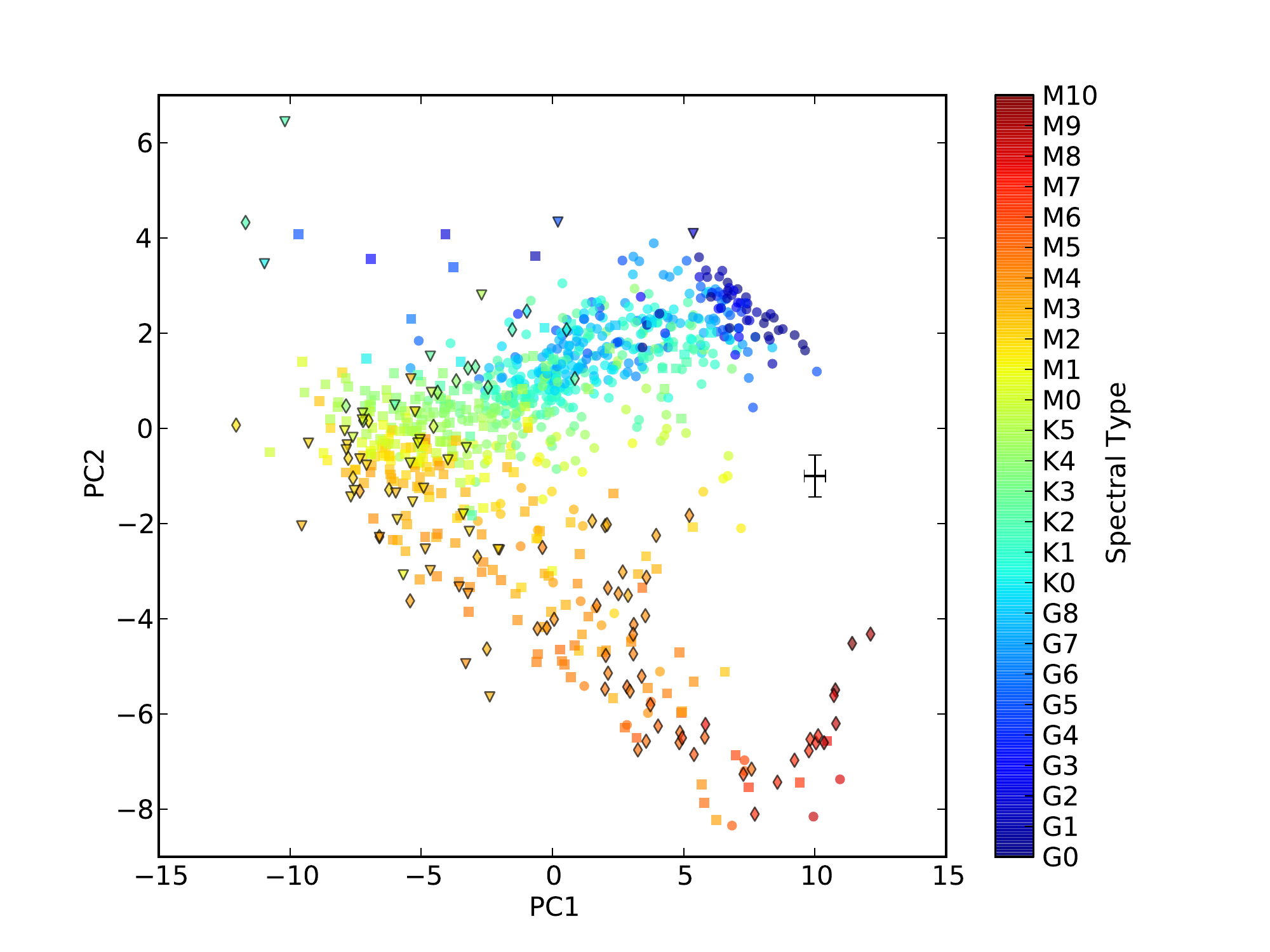}
   \includegraphics[trim=1cm 0.5cm 2cm 1.2cm,clip,width=9cm]{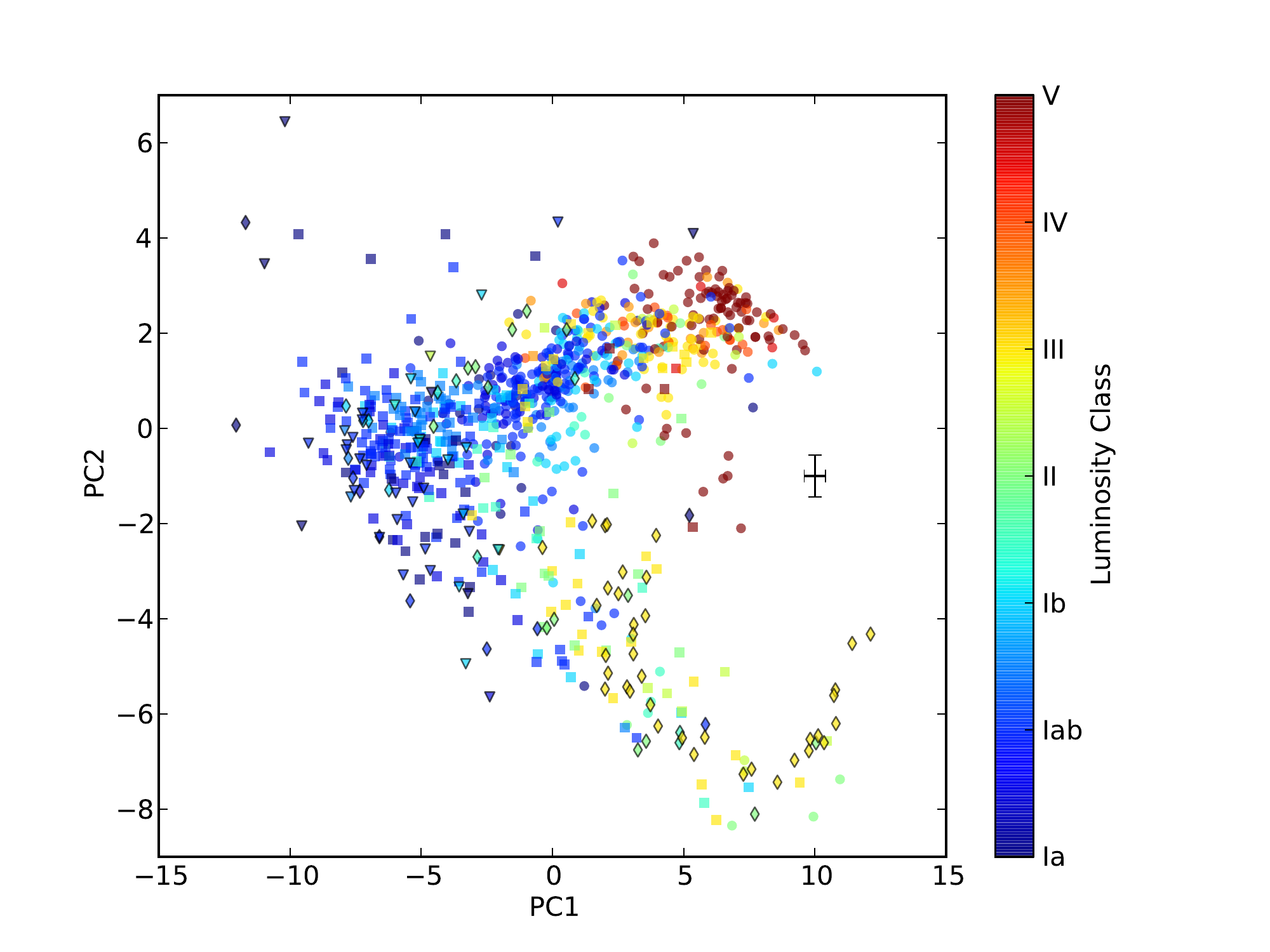}
   \caption{{\bf Left (\ref{PC1_PC2}a):} PC1 versus PC2 diagram. The colour indicates the SpT. The symbols used are the same as in Fig.~\ref{PC1_spt}. The cross indicates the median uncertainties, which have been calculated by propagating the uncertainties through the lineal combination of the input data (EWs and bandheads) with the coefficients calculated.
   {\bf Right (\ref{PC1_PC2}b):} the same as left figure, but here the colour indicates the luminosity class.}
   \label{PC1_PC2}
\end{figure*}

In Fig.~\ref{PC1_PC2} we show the (PC1, PC2) diagram. The data follow a clear sequence that depends mainly on SpT, because of the behaviours described in Sect.~\ref{pca_cal}. This diagram is very informative and it can give a rough estimate of the SpT and LC of a star, by simply looking at at the star's position on the diagram. In addition, we have to note that all the stars follow the same trends regardless of their galactic origin (i.e. their metallicity).

\begin{figure*}[ht!]
   \centering
   \includegraphics[trim=1cm 0.5cm 2.1cm 1.2cm,clip,width=9cm]{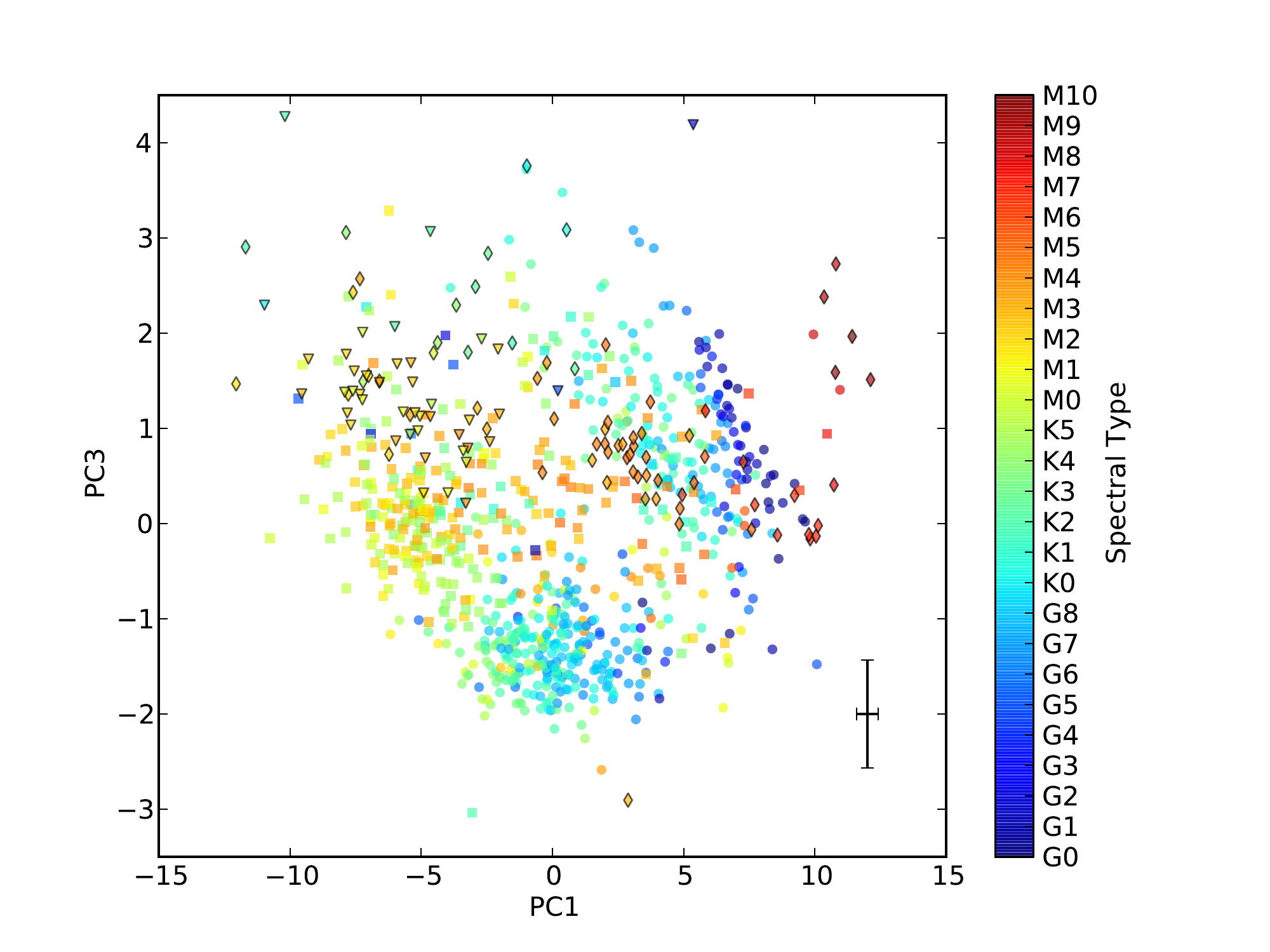}
   \includegraphics[trim=1cm 0.5cm 2.1cm 1.2cm,clip,width=9cm]{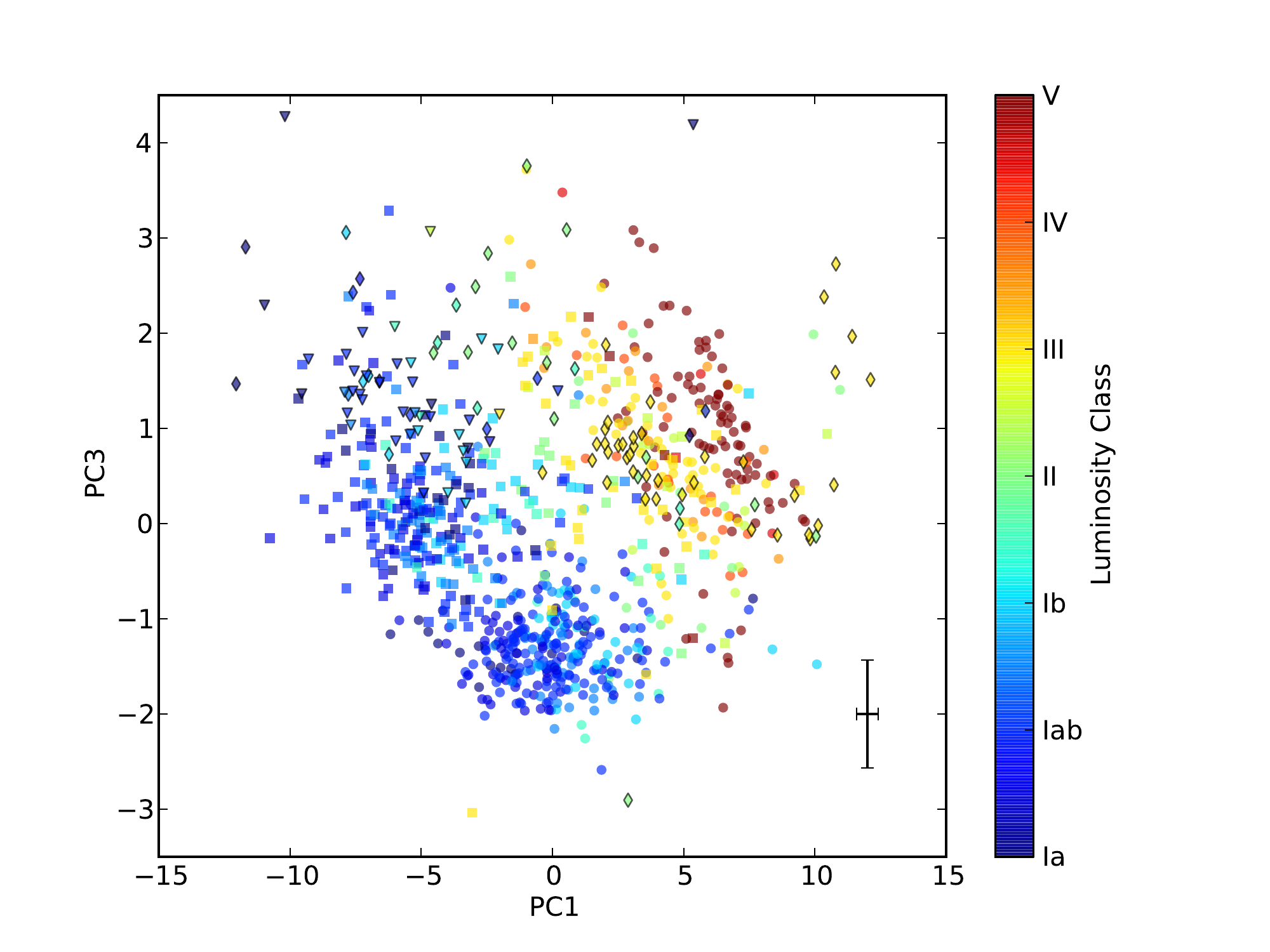}
   \caption{{\bf Left (\ref{PC1_PC3}a):} PC1 versus PC3 diagram. The colour indicates the SpT. The symbols used are the same as in Fig.~\ref{PC1_spt}. The cross indicates the median uncertainties calculated as explained in Fig.~\ref{PC1_PC2}.
   {\bf Right (\ref{PC1_PC3}b):} the same as left figure, but here the colour indicates the luminosity class.}
   \label{PC1_PC3}
\end{figure*}

The (PC1, PC3) diagrams (Fig.~\ref{PC1_PC3}) provide a very good way to separate the SGs from the non-SGs (see Fig.~\ref{PC1_PC3}), by simply tracing a line between both groups. If we apply this criterion, we find that almost all the stars that are not classified correctly have SpTs later than approximately M3 (see Fig.~\ref{PC1_PC3}, panel b). This is because the mid and late M~type SGs tend to be in the same areas of the diagram as the non-SGs. Therefore, the number of wrongly classified stars can be reduced by first using the (PC1, PC2) diagram to flag out these late stars.

In conclusion, these diagrams provide a quick and easy way to obtain a preliminary SpT and LC classification, especially for stars earlier than M3, but they do not represent a very accurate method.


\subsubsection{Support vector machine classification}
\label{SVM}

To calculate the optimal boundary between different groups of stars in the 15-dimensional space of the PCs, we used the Support Vector Machines (SVM) method, through the python package \textit{sklearn} \citep{scikit-learn}. This method requires a sample of stars labelled as belonging to each different group, to train the SVM. As we decided to use a linear version of the SVM, its output result is a vector of coefficients, which in linear combination with the 15~PCs produces a hyper-plane (in 15 dimensions) as a boundary between the two different groups labelled. We repeated the SVM through a bootstrapping process $10\,000$ times, each time giving as input a randomly selected sample with a size $N$ depending (about 50\% of the size) on the corresponding input sample.

As the mid and late M type stars have a different behaviour from the rest of the sample, they are the main source of SG misidentification. In order to minimise these errors, we split our sample through the SVM process into two sets, according to whether the stars are earlier or later than a given subtype (K5, M0, M1, M2, and M3), and then, separate the SGs from the non-SGs within each of these sets, using again the SVM method. This system requires us to determine which is the optimal M~subtype boundary that minimises the wrong identifications. We could not use subtypes later than M3 as the boundary because the late subsample then becomes too small to perform any meaningful analysis. The calculated coefficients that separate the indicated groups are given in Tables~\ref{SVM_M_red}, \ref{SVM_SGearly_red}, and \ref{SVM_SGlate_red}.

\subsection{Fitting spectral types}
\label{estimation}

The spectral differences between adjacent subtypes are too subtle to allow separation through the SVM method, because the dispersion in values of most spectral features within a given subtype is significantly high (see figures in Sect.~\ref{classical}). In view of this, we decided to complement the PC analysis with studies of the behaviour of the individual spectral features that show the stronger trends with SpT, and performed linear fits to their variation with SpT. We also performed linear fits to the behaviour of the PC2 with SpT. The linear regressions were performed using the Theil-Sen estimator (which is based on a median of the slope distribution obtained through random pairings of the sample) to avoid the effect of the outliers.

As with the PCs, most spectral features change their behaviour as we move to later SpTs, because of the rise and saturation of the TiO bands. Therefore, we divided the sample into two different subsamples, one composed of those stars considered early, and the other by those considered late, because we expected different behaviours for each one of them. The subsamples were obtained by using the hyper-plane calculated with the SVM, taking as boundary the M1 subtype (because the main TiO bands in the CaT spectral range become measurable at this subtype). Afterwards, we followed a different procedure for each group.

\subsubsection{Early subsample}

\begin{figure}[th!]
   \centering
   \includegraphics[trim=0.8cm 0.5cm 2.5cm 1.2cm,clip,width=9cm]{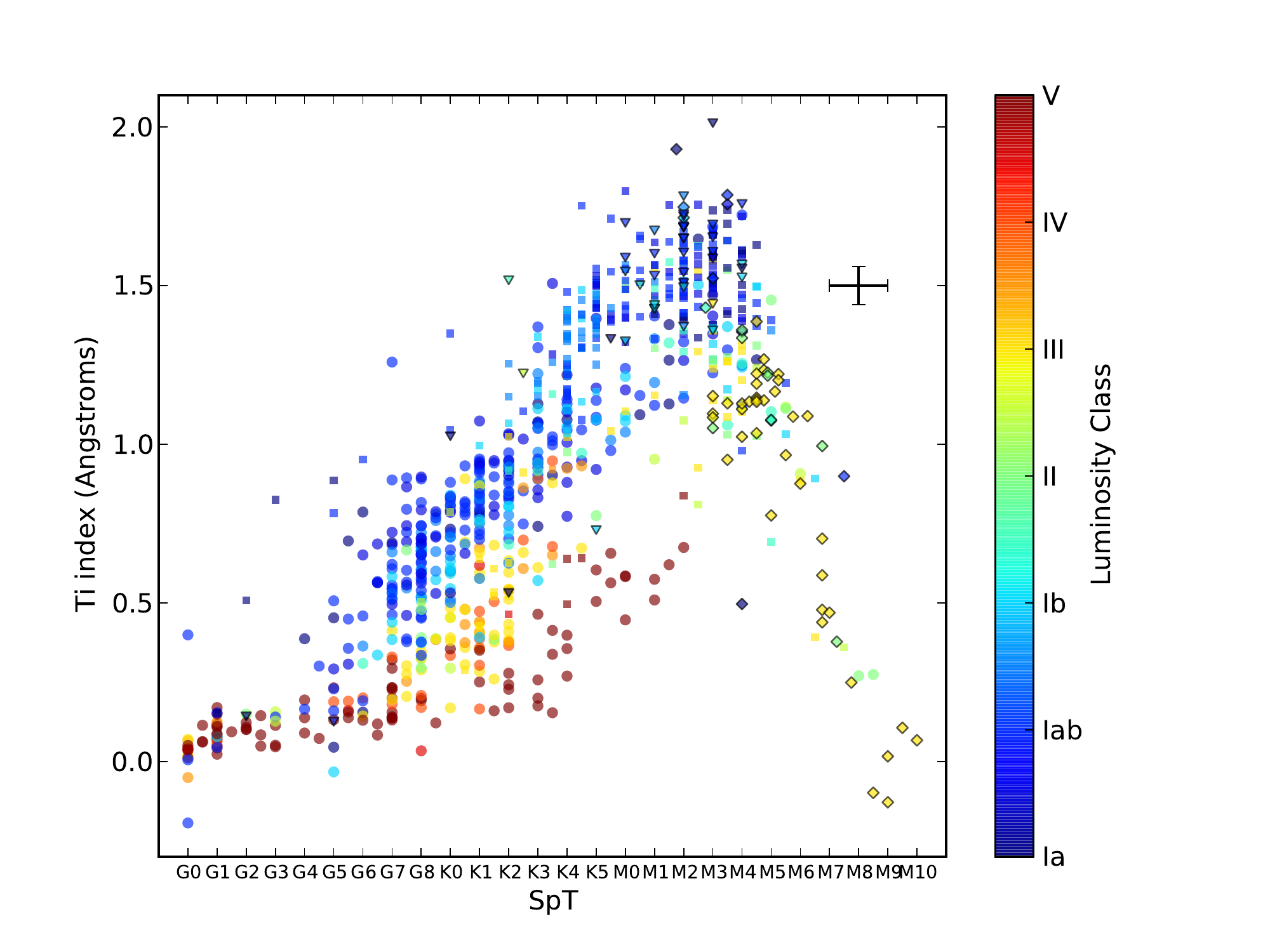}
   \caption{Spectral type as a function of the sum of the equivalent widths of the Ti\,{\sc{i}} lines. The symbols are the same as in Fig.~\ref{PC1_spt}. The error bars indicate the median uncertainty for the EW, and the estimated uncertainty for SpT from GDN2015.}
   \label{SpT_Ti}
\end{figure}

For the early subsample, we identified two indicators that correlate well with SpT. Firstly, the strength of Ti\,{\sc{i}} lines (at $8518.1$, $8683.0$, $8692.0$, $8730.5$, and~$8734.5$\:\AA{}), as seen in Fig.~\ref{SpT_Ti}), most likely because they are very sensitive to temperature \citep{dor2016}. Thus, we have added together their EWs in a single value that we name "Ti index". Secondly, PC2 shows a clear trend with SpT for early-type SGs. In both cases we have considered only the SGs as selected by the PCA-SVM method.

As the Ti index depends on the metallicity of the host galaxy, we have calculated two different fits, one for SGs from the SMC and the other for those of the LMC and the Galaxy. We have considered these together because the number of SGs earlier than M1 in these galaxies is low, and the effects on the observed spectra of the difference in metallicity between these two galaxies are small (see Fig.~\ref{sec_gal}). The coefficients of the lines calculated are given in Table~\ref{theil-sen_red}.

\subsubsection{Late subsample}

The late subsample was split again to remove the latest M stars, because their behaviour is different due to the saturation of the TiO bandheads. As their number is low, we were not able to use the SVM method. In its place, we separated them by simple cuts in the (PC1, PC2) diagram (see Fig.~\ref{PC1_PC2}a): PC1$> 8$ and PC2 $<-3$. As they are very few (only 16, and all non-SGs), no reliable analysis was possible for these very late-M stars. This left us with a sample of stars with spectral types between M1 and approximately M7.

Again, we found two indicators that can be used for the linear regression: PC2 (see Fig.~\ref{PC2_spt}), and the TiO bandhead at $8859$\:\AA{} (see Fig.~\ref{SpT_TiO}). Both SGs and non-SGs present the same behaviour, and we used both SGs and non-SGs for these calculations.

In both cases (PC2 and TiO bandhead), the data distribution suggests that the regression should have a higher order than a simple linear fit. Therefore, we tested polynomial fits with order two and three for both variables in addition to lines calculated through Theil-Sen. These fits naturally reproduce the data better, yet the method to avoid overfitting used by \cite{ram2006} shows that, given the uncertainties, these higher-order polynomials are not better fits than a first order polynomial. Therefore, in Table~\ref{theil-sen_red} we show the coefficients for the linear fits only.

\section{Results}
In Sect.~\ref{Analy}, we developed a method to separate SGs from non-SGs using the PCA and SVM algorithms, and calculated linear fits that can automatically indicate the SpT of cool supergiants. In this section, we discuss the efficiency of the methods proposed and compare them with traditional spectroscopic criteria.

\subsection{Efficiency in finding supergiants}
\label{effi}

We have developed a method to separate SGs from non-SGs using PCA and SVM algorithms, and calculated five sets of results. Each of these sets corresponds to the five different putative boundaries (at K5, M0, M1, M2, and M3; see Sect.~\ref{SVM}) selected to separate the stars into the early and late subsamples. The results of this process are shown in Table~\ref{efficiency_a}. We define the efficiency as the fraction of known SGs in the sample or subsample that are tagged as SGs by a given criterion. The contamination is the fraction of stars identified as SGs that in fact are not. As efficiency and contamination are fractions, their two-sigma uncertainties are equal to $1/\sqrt[]{n}$, where $n$  is the total number of SGs in the correspondingly sample.

The efficiency for the early subsample is extremely high: $0.99$ for all the five boundaries, with uncertainties between $\pm0.04$ and $\pm0.06$ depending on the boundary. The efficiency is also high for the late subsample, with values between $0.89$ and $0.97$. The uncertainties of these fractions depend (inversely) only on the number of objects forming the group considered. Because the number of objects with late subtypes is lower, the uncertainties for the late subgroup are higher. As a result, the difference between the efficiencies for the early and the late subsamples are not statistically significant. In fact, the efficiency for the whole sample is always $\sim0.98\pm0.04$ for any of the five putative boundaries studied.

We have also evaluated the contamination in the five sets. For the whole sample the contamination is very low ($0.02\pm0.04$). As in the case of the efficiency, the differences between the contaminations for both subsamples with all the boundaries used are not statistically significant.

These results show that the methods proposed are very effective, especially if we consider that they were obtained for a sample with a wide range of SpTs coming from three different galaxies. For completeness, we also checked if there are significant variations in the contamination and the efficiency for a mono-galactic sample. As we are planning to apply this method to a large Galactic sample, we measured the efficiency and contamination of our method using a sample composed only of stars from the Milky Way (i.e. our MK standards and stars from the Perseus arm). The results are shown in Table~\ref{efficiency_b}. Unfortunately, as this sample is smaller than the full sample, the uncertainties of the fractions are higher than those obtained for that sample. Thus, even though we find lower values for the efficiency ($0.90\pm0.13$) in this case than for the full sample, the difference cannot be considered significant. The contamination obtained ($0.03\pm0.13$) is also compatible within the uncertainties to that obtained for the full sample.

\begin{table}[th!]
\caption{Supergiant identification efficiency and contamination, and their errors, for our whole sample, depending on the  putative boundary used.}
\label{efficiency_a}
\centering
\begin{tabular}{c | c | c | c }
\hline\hline
\noalign{\smallskip}
Boundary&Subsample&Efficiency&Contamination\\
\noalign{\smallskip}
\hline
\noalign{\smallskip}
&Early&$0.99\pm0.06$&$0.00\pm0.06$\\
K5&Late&$0.97\pm0.06$&$0.04\pm0.06$\\
&All&$0.98\pm0.04$&$0.02\pm0.04$\\
\noalign{\smallskip}
\hline
\noalign{\smallskip}
&Early&$0.99\pm0.05$&$0.00\pm0.05$\\
M0&Late&$0.97\pm0.07$&$0.05\pm0.07$\\
&All&$0.98\pm0.04$&$0.02\pm0.04$\\
\noalign{\smallskip}
\hline
\noalign{\smallskip}
&Early&$0.99\pm0.05$&$0.00\pm0.05$\\
M1&Late&$0.95\pm0.08$&$0.05\pm0.08$\\
&All&$0.98\pm0.04$&$0.02\pm0.04$\\
\noalign{\smallskip}
\hline
\noalign{\smallskip}
&Early&$0.99\pm0.05$&$0.01\pm0.05$\\
M2&Late&$0.96\pm0.09$&$0.05\pm0.09$\\
&All&$0.98\pm0.04$&$0.02\pm0.04$\\
\noalign{\smallskip}
\hline
\noalign{\smallskip}
&Early&$0.99\pm0.04$&$0.01\pm0.04$\\
M3&Late&$0.89\pm0.11$&$0.07\pm0.12$\\
&All&$0.98\pm0.04$&$0.02\pm0.04$\\
\noalign{\smallskip}
\hline
\end{tabular}
\end{table}

\begin{table}[th!]
\caption{Supergiant identification efficiency and contamination, and their errors, for the galactic sample, depending on the  putative boundary used.}
\label{efficiency_b}
\centering
\begin{tabular}{c | c | c | c }
\hline\hline
\noalign{\smallskip}
Boundary&Subsample&Efficiency&Contamination\\
\noalign{\smallskip}
\hline
\noalign{\smallskip}
&Early&$1.00\pm0.30$&$0.00\pm0.30$\\
K5&Late&$0.92\pm0.14$&$0.04\pm0.14$\\
&All&$0.94\pm0.13$&$0.03\pm0.13$\\
\noalign{\smallskip}
\hline
\noalign{\smallskip}
&Early&$1.00\pm0.27$&$0.00\pm0.27$\\
M0&Late&$0.92\pm0.14$&$0.04\pm0.15$\\
&All&$0.94\pm0.13$&$0.03\pm0.13$\\
\noalign{\smallskip}
\hline
\noalign{\smallskip}
&Early&$0.95\pm0.22$&$0.05\pm0.22$\\
M1&Late&$0.91\pm0.15$&$0.03\pm0.16$\\
&All&$0.92\pm0.13$&$0.03\pm0.13$\\
\noalign{\smallskip}
\hline
\noalign{\smallskip}
&Early&$0.96\pm0.20$&$0.00\pm0.20$\\
M2&Late&$0.89\pm0.16$&$0.03\pm0.17$\\
&All&$0.92\pm0.13$&$0.02\pm0.13$\\
\noalign{\smallskip}
\hline
\noalign{\smallskip}
&Early&$0.97\pm0.16$&$0.03\pm0.16$\\
M3&Late&$0.78\pm0.21$&$0.05\pm0.23$\\
&All&$0.90\pm0.13$&$0.03\pm0.13$\\
\noalign{\smallskip}
\hline
\end{tabular}
\end{table}

\subsection{Spectral subtype estimation}

For each of the regressions discussed in Sect.~\ref{estimation}, we have calculated the distribution of the errors (i.e. the difference between the real value of the SpT and the value predicted by the regression). As the data have some outliers, the standard deviation was done through the median absolute deviation, a robust estimator. The results for each regression are shown in Table~\ref{theil-sen_red}.

We divided the sample into two subsamples, early and late, before carrying out the fits. These divisions were made by applying the SVM coefficients obtained in the previous section, and present a small degree of contamination. Therefore, we show each linear regression covering all the subtypes in the subsample used, even if they are beyond the limit of the split. When this method is applied to a problem sample, the same may happen: a few stars will be in the wrong subsample, and the behaviour of such stars affects the linear regressions calculated. Thus, despite the "split", each regression may give the classification of some points above or below the limit of the split. 

In addition, there are a few very luminous stars that are outliers in the PC2 diagrams. The linear regression gives these stars SpTs earlier than G0, while they are really G or K hypergiants (or very luminous supergiants). We recommend the use of the Ti index instead of the PC2 for them. To identify these stars, the CaT can be used because these extremely high-luminosity stars have very high values of the CaT (see Sect.~\ref{sec_CaT}). 

The SpTs of the late subsample are strongly correlated with both variables considered, the classical indicator (the strength of the TiO bandhead at $8859$\:\AA{}) and the PC2. The results prove that, for a quick spectral identification, PC2 is as good as the TiO bandhead. The standard deviations of these regressions are 0.7 subtypes, and so this classification is only slightly worse than the classical classification process by visual inspection of the whole spectrum (0.5 subtypes). The Ti~index cannot be used effectively with the late subsample, because at the typical SpTs the Ti lines are heavily affected by molecular bands.

For the early sample, the fit of the SpT to the PC2 has a standard deviation of 1.7 subtypes, significantly larger than for the late subsample. The Ti index provides similar results, but it is affected by metallicity, and thus the calculated fits can only be used for populations with similar metallicities to the SMC or the Galaxy/LMC. Thus we obtained two different linear regressions, one for stars from the SMC and another one for those from the LMC. The results are better than for the PC2 linear regression, but the standard deviation is still significantly higher than for the late subsample. Thus, these regressions are merely a quick estimator. A detailed classification of the stars from the early subsample can only be performed using the classical criteria, as they are not affected by the appearance of TiO bands. 

\begin{table*}[th!]
\caption{Results from the linear regressions done between different variables and spectral subtypes.}
\label{theil-sen_red}
\centering
\begin{tabular}{c | c | c | c | c}
\hline\hline
\noalign{\smallskip}
Variable used&Subsample&Slope&y-intercept (subtypes)&$\sigma(y_{\textrm{real}}-y_{\textrm{expec}})$ (subtypes)\\
\noalign{\smallskip}
\hline
\noalign{\smallskip}
PC2\tablefootmark{a}&Early&-3.5&13.4&1.7\\
Ti index\tablefootmark{b}&Early - only SMC&9.3&2.4&1.4\\
Ti index\tablefootmark{b}&Early - Galaxy and LMC&5.8&5.8&1.2\\
PC2\tablefootmark{a}&Late&-0.55&16.7&0.7\\
TiO \tablefootmark{a} $8859$\:\AA{}&Late&9.4&15.9&0.7\\
\noalign{\smallskip}
\hline
\end{tabular}
\tablefoot{For details about these calculations see section~\ref{estimation}.\\
\tablefoottext{a}{Dimensionless}\\
\tablefoottext{b}{Expressed in \AA{}}
}
\end{table*}

\subsection{Classical criteria revisited}
\label{classical}

The two main features used to identify SGs are the blend of Ti\,{\sc{i}}, Fe\,{\sc{i}}, and CN molecular bands at $8468$\AA{}, and the infared CaT (formed by the lines at $8498$, $8542$, and $8662\:$\AA{}). \cite{gin1994} show the typical EW values they obtained for these features, but their SG sample is small and only with SpTs earlier than K3. \cite{neg2011} also show both values for a small number of standards with SpTs from K0 to  M5. However, the number of SGs in their sample is limited. Therefore we have decided to take advantage of our very large sample, and we show here the behaviour of both spectral features, in Figures~\ref{CaT} and~\ref{EW8468}.

We also study one of the most useful ratios in this region, Fe\,{\sc{i}}~$8514\:$\AA{} to Ti\,{\sc{i}}~$8518\:$\AA{}, which was proposed as a luminosity indicator by \cite{kee1945}, and the behaviour of the TiO bandhead at $8859$\:\AA{} because this is the main TiO band in the region and has been widely used to obtain the SpT in the M sequence \citep[e.g.][]{ram1981,neg2011}.

\subsubsection{Calcium Triplet}
\label{sec_CaT}

The CaT depends strongly on luminosity \cite[e.g.][and references therein]{dia1989}, and it has been used extensively for luminosity classification \citep[e.g.][]{gin1994,car1997,neg2011}. However, our values are systematically lower than those of these researchers' \citep[up to $\sim2$\,--\,$3$\:\AA{} with respect to those obtained by][]{gin1994}. This is due to two reasons. Firstly, the continuum is determined differently. The method used by \cite{gin1994} results in values of EW higher than in any other work. Secondly, for the spectral range that we have in common with \cite{gin1994}, almost all our stars come from the MCs, while all their objects come from the Milky Way. Because of the lower metallicity in the Clouds, our values are systematically lower.

\begin{figure}[th!]
   \centering
   \includegraphics[trim=0.8cm 0.5cm 2.5cm 1.2cm,clip,width=9cm]{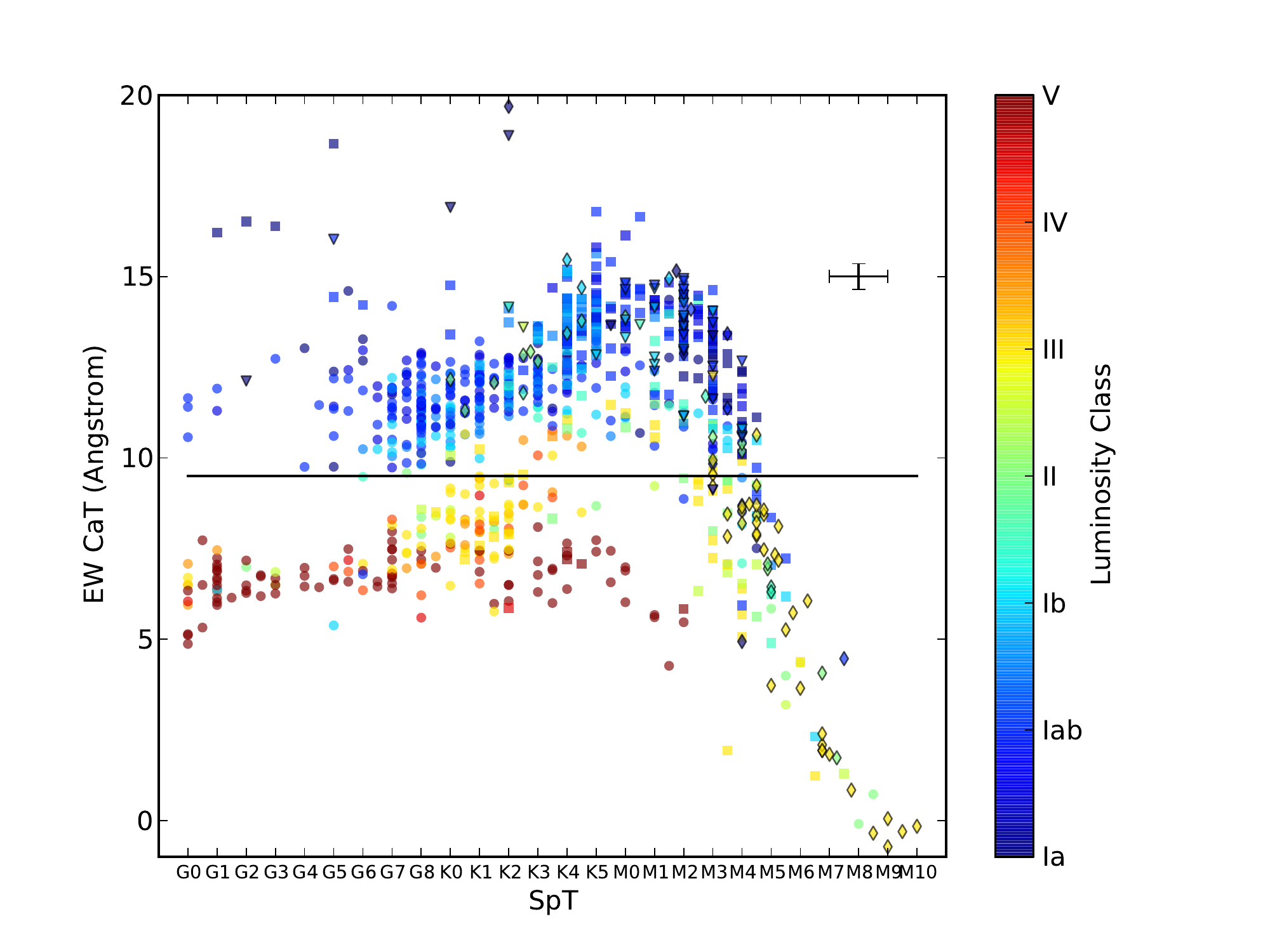}
   \caption{Spectral type versus total equivalent width of the Calcium Triplet ($8498$\:\AA{}, $8542$\:\AA{}, and $8662\:$\AA{}). The shapes are the same as in Fig.~\ref{PC1_spt}. The error bars indicate the median uncertainty for the sum of EWs, and the estimated uncertainty for SpT from GDN2015. The black line marks $EW=9\:$\AA{}, as this value separates optimally the SGs from the non-SGs.}
   \label{CaT}
\end{figure}

The EW of the CaT depends also on SpT, as was shown by \cite{gin1994}. We find the same behaviour here (see Fig.~\ref{CaT}), with the EW of the CaT growing from early-G subtypes, until it stops increasing around spectral type M0 due to line saturation at low temperatures. After this, it starts to decrease with SpT, due to the appearance of TiO bands, which quickly affect the continuum, erasing the wings of the CaT. In consequence, it becomes useless as a luminosity criterion for subtypes later than approximately M3.

In some works \citep{gin1994,car1997}, the sum of the EWs of the three Ca\,{\sc{ii}} lines were used, while in others \citep{dia1989,neg2011} only the two strongest lines (Ca\,{\sc{ii}}~8542\:\AA{} and Ca\,{\sc{ii}}~8662\:\AA{}) were considered. In both cases, in order to used the CaT as an LC discriminator, metallicity and the SpT range must be carefully considered, as the CaT changes noticeably with these two variables. We note that metallicity not only affects the strength of these lines, but also determines the typical SpT range of cool SGs. If we investigate the EW of the CaT for stars presenting a wide range of metallicities and SpTs, the separation between SGs and non-SGs seems to be around $\sim9$\:\AA{} (Fig.~\ref{CaT}), with all SGs except some of those later than M3 having CaT with values higher than this. With this boundary, the efficiency is $0.95\pm0.04$, and the contamination is $0.07\pm0.04$ (there are some bright and/or high-metallicity giants with CaT values as high as $11$\:\AA{}). While efficiency is statistically similar to that obtained with the PCA method (see Sect.~\ref{effi}), the contamination is slightly, but still significantly, higher. However, the simplicity of this criterion makes it a good tool for a quick and preliminary classification.

\begin{figure}[th!]
   \centering
   \includegraphics[trim=0.8cm 0.5cm 2.5cm 1.2cm,clip,width=9cm]{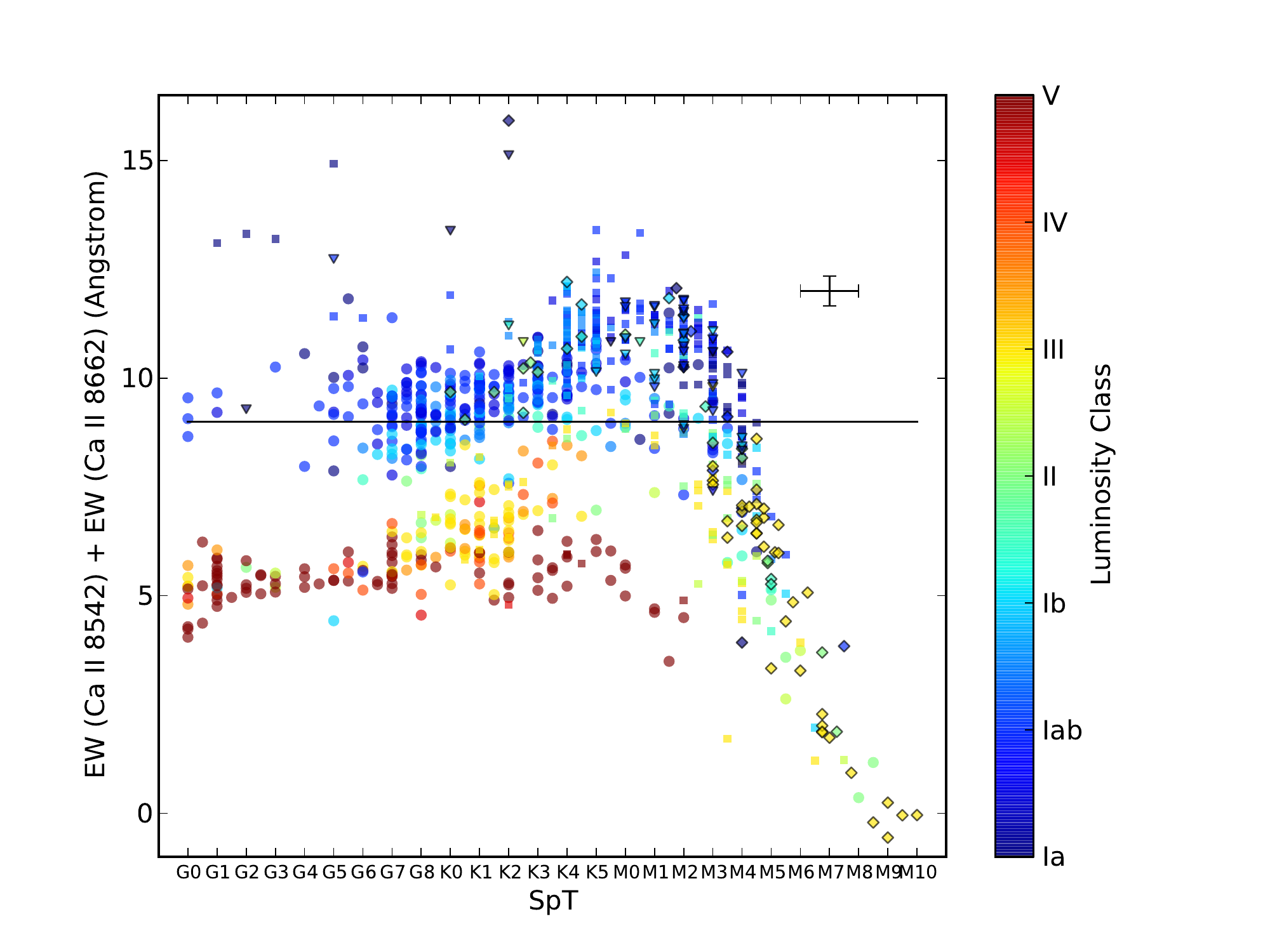}
   \caption{Spectral type as a function of the equivalent width of the two strongest lines of the Calcium Triplet ($8542$\:\AA{} and $8662$\:\AA{}). The shapes are the same as in Fig.~\ref{PC1_spt}. The error bars indicate the median uncertainty for the sum of EWs, and the estimated uncertainty for SpT from GDN2015. The black line marks $EW=9\:$\AA{}, the value proposed by \cite{neg2011} as boundary between RSGs and non-SGs at approximately solar metallicity (see text).}
   \label{CaT2}
\end{figure}

\citet{neg2011} proposed a boundary value of EW$=9\:$\AA{} for the sum of the two strongest lines of the CaT for stars from our Galaxy (at the typical metallicities and SpTs). We compare this boundary with our sample in Fig.~\ref{CaT2}. This criterion keeps the contamination at similar value as the PCA/SVM method ($0.02\pm0.05$), but with a much worse efficiency, only $0.79\pm0.04$.

Finally, we checked those stars whose CaT values are much higher than typical for their SpTs. These extreme values are not caused by the measurement method, but are genuine physical features. Most of the stars with very high values are classified as Ia, but there is also one star (RW~Cep) that is a known hypergiant (0\,--\,Ia). Thus, their extraordinarily high CaT values are likely due to their extreme luminosities alone. Moreover, most of these stars have G or early-K subtypes, significantly earlier than those usual for their galaxy. Therefore, this group seems to be composed of yellow hypergiants, whose properties are expected to be different from those of the typical RSGs. We will explore the behaviour of these stars in future works.

\subsubsection{Blend at $8468\:$\AA{}}

The spectral feature at $8468$\AA{} is a blend of many lines (mainly Ti\,{\sc{i}} and Fe\,{\sc{i}}) and multiple molecular bandheads of CN \citep{gin1994,car1997}. It has been used as a luminosity indicator by \cite{gin1994} and \cite{neg2011}, but with very limited samples.

\begin{figure}[th!]
   \centering
   \includegraphics[trim=0.8cm 0.5cm 2.5cm 1.2cm,clip,width=9cm]{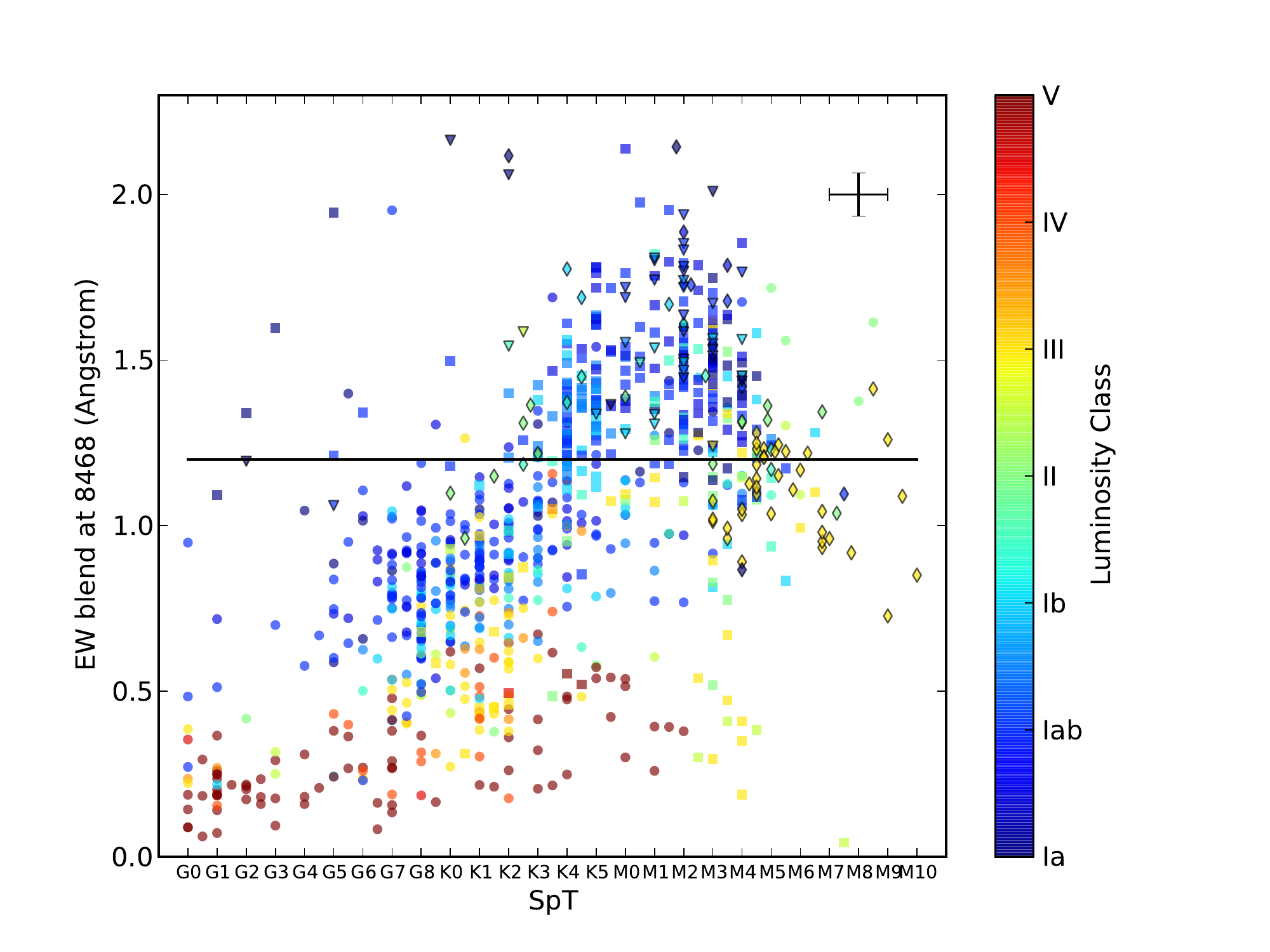}
   \caption{Spectral type to the equivalent width of $8468$\:\AA{} blend. The shapes are the same as in Fig.~\ref{PC1_spt}. The error bars indicate the median uncertainty for the EW, and the estimated uncertainty for SpT from GDN2015. The black line marks $EW=1.2$\:\AA{}, because this value has been classically used as the boundary between SGs and non-SGs (see text).}
   \label{EW8468}
\end{figure}

In \cite{gin1994} the blend shows a strong dependence with SpT from G0 to early-K, both for giant and SG stars, that becomes less pronounced for K stars. The blend in \cite{neg2011} shows the same trend for K subtypes, and then becomes flat for early-M subtypes, surely because of the effect of TiO bands. The behaviour we show here (Fig.~\ref{EW8468}) is consistent with these two studies. However, our values are systematically lower than those from \cite{gin1994}, as was the case for the CaT, and for the same reasons. 

In \cite{neg2011}, the value of $EW=1.2$\:\AA{} was used as a boundary between giants and SGs for Milky Way stars. As we show, this value seems appropriate for subtypes between mid-K and early-M only. However, this criterion should be carefully reconsidered for stars from low-metallicity galaxies, because their atomic features are weaker and also because the SG-typical SpTs are earlier.

Using the EW value as discriminator has a global efficiency of $0.46\pm0.04$ and a contamination of $0.13\pm0.06$. If we only consider those stars that are K4 or later, the contamination does not change ($0.13\pm0.06$), but the efficiency becomes higher ($0.76\pm0.06$). However, these values are still clearly worse than utilising the CaT or our PCA method (see Sect.~\ref{effi} and  Sect.~\ref{sec_CaT}).

\subsubsection{Fe\,{\sc{i}}~$8514\:$\AA{} and Ti\,{\sc{i}}~$8518\:$\AA{}}

The ratio between the Fe\,{\sc{i}}~$8514\:$\AA{} and Ti\,{\sc{i}}~$8518\:$\AA{} lines was proposed as a criterion for LC classification by \cite{kee1945}, and it has been used with good results by \cite{neg2012} for K and early-M stars. However, it has not been properly characterised with a statistically significant sample. Threshold values of the ratio have not been given before, either.

\begin{figure*}[th!]
   \centering
   \includegraphics[trim=0.8cm 0.5cm 2cm 1.2cm,clip,width=9cm]{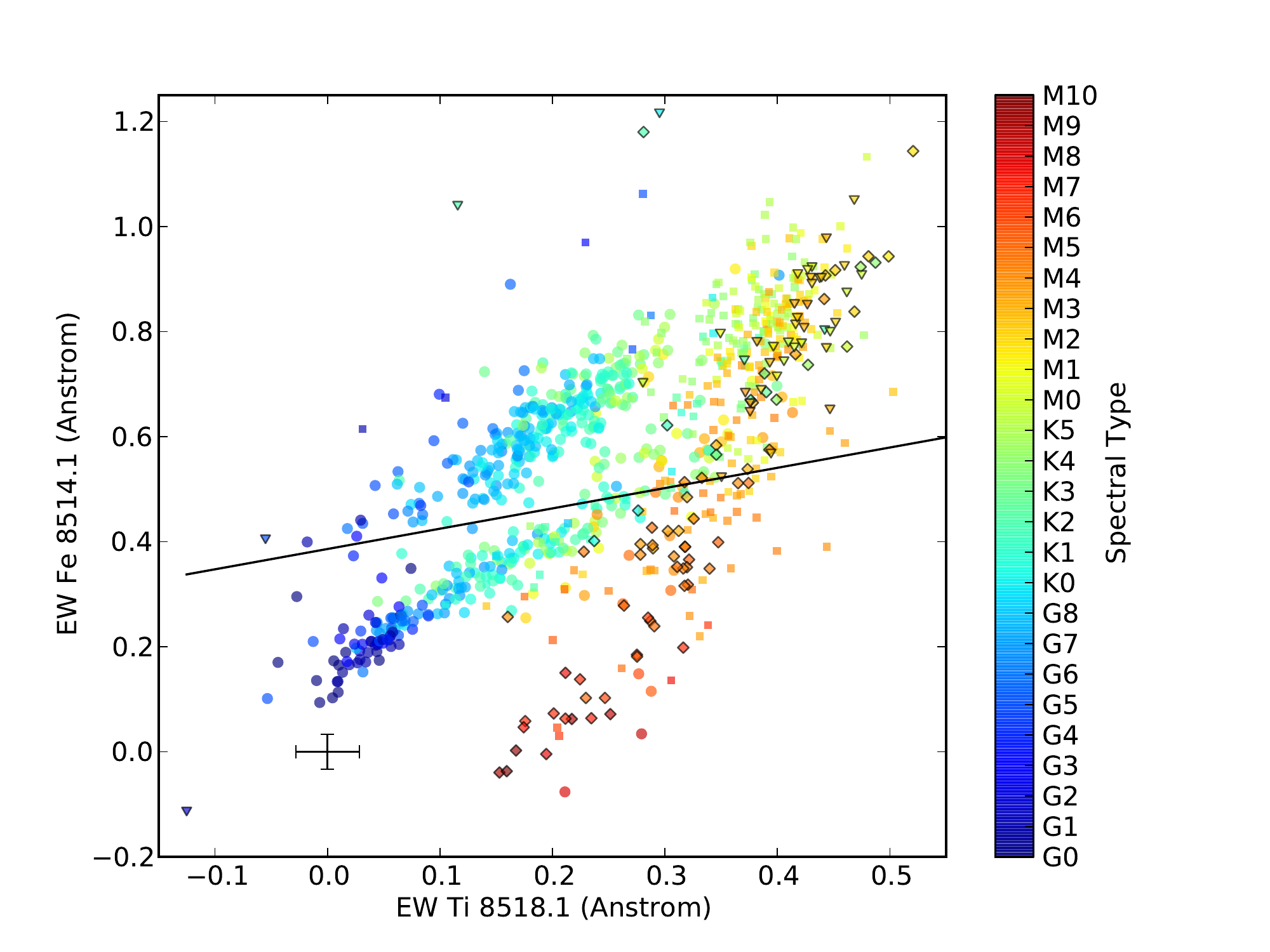}
   \includegraphics[trim=0.8cm 0.5cm 2cm 1.2cm,clip,width=9cm]{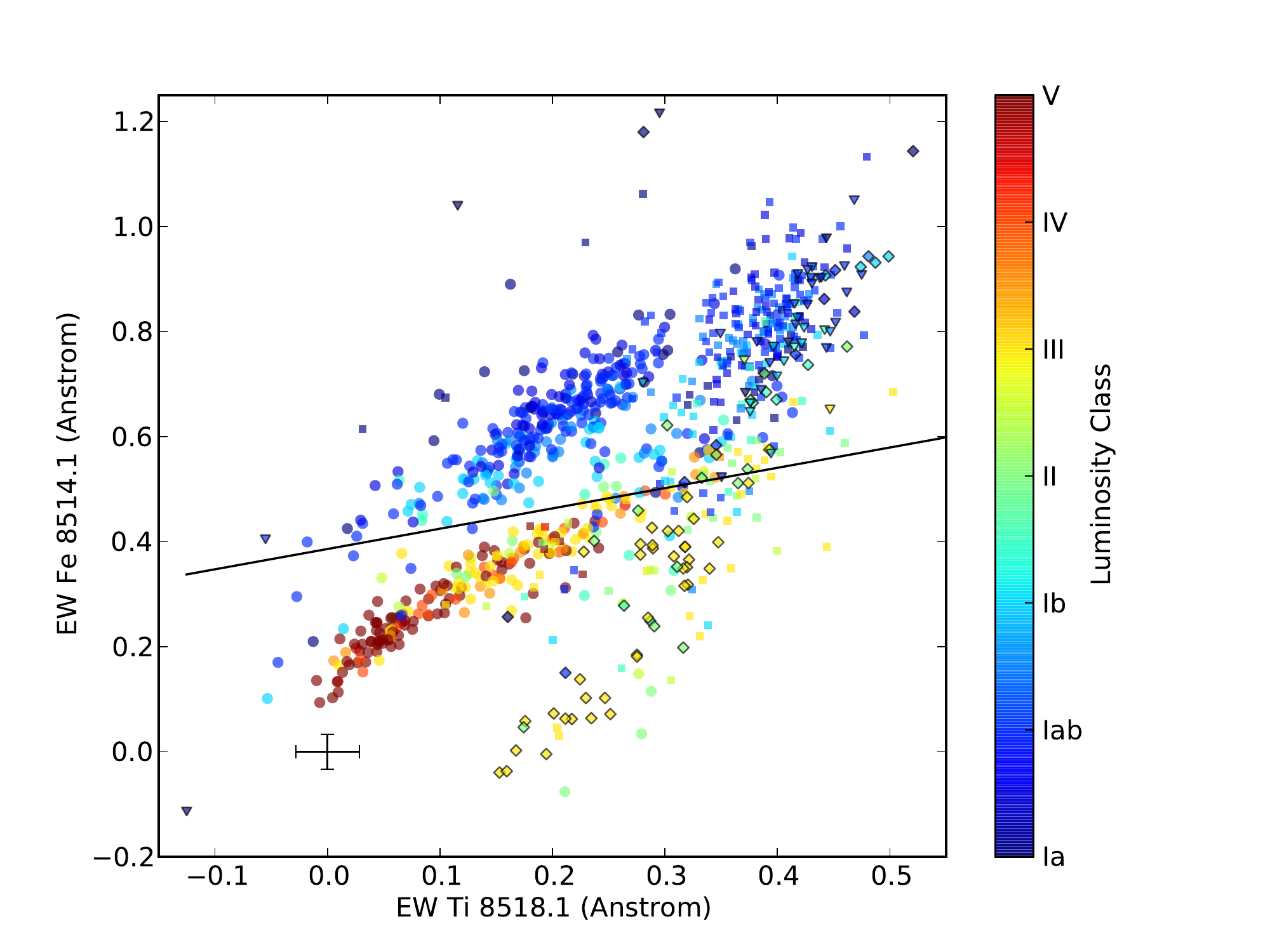}
   \caption{{\bf Left (\ref{14_18}a):} EWs of the lines Fe\,{\sc{i}}~$8514\:$\AA{} and Ti\,{\sc{i}}~$8518\:$\AA{}. The colour indicates the SpT. The shapes are the same as in Fig.~\ref{PC1_spt}. The cross indicates the median uncertainties. The black line is the calculated separation between SGs and non-SGs (see text).
   {\bf Right (\ref{14_18}b):} the same as in left figure, but the colour indicates the luminosity class.}
   \label{14_18}
\end{figure*}

Figure~\ref{14_18} shows the relation between both lines, and how their values separate very effectively different LCs, though with a strong dependence on SpT. For early G types these lines are not useful because they are too weak, and for later stars these lines appear on the red side of a TiO bandhead (at $8513$\:\AA{}), and become useless as luminosity indicators around M2. Because of these reasons, we measure negative values for the EWs of these lines in some stars beyond these limits.

We have used the SVM method explained above (with $10\,000$ random subsamples of 500 stars each) to calculate the optimised line (and its uncertainties) that separates the SGs from the non-SGs based solely on these two lines. The coefficients obtained for this line $(EW(8514.1)=m\cdot~EW(8518.1)+n)$ are $m=0.37\pm0.05$ and $n=0.388\pm0.011$. The efficiency of this boundary is $0.93\pm0.04$ and the contamination is $0.07\pm0.04$: both slightly, but significantly, worse than those obtained through the PCA method (see Sect.~\ref{effi}). However, the use of this line ratio has a disadvantage that we need to keep in mind: it depends only on two lines, neither especially strong. So, if either of these two is slightly anomalous, it will surely result in a wrong classification. The PCA method is more robust, as it depends on a large number of spectral features.
 
\subsubsection{TiO bandhead at $8859\:$\AA{}}
\label{tio_bandhead}

The spectral sequence for K and M subtypes in SGs is defined by the appearance and depth of TiO bandheads in the optical range (see GDN2015 and references therein), which are correlated with TiO bands in the CaT spectral range. The triple TiO bandhead at $8432+8442+8452\:$\AA{} together with the TiO bandhead at $8859$\:\AA{} are the strongest bandheads near the CaT, and hence the main SpT markers for the M sequence in this spectral region.

\begin{figure}[th!]
   \centering
   \includegraphics[trim=0.8cm 0.5cm 2.5cm 1.2cm,clip,width=9cm]{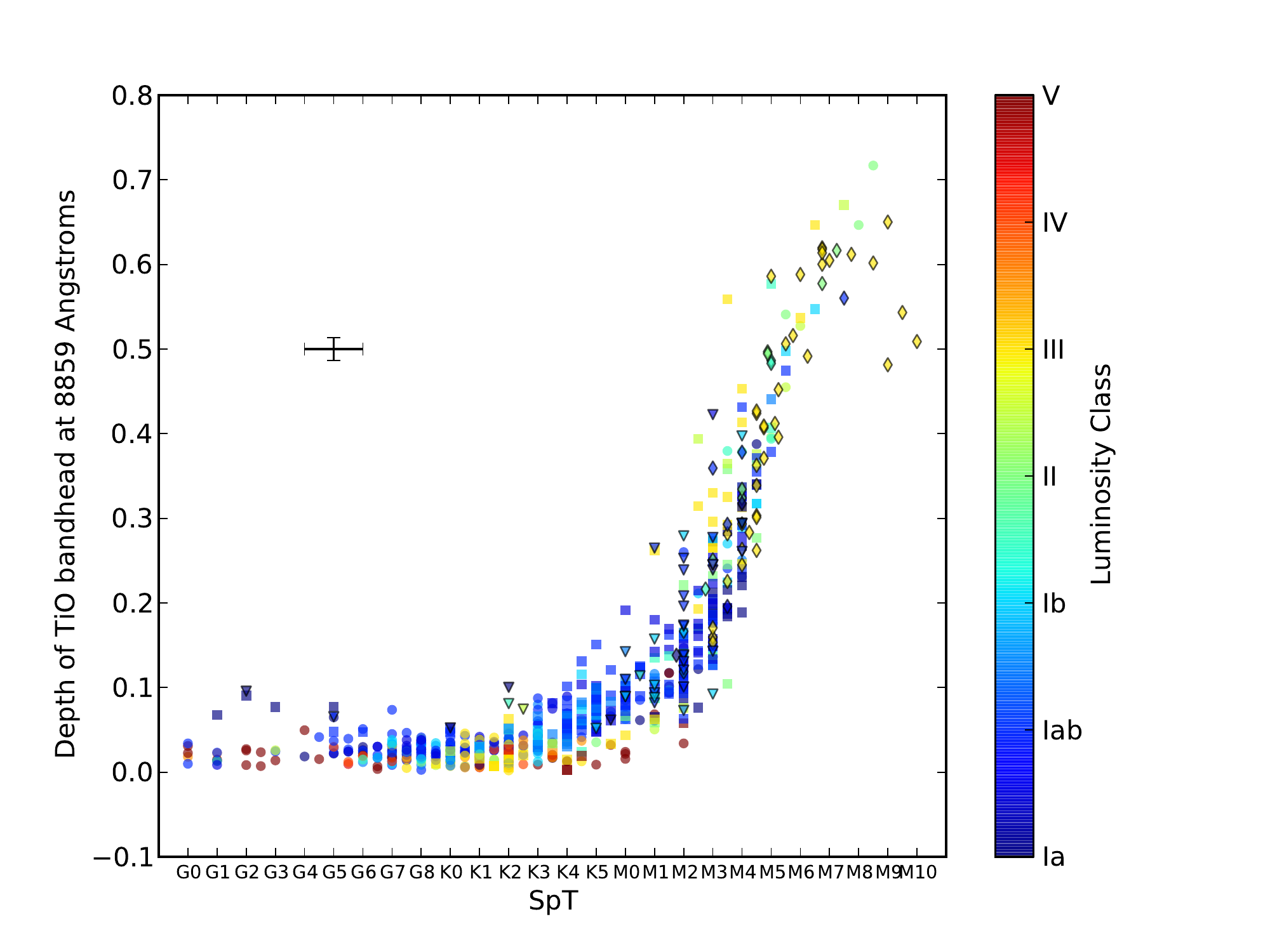}
   \caption{Spectral type to the bandhead depth of TiO $8859\:$\AA{}. The colour indicates the luminosity class. The shape indicates their origin, circles are from the SMC survey, squares are from the LMC survey, diamonds are standards, inverted triangles are from the Perseus arm survey. This bandhead rises at M1, growing with increasing SpT. However, at approximately M7 the bandhead is saturated, and starts to decrease with increasing SpT because of the erosion caused by the TiO bandhead at $8432\:$\AA{}.The error bars indicate the median uncertainty for the bandhead depth, and the estimated uncertainty for SpT from GDN2015}
   \label{SpT_TiO}
\end{figure}

In Fig.~\ref{SpT_TiO}, we can see that stars earlier than K5 present values close to zero. The TiO band first appears at approximately M1, and its strength increases with SpT until approximately M7, where it reaches a maximum. For later types, the band is saturated and the apparent depth of the bandhead starts to decrease, as the continuum is eroded by the TiO bands at shorter wavelengths. This behaviour does not seem to depend on LC, at least not between I and III, even though the number of confirmed RSGs with SpTs later than M6 is very small.

Although $8869$\:\AA{}~TiO band provides information about the SpT of the stars, it cannot be used alone to assign a SpT. Late-M types may be confused with mid-M stars if the only SpT criterion is the strength of the TiO band because of the saturation of the band. Even more importantly, such stars are scarce in our sample, but they are quite common in the Galactic Plane (where many AGB stars have spectral types in the M7\,--\,M10 range). In addition, all the G and K stars could be confused with the early-M types, especially for noisy spectra.

These problems may be solved easily by attending to the atomic lines. For example,  stars later than M8 hardly have any atomic line in the whole CaT spectral region, and they present a strong VO~$8624\:$\AA{} bandhead. However, these criteria must be considered carefully, case by case, because the strength of the lines depends on their luminosity, and the VO bandhead may be contaminated by the nearby DIB at $\sim8621$\:\AA{}. In consequence, for a quick and reliable analysis of a large number of stars we suggest using the PCA detailed in Sect.~\ref{estimation}.

\subsection{Detecting extreme red supergiants affected by veiling}

The veiling effect was first reported by \cite{mer1940} in late stars, but \cite{hum1974} studied it carefully in SGs. The cause of the veiling is still unclear, but it has been suggested that it might be caused by free-bound emission \citep{hum1974} or by scattering through an expanding circumstellar dust shell \citep{mas2009}. The veiling weakens all absorption atomic lines significantly, but \cite{hum1974} showed that this weakening is wavelength-dependent for RSGs, with the maximum weakening around the CaT region. The EW of the CaT lines decreases below the typical EW values of red giants, reaching values similar to those of dwarf stars. However, the strength of the TiO bands does not seem to be affected. 

Among SGs, it has only been reported for some extreme RSGs (ERSGs). These stars, of which only a handful are known, are characterised by their extremely high luminosities (close to the Eddington Limit, $\log(L/L_{\sun})\sim5.8$), very high mass loss, late SpTs, and extreme SpT variations (see \citealt{hum1974}; \citealt{sch2006}; and see discussion on variability in \citealt{dor2016}). Finally, we have to note that the veiling effect is not constant. It increases and disappears along the spectral variation period of these stars \citep{hum1974}.

\begin{figure*}[th!]
   \centering
   \includegraphics[trim=1cm 0.5cm 2cm 1.2cm,clip,width=9cm]{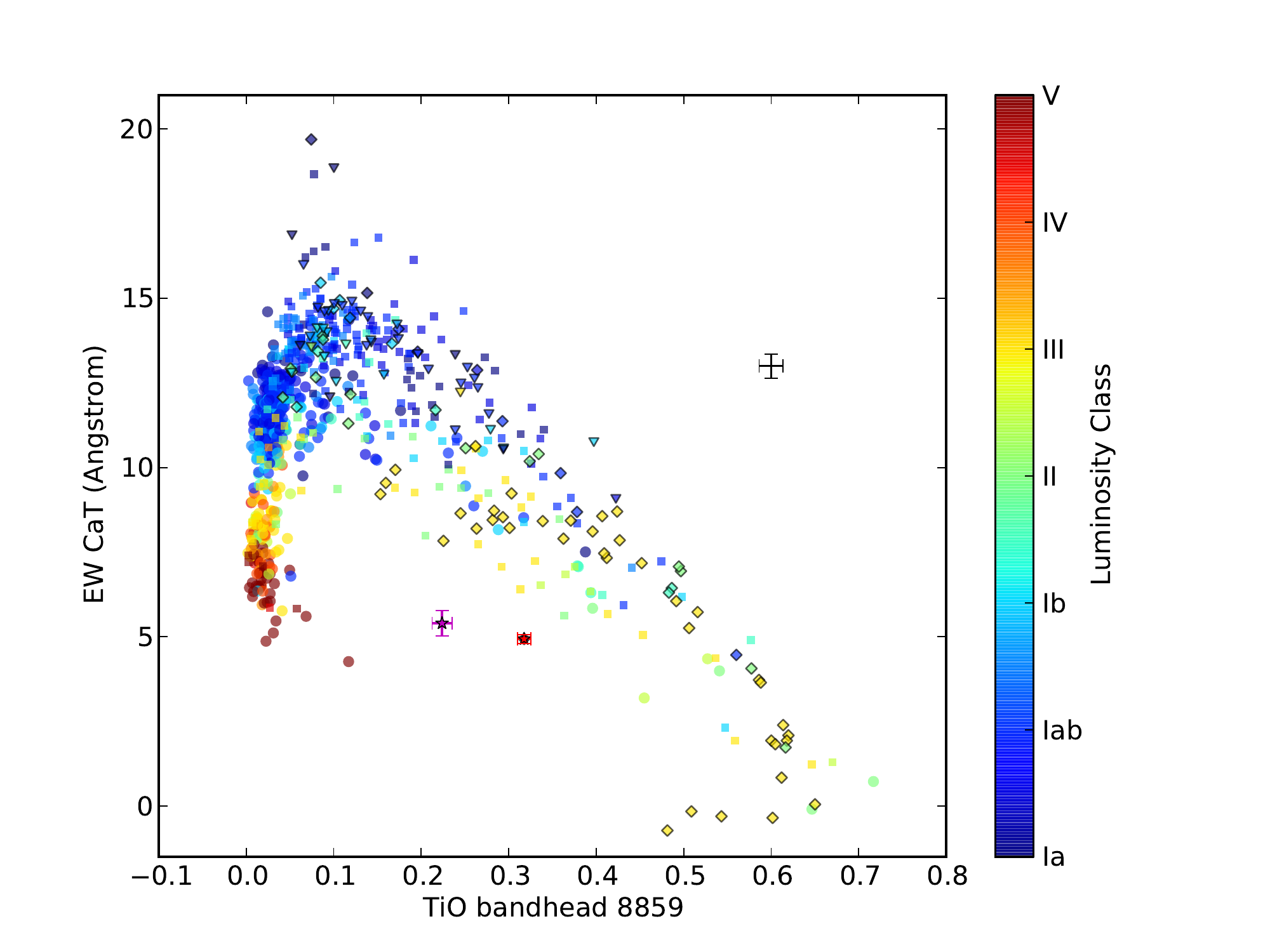}
   \includegraphics[trim=1cm 0.5cm 2cm 1.2cm,clip,width=9cm]{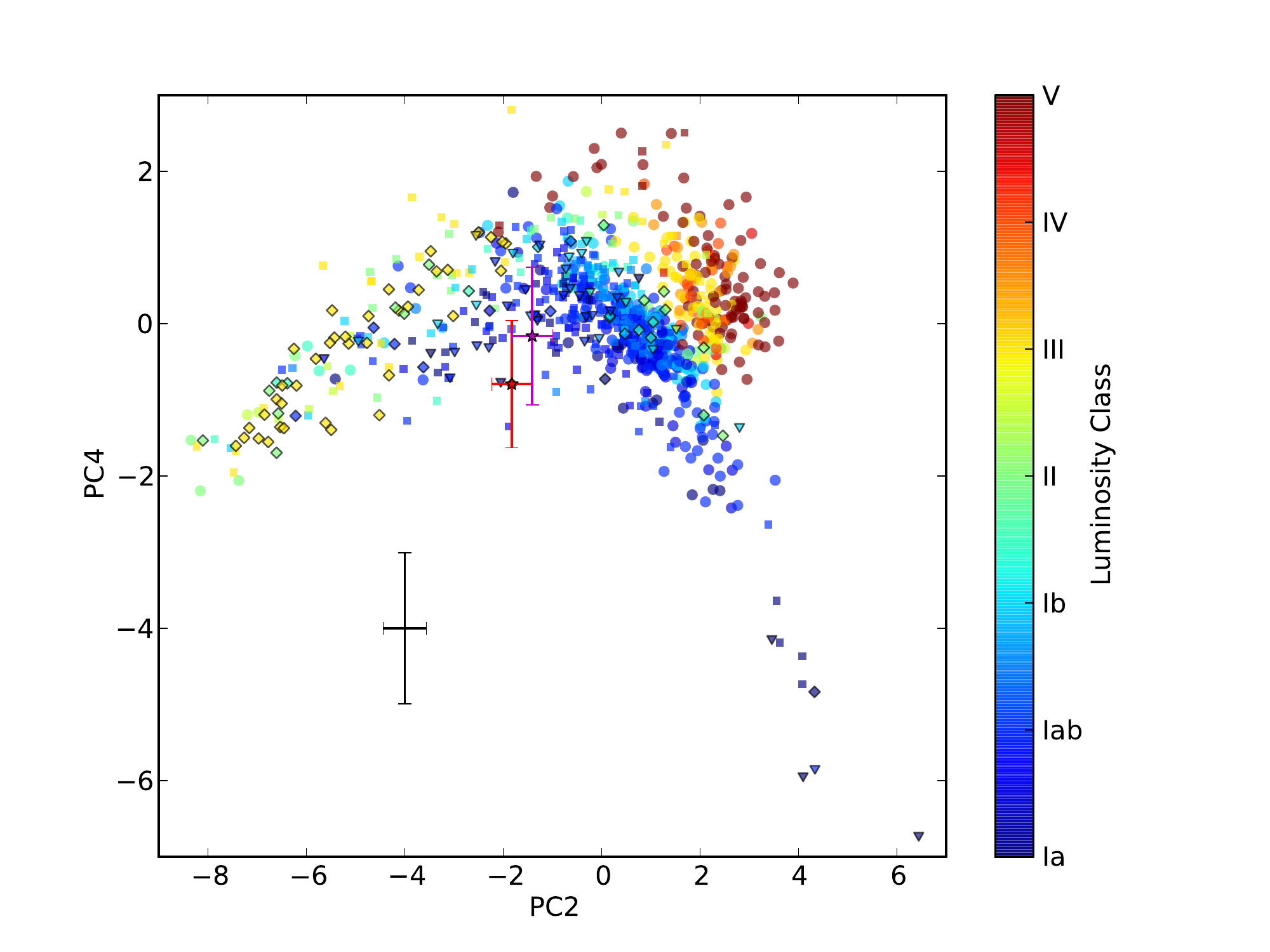}
   \caption{{\bf Left (\ref{veiling_reduc}a):} Depth of the TiO bandhead at $8859$\:\AA{} with respect to the sum of the EWs of the CaT lines. The shapes are the same as in Fig.~\ref{PC1_spt}. The black cross indicates the median uncertainties. The magenta star is the ERSG S~Per, and the red star is UY~Sct, both  with their own error bars.
   {\bf Right (\ref{veiling_reduc}b):} PC2 versus PC4 diagram. The symbols are the same as in the left figure. The black cross indicates the median uncertainties, which have been calculated by propagating the uncertainties through the lineal combination of the input data (EWs and bandheads) with the coefficients calculated. The magenta and red stars are the same as in the left panel.}
   \label{veiling_reduc}
\end{figure*}

Because of the peculiarity and scarcity of RSGs affected by veiling, we propose here a few diagrams to identify them (see Figs.~\ref{veiling_reduc} and \ref{veiling_gaia}). To locate the position of veiled RSGs, we have added to these diagrams our measurements of a spectrum of the ERSG S~Per, which was observed with the INT, in the same conditions as the stars from the galactic sample. This star was not included in the PCA calculations precisely because of its well-known spectral variability. 

In Fig.~\ref{veiling_reduc}a, S~Per (magenta star symbol) presents a sum of the CaT EWs of around 5\:\AA{}, clearly lower than the giants. However, the strength of the TiO band is small, too small to weaken the CaT significantly in a supergiant. In fact, any giant with a similar TiO bandhead has a higher CaT value. Therefore it is isolated from any other star, except another RSG (with TiO bandhead of $\sim$0.32 and CaT of $\sim$5\:\AA). This star is UY~Sct, which was classified as M2\,--\,M4~Ia by \cite{sol1978}. Even though its spectrum has not been reported as veiled before, UY~Sct has been identified as the largest star known to date \citep{arr2013}. Therefore, it seems only natural to count it as an ERSG.

These two ERSGs appear isolated in Fig.~\ref{veiling_reduc}a. This is due to the nature of our sample, because we should expect mid- and late-M dwarfs (that we have not measured) to lie close to our ERSGs. As a consequence of veiling, our ERSGs are not identified as SGs by the PCA criteria developed above. The PC2 versus PC4 diagram (Fig.~\ref{veiling_reduc}b) seems to be a good way to reveal the nature of these stars, as they lay among other RSGs. We have to caution, that we do not know where late-M dwarfs should fall in this diagram, even though the behaviour of other dwarfs suggests that they would occupy a different position. However, it is unlikely that red dwarfs and RSGs would be confused in the Galactic plane if IR photometry is available because the expected reddening is completely different in each case.

In summary, stars lying close to the position of our two ERSGs in the diagrams shown in Fig.~\ref{veiling_reduc} can be considered as candidates for veiled ERSGs. To confirm their nature, they should be observed throughout their spectral variation periods, because the veiling is expected to change along them. Thus, these stars would vary their position in the diagrams, and at some point they would be observed as normal (not veiled) RSGs. 

\section{Conclusions}

In this work, we have developed criteria based on PCA and SVM methods to separate SGs from non-SGs through their spectral features in the CaT spectral region, using a statistically significant sample from the SMC, the LMC, and the Galaxy. We obtained an efficiency identifying SGs of $0.98\pm0.04$, and a contamination of $0.02\pm0.04$.

We also revisited those criteria used in the past to identify SGs (the sum of the EWs of the CaT, the blend at $8468$\:\AA{}, and the ratio between Fe\,{\sc{i}}~$8514\:$\AA{} and Ti\,{\sc{i}}~$8518\:$\AA), studying their behaviour for a significantly larger sample. We have evaluated their limitations and compared their efficiency and contamination with those obtained through our PCA/SVM method. We show that all classical methods present efficiencies similar to (strength of the CaT and the Fe\,{\sc{i}}~$8514\:$\AA{} to Ti\,{\sc{i}}~$8518\:$\AA{} ratio) or significantly lower (the blend at $8468$\:\AA{}) than the PCA method. However, their contaminations are significantly worse.

In conclusion, the PCA method is more reliable than the classical ones. Furthermore, as the PCA uses information from many different spectral features, it is more robust than those criteria which are based on a few lines.

The PCA method can also be applied to spectra taken with the RVS on board \textit{Gaia}. These spectra cover a spectral region shorter than our spectra, but in Appendix~\ref{gaia} we repeat the same analysis, using only those lines that lie inside the RVS spectral range.

In spite of the good results obtained, we must highlight the fact that the efficiency and contamination depend on the typical SpTs in the sample (the mid- and late-M stars are hard to identify even with our PCA method), which depends on the metallicity of the population. We tested the PCA method for a sample from the Galaxy alone, obtaining a good efficiency ($0.94\pm0.13$) and a low contamination ($0.03\pm0.13$), statistically equivalent to the results from the three-galaxy sample.

In addition, we also investigated the behaviour of a number of features depending on SpT (TiO bandhead at $8859$\:\AA{}, the EW of the Ti lines, and the calculated PC2). From this we developed a method to estimate the SpT of CSGs.

Finally, we have also developed criteria to identify veiled RSGs, or at least good candidates for being such objects.

\acknowledgements
The observations have been partially supported by the OPTICON project (observing proposals 2010B/01, 2011A/014 and 2012A/015), which is funded by the European Commission under the Seventh Framework Programme (FP7). Part of the observations have been taken under service mode, (service proposal AO171) and the authors gratefully acknowledge the help of the AAO support astronomers. The INT and its service programme are operated on the island of La Palma by the Isaac Newton Group in the Spanish Observatorio del Roque de los Muchachos of the Instituto de Astrof\'{\i}sica de Canarias (observing proposals 024-INT2/11B and 97-INT8/12B, and the CAT service programme). This research is partially supported by the Spanish Ministerio de Econom\'{\i}a y Competitividad (Mineco) under grants AYA2012-39364-C02-02, AYA2015-68012-C2-2-P (MINECO/FEDER), and FPI BES-2011-049345. The work reported in this publication has been partially supported by the European Science Foundation (ESF), in the framework of the GREAT Research Networking Programme. This research made use of the Simbad, Vizier, and Aladin services developed at the Centre de Donn\'ees Astronomiques de Strasbourg, France. This publication makes use of data products from the Two Micron All Sky Survey, which is a joint project of the University of Massachusetts and the Infrared Processing and Analysis Center/California Institute of Technology, funded by the National Aeronautics and Space Administration and the National Science Foundation.
\endacknowledgements

\bibliographystyle{aa}
\bibliography{general}

\appendix

\section{Application for \textit{Gaia}}
\label{gaia}
\subsection{The \textit{Gaia} spectral range}

The spectral range covered by the \textit{Gaia} spectrograph, RVS, is narrower than the range used in this work. Therefore, to apply our method to spectra from the RVS, we repeated the PCA calculations using only the lines inside its spectral range, from $8470\:$\AA{} to $8740\:$\AA{}.

We have built the \textit{Gaia} input list by taking the shortened input list (see Section~\ref{pca_cal}) and removing all the lines lying outside the \textit{Gaia} spectral range. In addition, the lines Ti\,{\sc{i}}~8734.5\:\AA{} and Mg\,{\sc{i}}~8736.0\:\AA{} were not used either, even if they are inside the \textit{Gaia} range, because their red continuum is beyond $8740\:$\AA{} and so their EW cannot be measured in \textit{Gaia} spectra. In total, the \textit{Gaia} input list contains three bandheads and 19 atomic lines.

\subsection{PCA and SVM for the \textit{Gaia} input list}

\begin{figure*}[th!]
   \centering
   \includegraphics[trim=1cm 0.5cm 2cm 1.2cm,clip,width=9cm]{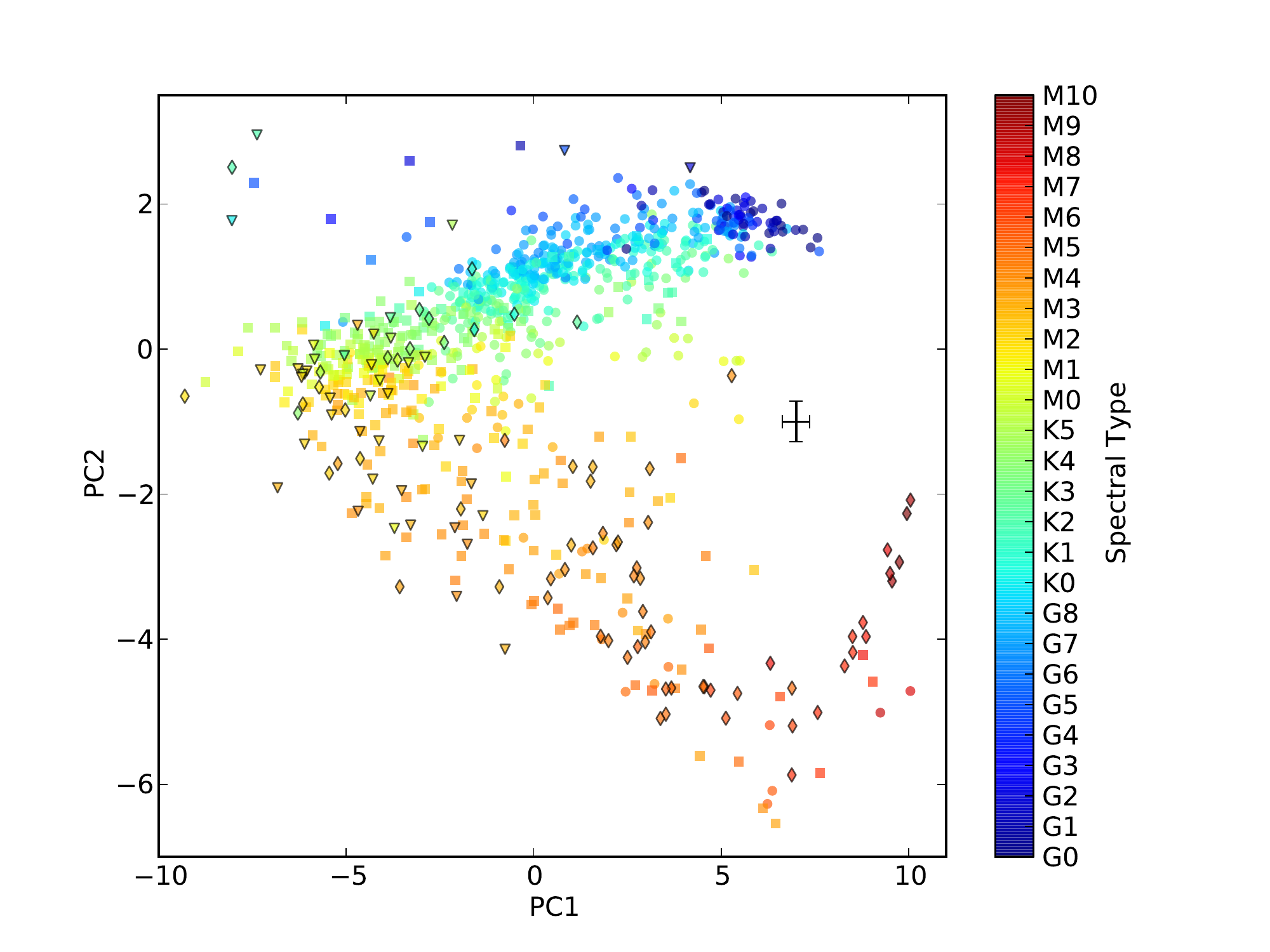}
   \includegraphics[trim=1cm 0.5cm 2cm 1.2cm,clip,width=9cm]{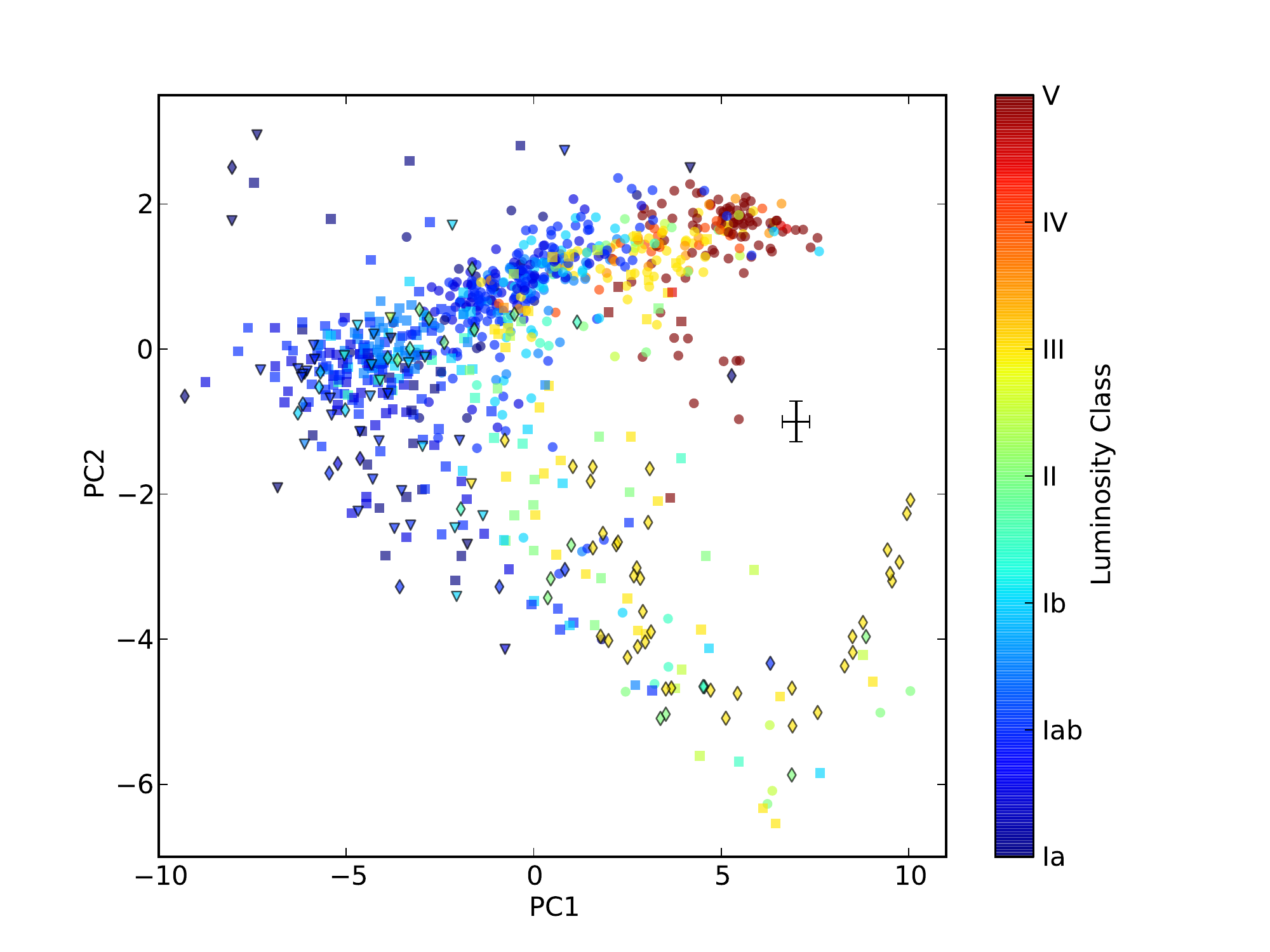}
   \caption{{\bf Left (\ref{PC1_PC2_gaia}a):} PC1 versus PC2 diagram. Both were calculated from the \textit{Gaia} input list. The colour indicates the SpT. The symbols used are the same as in Fig.~\ref{PC1_spt}. The cross indicates the median uncertainties, which have been calculated by propagating the uncertainties through the lineal combination of the input data (EWs and bandheads) with the coefficients calculated.
   {\bf Right (\ref{PC1_PC2_gaia}b):} the same as left figure, but here the colour indicates the luminosity class.}  
   \label{PC1_PC2_gaia}
\end{figure*}

\begin{figure*}[th!]
   \centering
   \includegraphics[trim=1cm 0.5cm 2cm 1.2cm,clip,width=9cm]{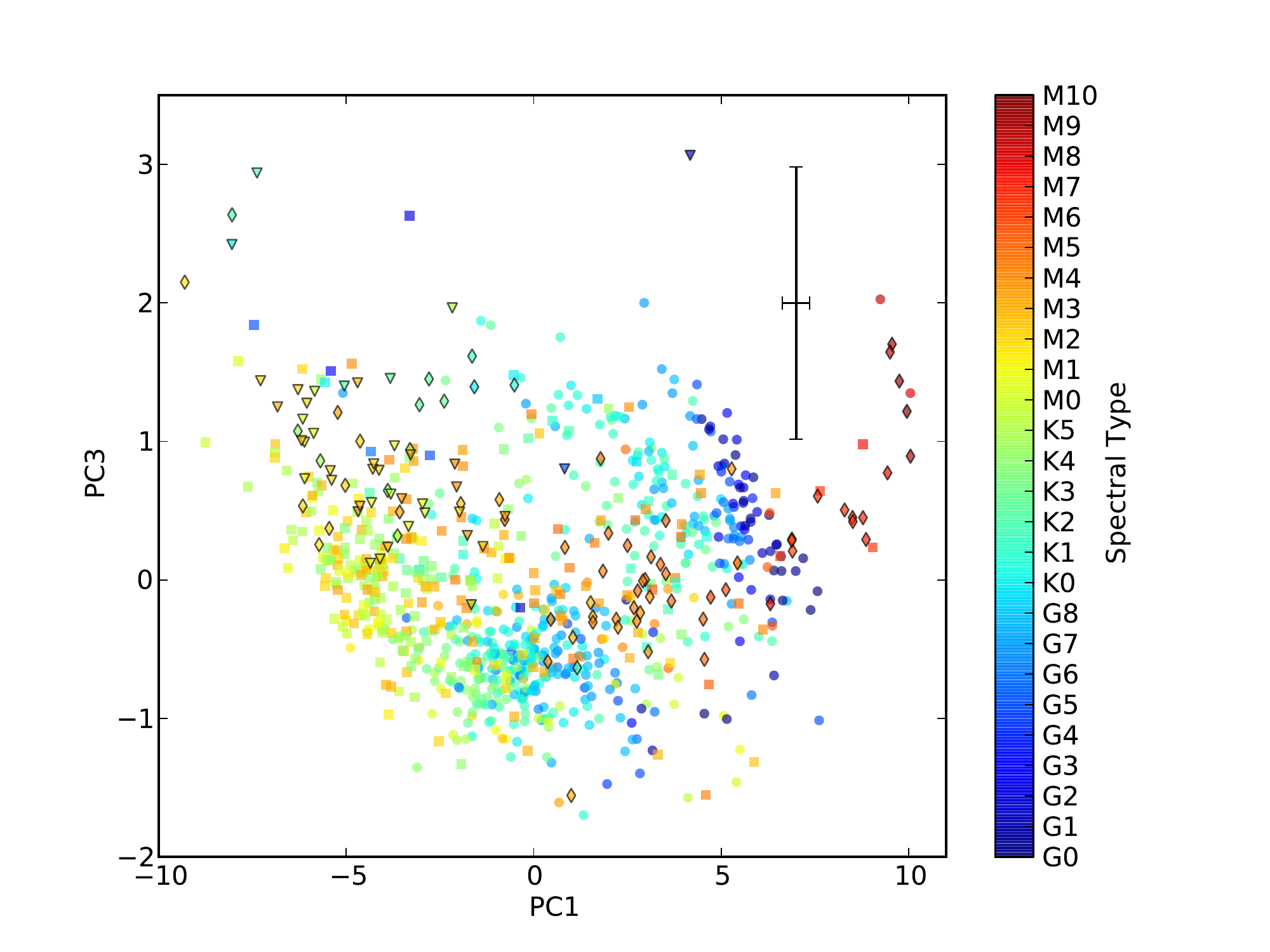}
   \includegraphics[trim=1cm 0.5cm 2cm 1.2cm,clip,width=9cm]{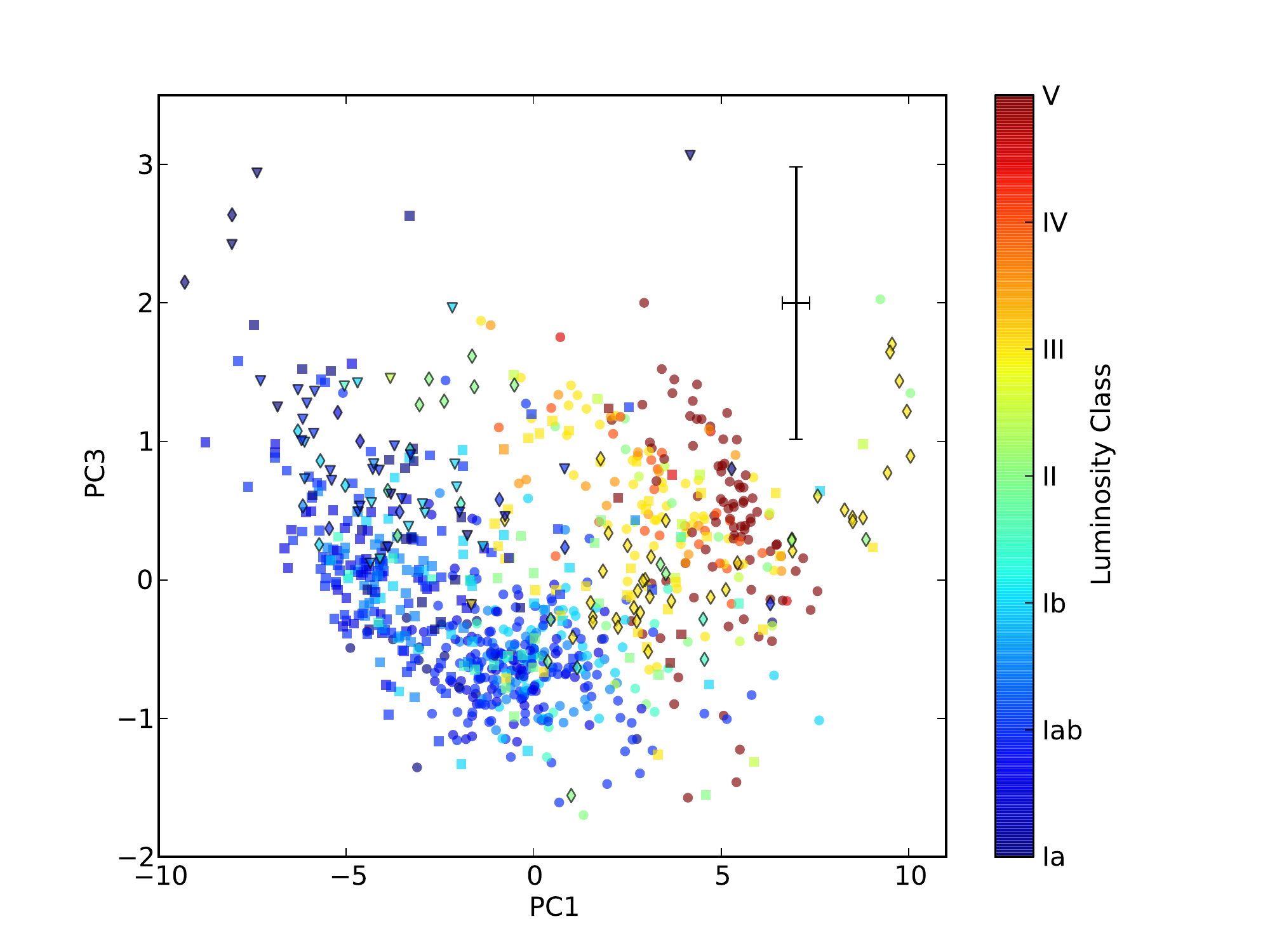}
   \caption{{\bf Left (\ref{PC1_PC3_gaia}a):} PC1 versus PC3 diagram. Both were calculated from the \textit{Gaia} input list. The colour indicates the SpT. The symbols used are the same as in Fig.~\ref{PC1_spt}. The cross indicates the median uncertainties, which have been calculated by propagating the uncertainties through the lineal combination of the input data (EWs and bandheads) with the coefficients calculated.
   {\bf Right (\ref{PC1_PC3_gaia}b):} the same as in the left figure, but here the colour indicates the luminosity class.}
   \label{PC1_PC3_gaia}
\end{figure*}

\begin{figure}[ht!]
   \centering
   \includegraphics[trim=1cm 0.5cm 0.5cm 1.2cm,clip,width=9cm]{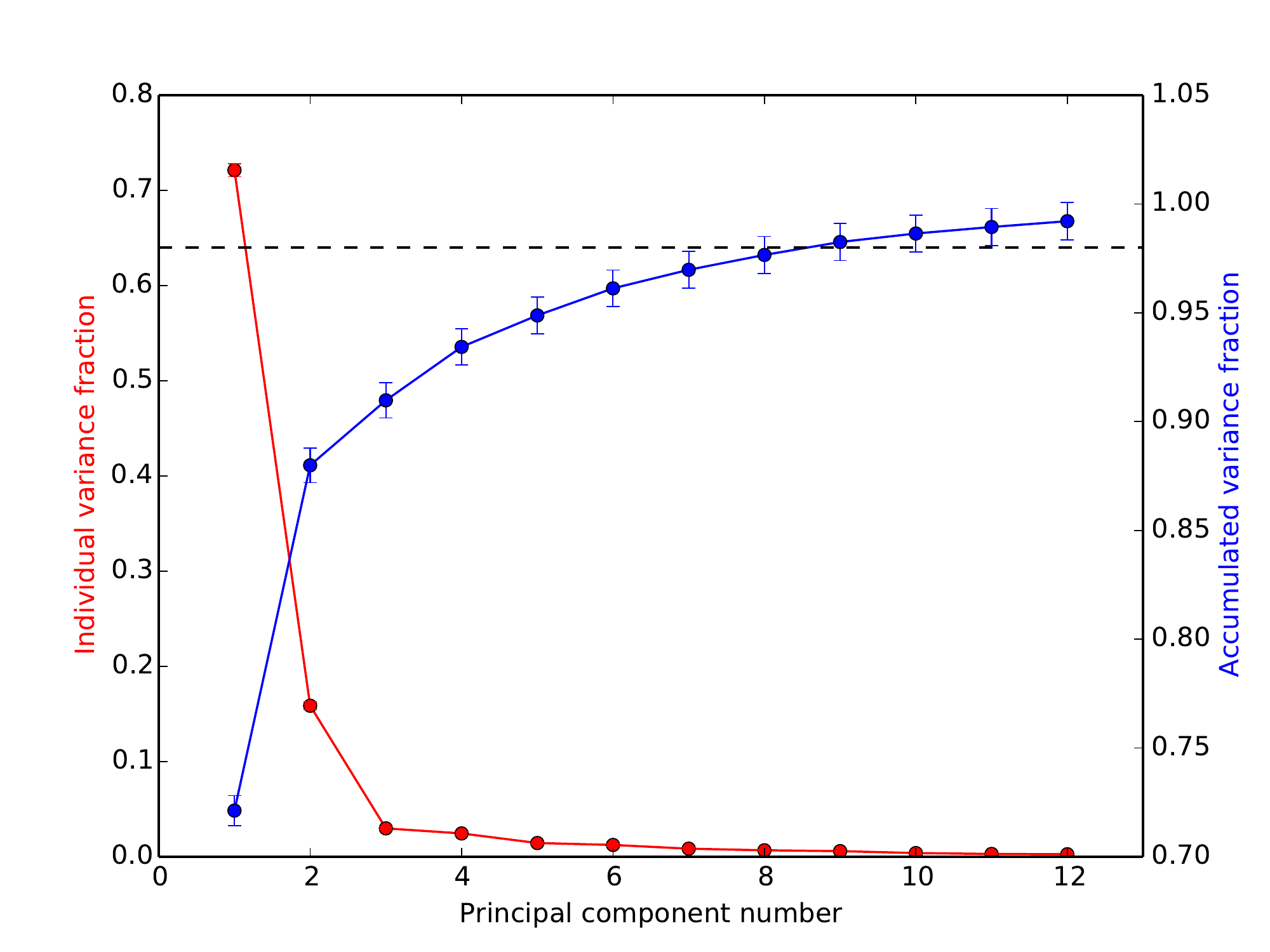}
   \caption{Summary of variance fractions of the principal components calculated from the \textit{Gaia} input list. The red circles are the individual variance (left vertical axis). The blue circles are the accumulated variance (right vertical axis). As the PCA calculations were done for $10\,000$ random samples, each circle is the median of the $10\,000$ variances obtained for each PC. The error bar in each point is its correspondent standard deviation. The circles without error bars have errors smaller than the circle itself. Only the first 12 PCs are displayed here. The horizontal discontinued line marks the 98\% of the accumulated variance.}
   \label{variance_gaia}
\end{figure}

We repeated the same PCA method described in sections~\ref{sect_PCA} and \ref{SVM}, with the \textit{Gaia} input list. Although it contains less features than our original list, the PCs obtained are similar to those presented before. The behaviour of the first three PCs is almost the same (see Figures~\ref{PC1_PC2_gaia} and \ref{PC1_PC3_gaia}), and they together contain more than 90\% of the accumulated variance. As has been explained before, the later PCs contain progressively less variance (see Fig.~\ref{variance_gaia}), and so, to reach 98\% of the accumulated variance of the \textit{Gaia} PCA, the first nine PCs are needed. We used these nine PCs were for the subsequent SVM method. For the same reason, the standard deviations of the coefficients for PC8 and PC9 are significantly high.

We calculated the SVM boundaries following the same procedures presented in Sect~\ref{SVM}. The coefficients obtained are given in Tables~\ref{SVM_M_gaia}, \ref{SVM_SGearly_gaia}, and \ref{SVM_SGlate_gaia}.

\subsubsection{Linear regressions}
\label{estimation_gaia}

As the \textit{Gaia} input list has given slightly different PC2 coefficients, we have recalculated the estimations of SpT through linear regression. In addition, one of the lines of the Ti index (Ti\,{\sc{i}}~$8734.5\:$\AA{}) is not included in the \textit{Gaia} input list. Thus, we have calculated a new Ti index for the \textit{Gaia} input list, excluding this line. Finally, the TiO bandhead at $8859$\:\AA{} is outside the \textit{Gaia} range, while all the other TiO bands inside this range are too weak to provide a clear relation for the whole M sequence. Therefore we have not used any TiO band to estimate the SpT.

The procedure has been similar to that outlined in Sect.~\ref{estimation}, with some small differences. Firstly, the separation between SGs and non-SGs was done through the PCA method using the data for the M0 boundary, because it has the best efficiency and contamination in the identification of SGs in the case of the \textit{Gaia} input list (see Sect.~\ref{effi_gaia}). Secondly, in Sect.~\ref{estimation} we used the M1 type as the boundary between early and late subsamples, because the main TiO bandhead (at $8859$\:\AA{}) becomes noticeable at this subtype, changing the behaviour of the PC. This band is not inside the \textit{Gaia} spectral range. Thus the PC2 behaviour does not change before M2, when the other lesser TiO bands start to be noticeable. In consequence we have used M2 as the boundary between early and late subsamples for these calculations. The coefficients obtained for the linear regressions are in Table~\ref{theil-sen_gaia}.

\begin{table*}[th!]
\caption{Results from the linear regressions done between different variables and spectral subtypes, using the \textit{Gaia} input list.}
\label{theil-sen_gaia}
\centering
\begin{tabular}{c | c | c | c | c}
\hline\hline
\noalign{\smallskip}
Variable used&subsample&Slope&$y$-intercept (subtypes)&$\sigma(y_{\textrm{real}}-y_{\textrm{expec}})$ (subtypes)\\
\noalign{\smallskip}
\hline
\noalign{\smallskip}
PC2\tablefootmark{a}&Early&-4.9&14.0&1.2\\
Ti index\tablefootmark{b}&Early - only SMC&11&2.4&1.3\\
Ti index\tablefootmark{b}&Early - Galaxy and LMC&9.1&3.7&1.2\\
PC2\tablefootmark{a}&Late&-0.65&16.9&0.7\\

\noalign{\smallskip}
\hline
\end{tabular}
\tablefoot{For details about their calculation see section~\ref{estimation_gaia}.\\
\tablefoottext{a}{Dimensionless}\\
\tablefoottext{b}{Expressed in \AA{}}
}
\end{table*}

\subsubsection{Efficiency}
\label{effi_gaia}

In Table.~\ref{efficiency_gaia} we compare the efficiency (i.e. the fraction of known SGs in the sample or subsample that are tagged as SGs by a given criterion) of the boundaries obtained for the \textit{Gaia} input list in each boundary subtype for the \textit{Gaia} input list.

With this spectral range, the efficiency is $0.97\pm0.04$ for all the five boundaries used, equivalent to that obtained for the complete line list in Sect.~\ref{effi}. The efficiencies for the early and late subsamples are statistically equivalent in all the cases, except for the boundaries M2 and M3, where we find significantly lower values. This is probably due to two factors. Firstly, the atomic lines less affected by the rise of TiO bands are those lying between $8700$ and $8859\:$\AA{}, and thus are not included in this analysis because they fall outside the \textit{Gaia} spectral range. Secondly, the main TiO bandhead of the CaT region (at $8859$\:\AA{}) is also outside this window. On the other hand, the contamination is about $0.03\pm0.04$ for all the five boundaries, statistically equivalent to the result obtained for the complete line list in Sect.~\ref{effi}.

\begin{table}[th!]
\caption{Supergiant identification efficiency and contamination of the PCA method for the \textit{Gaia} line list, and their errors, for our whole sample, depending on the  putative boundary used.}
\label{efficiency_gaia}
\centering
\begin{tabular}{c | c | c | c }
\hline\hline
\noalign{\smallskip}
Boundary&Subsample&Efficiency&Contamination\\
\noalign{\smallskip}
\hline
\noalign{\smallskip}
&Early&$0.98\pm0.06$&$0.01\pm0.06$\\
K5&Late&$0.95\pm0.06$&$0.05\pm0.06$\\
&All&$0.97\pm0.04$&$0.03\pm0.04$\\
\noalign{\smallskip}
\hline
\noalign{\smallskip}
&Early&$0.99\pm0.05$&$0.01\pm0.05$\\
M0&Late&$0.93\pm0.07$&$0.06\pm0.07$\\
&All&$0.97\pm0.04$&$0.03\pm0.04$\\
\noalign{\smallskip}
\hline
\noalign{\smallskip}
&Early&$0.99\pm0.05$&$0.01\pm0.05$\\
M1&Late&$0.91\pm0.08$&$0.07\pm0.08$\\
&All&$0.97\pm0.04$&$0.03\pm0.04$\\
\noalign{\smallskip}
\hline
\noalign{\smallskip}
&Early&$0.99\pm0.05$&$0.02\pm0.05$\\
M2&Late&$0.89\pm0.09$&$0.08\pm0.09$\\
&All&$0.97\pm0.04$&$0.03\pm0.04$\\
\noalign{\smallskip}
\hline
\noalign{\smallskip}
&Early&$0.99\pm0.04$&$0.02\pm0.04$\\
M3&Late&$0.77\pm0.12$&$0.07\pm0.13$\\
&All&$0.97\pm0.04$&$0.02\pm0.04$\\
\noalign{\smallskip}
\hline
\end{tabular}
\end{table}

\subsubsection{Veiled RSGs and the \textit{Gaia} range.}

\begin{figure*}[th!]
   \centering
   \includegraphics[trim=1cm 0.5cm 2cm 1.2cm,clip,width=9cm]{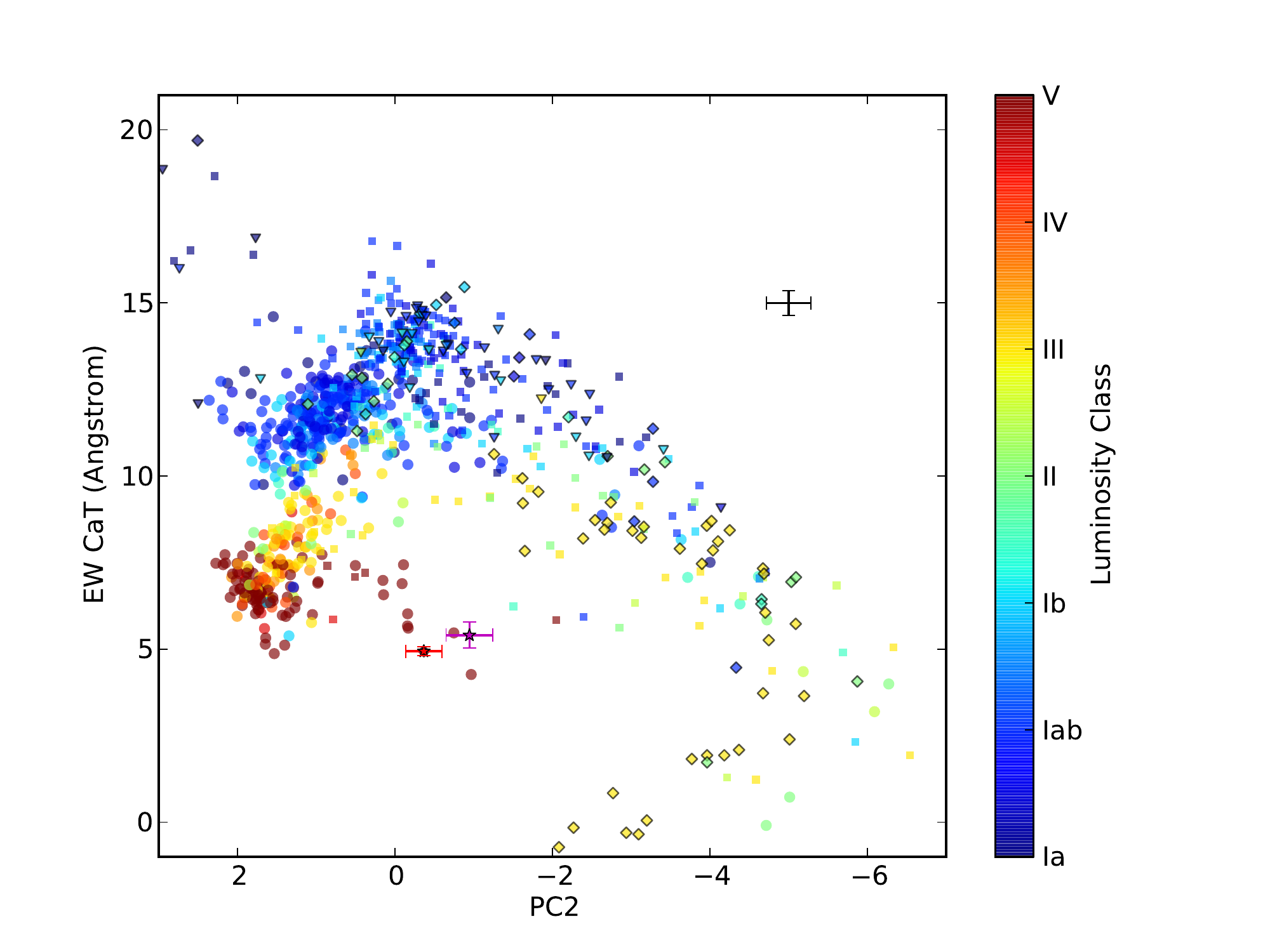}
   \includegraphics[trim=1cm 0.5cm 2cm 1.2cm,clip,width=9cm]{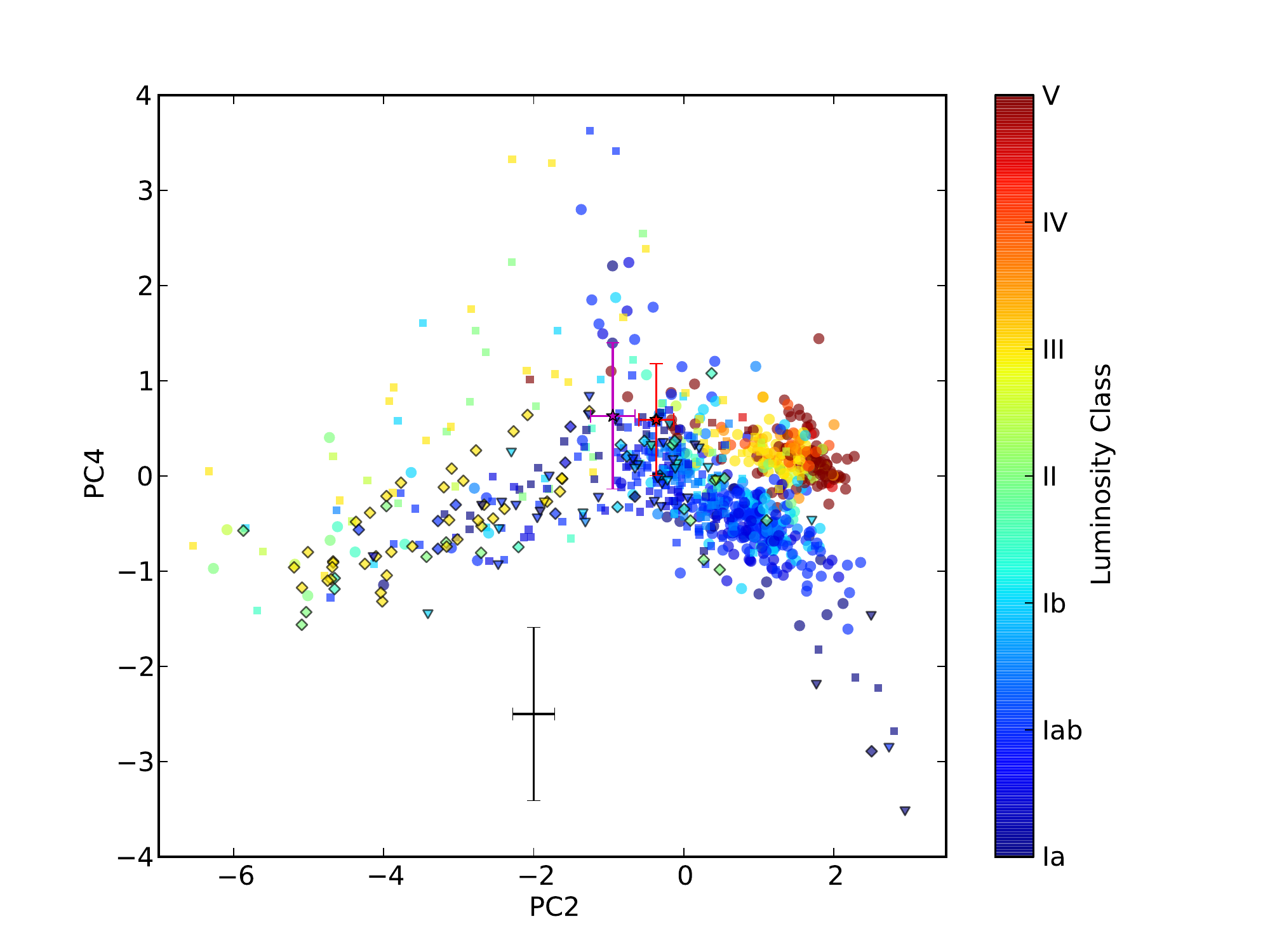}
   \caption{{\bf Left (\ref{veiling_gaia}a):} PC2 (calculated from the \textit{Gaia} input list) to the sum of EWs of the CaT lines. The display is the same as in the Figure~\ref{veiling_reduc}
   {\bf Right (\ref{veiling_gaia}b):} PC2 versus PC4 diagram, calculated from \textit{Gaia} input list. The display is the same as in the Figure~\ref{veiling_reduc}}
   \label{veiling_gaia}
\end{figure*}

\textit{Gaia} spectra do not include the TiO bandhead at $8859$\:\AA{}. Therefore, we propose an alternative diagram to identify ERSGs (see Fig.~\ref{veiling_gaia}), using PC2, because its behaviour is similar to that of the TiO bandhead. In the present case, the veiled RSGs occupy the same region as the early-M dwarfs in the sum of the CaT EWs versus PC2 diagram, but they may be identified as SGs by comparing their positions in the PC2 versus PC4 diagram (Fig.~\ref{veiling_gaia}b). Moreover, since \textit{Gaia} will provide multi-epoch spectra, they may be used to check the different positions of a given star in these diagrams, as their veilings change along the spectral variation of these stars.

\onecolumn

\section{List of Milky Way stars}

\begin{longtab}
\begin{longtable}[h]{c | c | c | c | c | c}
\caption{List of Milky Way stars that we have used in this work. This list includes MK standards from \cite{kee1989} (with LCs from I to III) and also other well known CSGs. All the SpTs and LCs are taken from the literature. Some stars have been observed more than once. As these stars (even the MK standards) may present spectral variability, we have treated the different measurements as different objects in our calculations. Note also that the stars observed in Sept. and Oct. 2012 are considered from the same epoch, as all of them were observed in the last days of September or in the first ones of October. \label{stdr}}\\
\hline\hline
\noalign{\smallskip}
Name&MK standard&RA J2000&DEC J2000&Type&Epoch\\
\noalign{\smallskip}
\hline
\noalign{\smallskip}
\endfirsthead
\caption{continued.}\\
\noalign{\smallskip}
Name&MK standard&RA J2000&DEC J2000&Type&Epoch\\
\noalign{\smallskip}
\hline
\noalign{\smallskip}
\endhead

KN Cas&no&00:09:36.3&+62:40:03.9&M1ep\:Ib&Sep. 2012\\
MZ Cas&no&00:21:24.2&+59:57:11.0&M2\:Iab&Sep. 2012\\
HD 236449&no&00:31:38.3&+60:15:19.2&K2.5\:II\,--\,III&Sep. 2012\\
HD 4817&no&00:51:16.4&+61:48:19.6&K2\:Ib\,--\,II&Sep. 2012\\
HS Cas&no&01:08:19.9&+63:35:11.6&M4\:Ia&Sep. 2012\\
AW Psc&yes&01:11:16.1&+30:38:06.0&M9\:(III)&Jun. 2012\\
BD +59 274&no&01:33:29.2&+60:38:47.8&M0.5\:Ib\,--\,II&Sep. 2012\\
BD +60 265&yes&01:33:35.1&+61:33:03.44&M1.5\:Ib&Jan. 2015\\
HD 236835&no&01:43:02.7&+56:30:46.1&M2\:Ib&Sep. 2012\\
V589 Cas&no&01:46:05.4&+60:59:36.5&M3\:Iab&Sep. 2012\\
WX Cas&no&01:54:03.7&+61:06:33.0&M2\:Iab\,--\,Ib&Sep. 2012\\
V778 Cas&no&01:58:28.9&+59:16:08.6&M2\:Iab&Sep. 2012\\
BD +59 372&no&01:59:39.6&+60:15:01.8&K5\,--\,M0\:Ia&Sep. 2012\\
XX Per&no&02:03:09.3&+55:13:56.6&M4\:Ib&Sep. 2012\\
KK Per&no&02:10:15.7&+56:33:32.7&M2\:Iab\,--\,Ib&Sep. 2012\\
HD 13658&no&02:15:13.3&+58:08:32.2&M1\:Iab&Sep. 2012\\
HD 13686&yes&02:15:56.6&+63:13:46.90&K2.5 Ib\,--\,II&Jan. 2015\\
PP Per&no&02:17:08.2&+58:31:46.9&M0\:Iab\,--\,Ib&Sep. 2012\\
BU Per&no&02:18:53.2&+57:25:16.8&M4\:Ib&Sep. 2012\\
T Per&no&02:19:21.8&+58:57:40.3&M2\:Iab&Sep. 2012\\
HD 14242&no&02:20:22.4&+59:40:16.8&M2\:Iab&Sep. 2012\\
AD Per&no&02:20:29.0&+56:59:35.2&M3\:Iab&Sep. 2012\\
PR Per&no&02:21:42.4&+57:51:46.0&M1-\:Iab\,--\,Ib&Sep. 2012\\
SU Per&yes&02:22:06.9&+56:36:14.9&M3\,--\,M4\:Iab&Jun. 2012\\
V439 Per&no&02:23:11.0&+57:11:57.9&M0\:Iab&Sep. 2012\\
HD 14580&no&02:23:24.0&+57:12:43.0&M0\:Iab&Sep. 2012\\
HD 14826&no&02:25:21.8&+57:26:14.0&M2\:Iab&Sep. 2012\\
YZ Per&no&02:38:25.4&+57:02:46.2&M2\:Iab&Sep. 2012\\
GP Cas&no&02:39:50.4&+59:35:51.3&M2\:Iab&Sep. 2012\\
HD 237006&no&02:49:08.8&+58:00:48.2&M1\:Ib&Sep. 2012\\
HD 237010&no&02:51:03.9&+57:51:19.8&M2\:Iab&Sep. 2012\\
HD 17958&no&02:56:24.6&+64:19:56.3&K3\:Ib&Sep. 2012\\
HD 18391&no&02:59:48.7&+57:39:47.6&G5\:Ia\,--\,Ib&Sep. 2012\\
BD +59 580&no&03:00:39.9&+59:57:59.6&M1\:Ib&Sep. 2012\\
V411 Per&no&03:15:08.4&+54:53:03.0&M1\:Iab&Sep. 2012\\
BD +55 780&no&03:25:21.6&+55:33:10.1&K5\:Ib&Sep. 2012\\
HD 25725&yes&04:04:18.8&-15:43:30.5&M7+\:II&Sep. 2012\\
DG Eri&yes&04:20:41.4&-16:49:47.9&M4\:III&Sep. 2012\\
HD 28487&yes&04:29:38.9&+05:09:51.4&M3.5\:III&Sep. 2012\\
CE Tau&yes&05:32:12.8&+18:35:39.3&M2\:Iab\,--\,Ib&Sep. 2012\\
HD 39045&yes&05:51:25.8&+32:07:28.9&M3\:III&Sep. 2012\\
$\pi$ Aur&yes&05:59:56.1&+45:56:12.3&M3\:IIb&Sep. 2012\\
VY Leo&yes&10:56:01.4&+06:11:07.3&M5.5\:III&Jun. 2012\\
BK Vir&yes&12:30:21.0&+04:24:59.2&M7-\:III&Jun. 2012\\
TU Cvn&yes&12:54:56.5&+47:11:48.2&M5-\:III\,--\,IIIa&Jun. 2012\\
SW Vir&yes&13:14:04.3&-02:48:25.1&M7\:III&Jun. 2012\\
BY Boo&yes&14:07:55.7&+43:51:16.0&M4.5\:III&Jun. 2012\\
RX Boo&yes&14:24:11.6&+25:42:13.4&M7.5\,--\,M8\:III&Jun. 2012\\
RR Umi&yes&14:57:35.0&+65:55:56.9&M4.5\:III&Jun. 2012\\
17 Ser&yes&15:36:28.1&+15:06:05.1&M5\:IIIa&Jun. 2012\\
30 Her&yes&16:28:38.5&+41:52:54.0&M6-\:III&Jun. 2012\\
HD 151061&yes&16:45:11.4&-03:05:05.8&M5\,--\,M5.5\:IIIb:&Jun. 2012\\
HR 6242&yes&16:47:19.7&+42:14:20.1&M4.5\:III&Jun. 2012\\
IRC +10313&yes&16:51:05.9&+10:20:51.6&M7\,--\,M9III&Jun. 2012\\
IRC +10322&yes&17:14:19.3&+08:56:02.6&M7\,--\,M10III&Jun. 2012\\
$\alpha$ Her&yes&17:14:38.8&+14:23:25.2&M5\:Ib\,--\,II&Jun. 2012\\
MW Her&yes&17:35:40.0&+15:35:12.2&M8\,--\,M9\:III&Jun. 2012\\
HD 164349&yes&18:00:02.9&+16:45:22.26&K0.5\:IIb&May 2015\\
HD 167006&yes&18:11:54.1&+31:24:19.3&M3\:III&Sep. 2012\\
UY Sct&yes&18:27:36.5&-12:27:58.9&M2\,--\,M4\:Ia&Jun. 2012\\
HK Dra&yes&18:34:30.8&+51:46:56.1&M4\:III\,--\,IIIb&Jun. and Sep. 2012\\
XY Lyr&yes&18:38:06.4&+39:40:06.0&M4.5\,--\,M5+\:II&Jun. and Sep. 2012\\
HD 175309&yes&18:54:28.9&+10:37:57.1&M5-\,--\,M5.5\:(III)&Sep. 2012\\
$\delta^{2}$ Lyr&yes&18:54:30.2&+36:53:55.0&M4\:II&Jun. and Sep. 2012\\
UW AQL&yes&18:57:33.5&+00:27:37.97&M2+\:Iab&May 2015\\
HD 179820&yes&19:13:54.4&+02:37:30.8&M6+\:III&Sep. 2012\\
HD 180809&yes&19:16:22.4&+38:08:25.84&K0\:II&May 2015\\
HD 181475&yes&19:20:48.5&-04:29:53.38&K7\:IIa&May 2015\\
HD 184313&yes&19:33:46.0&+05:27:56.5&M5\,--\,M5.5III&Sep. 2012\\
HD 185622&yes&19:39:25.6&+16:34:37.23&K4\:Ib&May 2015\\
HD 186776&yes&19:44:49.0&+40:43:00.5&M3.5\:III&Sep. 2012\\
HD 186791&yes&19:46:15.9&+10:37:05.82&K3\:II&May 2015\\
HD 190788&yes&20:05:50.3&+25:36:03.2&M3-\:Ib\,--\,II&Sep. 2012\\
AC Dra&yes&20:20:06.0&+68:52:49.1&M4.5\,--\,M5\:III&Jun. and Sep. 2012\\
BI Cyg&yes&20:21:21.8&+36:55:55.8&M2\,--\,M4\:I&Jun. 2012\\
BC Cyg&yes&20:21:38.5&+37:31:58.9&M4\:I&Jun. 2012\\
RW Cyg&yes&20:28:50.5&+39:58:54.4&M3\,--\,M4\:Ia\,--\,Iab&Jun. 2012\\
EU Del&yes&20:37:54.7&+18:16:06.9&M6\:III&Jun. 2012\\
HD 196819&yes&20:38:17.2&+42:04:45.04&K2.5\:IIb&May 2015\\
DG Cyg&yes&20:43:25.4&+43:11:50.7&M9\:(III)&Jun. 2012\\
HD 198026&yes&20:47:44.2&-05:01:39.7&M3\:III&Sep. 2012\\
HD 200527&yes&21:02:24.2&+44:47:27.5&M4.5\:III&Sep. 2012\\
HD 200905&yes&21:04:55.9&+43:56:00.17&K4.5\:Ib\,--\,II&May 2015\\
HD 201251&yes&21:06:35.9&+47:39:14.48&K4\:Ib\,--\,IIa&May 2015\\
HD 202380&yes&21:12:47.2&+60:05:52.8&M2\:Ib&Sep. 2012\\
NV Peg&yes&21:28:59.7&+22:10:46.0&M4.5\:IIIa&Jun. and Sep. 2012\\
$\mu$ Cep&yes&21:43:30.4&+58:46:48.2&M2-\:Ia&Sep. 2012\\
HD 207328&no&21:46:16.6&+58:03:45.0&M3\:IIIa&Sep. 2012\\
EP Aqr&yes&21:46:31.8&-02:12:45.9&M7-\:III&Jun. and Sep. 2012\\
DZ Aqr&yes&22:21:41.8&-07:36:30.1&M7-\:(III)&Jun. 2012\\
RW Cep&yes&22:23:06.8&+55:58:11.25&K2\:0\,--\,Ia&Oct. 2012 and May 2015\\
HD 239978&yes&22:30:10.5&+57:00:26.11&M2\:Ia\,--\,Iab&May 2015\\
W Cep&no&22:36:27.5&+58:25:34.0&K0ep\:Ia&Sep. 2012\\
HR8621&yes&22:38:37.9&+56:47:44.3&M4+\:III&Jun. 2012\\
U Lac&no&22:47:43.4&+55:09:30.3&M4\:Iab&Sep. 2012\\
MY Cep&yes&22:54:31.7&+60:49:38.9&M7.5\:I&Jun. 2012\\
HD 217673&yes&23:01:30.4&+57:06:42.67&K1.5\:II&May 2015\\
GU Cep&no&23:10:10.8&+61:14:29.6&M2\:Iab&Sep. 2012\\
SS And&yes&23:11:30.0&+52:53:12.5&M7-\:II&Jun. 2012\\
V356 Cep&no&23:13:31.5&+60:30:18.7&M2\:Iab&Sep. 2012\\
HD 219978&yes&23:19:23.4&+62:44:48.23&K4.5\:Ib&May 2015\\
V358 Cas&no&23:30:27.3&+57:58:33.4&M3\:Ia\,--\,Iab&Sep. 2012\\
PZ Cas&no&23:44:03.3&+61:47:22.1&M3\:Ia&Sep. 2012\\
HD 223173&yes&23:47:01.5&+57:27:30.78&K3-\:IIb&May 2015\\
TZ Cas&no&23:52:56.2&+61:00:08.3&M3\:Iab&Sep. 2012\\
BD +63 2073&no&23:53:58.0&+64:15:02.7&M0\:Ib&Sep. 2012\\
$\rho$ Cas&no&23:54:23.0&+57:29:57.8&G2\:0&Sep. 2012\\
XZ Psc&yes&23:54:46.6&+00:06:33.5&M5\:III&Jun. 2012\\

\noalign{\smallskip}
\hline
\end{longtable}
\end{longtab}

\section{List of lines and bandheads measured}

\begin{table*}[th!]
\caption{Atomic lines measured.}
\label{atomic_ranges}
\centering
\begin{tabular}{c c | c c | c c | c}
\hline\hline
\noalign{\smallskip}
\multicolumn{2}{c|}{Atomic Lines}&\multicolumn{2}{c|}{Range of EW }&\multicolumn{2}{c|}{Pseudocontinuum ranges (\AA{})}&\\
&&\multicolumn{2}{c|}{measurement (\AA{})}&&&\\
&&&&&&\\
Wavelength&Chemical&Lower&Upper&At Blue&At Red&Reference\\
(\AA{})&species&limit&limit&&&\\
\noalign{\smallskip}
\hline
\noalign{\smallskip}
8468 (a)\tablefootmark{s}&Ti\,{\sc{i}}+Fe\,{\sc{i}}+CN&8462.2&8474.3&8451.6-8452.6&8474.4-8475.4&\cite{gin1994}\\
8468 (b)\tablefootmark{s}&Ti\,{\sc{i}}+Fe\,{\sc{i}}+CN&8462.2&8474.3&8448.3-8449.3&8474.4-8475.4&\cite{gin1994}\\
8498.0\tablefootmark{s}\tablefootmark{g}&Ca\,{\sc{ii}}&8492.5&8503.5&8489.2-8490.4&8507.6-8509.6&\cite{sol1978}\\
8514.1\tablefootmark{s}\tablefootmark{g}&Fe\,{\sc{i}}&8512.5&8516.3&8507.6-8509.6&8557.5-8559.0&\cite{car1997}\\
8518.1\tablefootmark{s}\tablefootmark{g}&Ti\,{\sc{i}}&8516.8&8519.8&8507.6-8509.6&8557.5-8559.0&\cite{car1997}\\
8542.0\tablefootmark{s}\tablefootmark{g}&Ca\,{\sc{ii}}&8532.0&8553.0&8507.6-8509.6&8557.5-8559.0&\cite{sol1978}\\
8582.0\tablefootmark{s}\tablefootmark{g}&Fe\,{\sc{i}}&8581.0&8583.7&8579.5-8580.8&8600.0-8602.0&\cite{car1997}\\
8611.0\tablefootmark{s}\tablefootmark{g}&Fe\,{\sc{i}}&8610.9&8612.7&8600.0-8602.0&8619.6-8620.6&\cite{car1997}\\
8621.5&Fe\,{\sc{i}}&8620.6&8622.3&8619.6-8620.6&8634.5-8640.4&\cite{car1997}\\
8623.0&Ti\,{\sc{i}} ?&8622.3&8623.9&8619.6-8620.6&8634.5-8640.4&\cite{kup2000}\\
8641.6&Ti\,{\sc{i}} ?&8640.6&8642.3&8634.5-8640.4&8684.4-8686.0&\cite{kup2000}\\
8643.0&Cr\,{\sc{i}} ?&8642.3&8643.9&8634.5-8640.4&8684.4-8686.0&\cite{kup2000}\\
8662.0\tablefootmark{s}\tablefootmark{g}&Ca\,{\sc{ii}}&8651.0&8673.0&8634.5-8640.4&8684.4-8686.0&\cite{sol1978}\\
8675.0\tablefootmark{s}\tablefootmark{g}&Fe\,{\sc{i}}+Ti\,{\sc{i}}&8673.3&8676.6&8634.5-8640.4&8684.4-8686.0&\cite{mun1999}\\
8679.4\tablefootmark{s}\tablefootmark{g}&Fe\,{\sc{i}}&8676.9&8681.1&8634.5-8640.4&8684.4-8686.0&\cite{gin1994}\\
8683.0\tablefootmark{s}\tablefootmark{g}&Ti\,{\sc{i}}&8681.6&8684.2&8634.5-8640.4&8684.4-8686.0&\cite{gin1994}\\
8688.5\tablefootmark{s}\tablefootmark{g}&Fe\,{\sc{i}}&8687.3&8690.6&8684.4-8686.0&8695.5-8698.0&\cite{car1997}\\
8692.0\tablefootmark{s}\tablefootmark{g}&Ti\,{\sc{i}}&8691.0&8693.0&8684.4-8686.0&8695.5-8698.0&\cite{mun1999}\\
8699.1\tablefootmark{s}\tablefootmark{g}&Mn\,{\sc{i}}&8698.0&8700.2&8695.5-8698.0&8704.2-8706.3&\cite{kup2000}\\
8702.3&Ce\,{\sc{ii}} ?&8701.1&8704.2&8695.5-8698.0&8704.2-8706.3&\cite{kup2000}\\
8710.2\tablefootmark{s}\tablefootmark{g}&Fe\,{\sc{i}}&8708.5&8711.3&8704.2-8706.3&8714.5-8715.5&\cite{kir1991}\\
8712.8\tablefootmark{s}\tablefootmark{g}&Fe\,{\sc{i}}&8711.3&8714.5&8704.2-8706.3&8714.5-8715.5&\cite{kir1991}\\
8717.5&Fe\,{\sc{i}}+Mn\,{\sc{i}} ?&8716.9&8719.2&8714.5-8715.5&8721.7,8723.4&\cite{kup2000}\\
8729.0\tablefootmark{s}\tablefootmark{g}&Fe\,{\sc{i}}+Si\,{\sc{i}}&8727.2&8729.8&8721.7-8723.4&8731.7-8733.8&\cite{gin1994}\\
8730.5\tablefootmark{s}\tablefootmark{g}&Ti\,{\sc{i}}&8729.8&8731.7&8721.7-8723.4&8731.7-8733.8&\cite{kup2000}\\
8734.5\tablefootmark{s}&Ti\,{\sc{i}}&8733.5&8735.5&8731.7-8733.8&8753.5-8755.6&\cite{gin1994}\\
8736.0\tablefootmark{s}&Mg\,{\sc{i}}&8735.5&8737.0&8731.7-8733.8&8753.5-8755.6&\cite{mun1999}\\
8740.7\tablefootmark{s}&Mn\,{\sc{i}}&8739.6&8741.5&8731.7,8733.8&8753.5-8755.6&\cite{mun1999}\\
8742.2\tablefootmark{s}&Si\,{\sc{i}}&8741.5&8743.0&8731.7,8733.8&8753.5-8755.6&\cite{mun1999}\\
8747.4&Fe\,{\sc{i}} ?&8746.3&8748.4&8731.7,8733.8&8753.5-8755.6&\cite{kup2000}\\
8751.7\tablefootmark{s}&Fe\,{\sc{i}}+Ti\,{\sc{i}}+Si\,{\sc{i}}&8749.7&8753.5&8731.7,8733.8&8753.5-8755.6&\cite{kup2000}\\
8757.0\tablefootmark{s}&Fe\,{\sc{i}}&8755.6&8758.8&8753.5-8755.6&8758.8-8761.0&\cite{kir1991}\\
8764.0\tablefootmark{s}&Fe\,{\sc{i}}&8762.5&8765.0&8758.8-8761.0&8775.0-8777.0&\cite{kir1991}\\
8766.0&Fe\,{\sc{ii}} ?&8765.2&8767.6&8758.8-8761.0&8775.0-8777.0&\cite{kup2000}\\
8772.7&Al\,{\sc{i}}+Cr\,{\sc{i}} ?&8771.5&8773.2&8758.8-8761.0&8775.0-8777.0&\cite{kup2000}\\
8773.7&Al\,{\sc{i}}+Cr\,{\sc{i}} ?&8773.2&8775.0&8758.8-8761.0&8775.0-8777.0&\cite{kup2000}\\
8778.8&Ti\,{\sc{i}} ?&8777.3&8780.0&8775.0-8777.0&8786.0-8788.5&\cite{kup2000}\\
8784.5&Fe\,{\sc{i}} ?&8783.4&8785.9&8775.0-8777.0&8786.0-8788.5&\cite{kup2000}\\
8790.2&Si\,{\sc{i}} ?&8788.9&8791.5&8786.0-8788.5&8810.0-8812.0&\cite{kup2000}\\
8793.2\tablefootmark{s}&Fe\,{\sc{i}}&8791.5&8794.2&8786.0-8788.5&8810.0-8812.0&\cite{kir1991}\\
8796.5&Fe\,{\sc{i}}+Cr\,{\sc{i}} ?&8794.9&8797.3&8786.0-8788.5&8810.0-8812.0&\cite{kup2000}\\
8800.4&Y\,{\sc{i}}&8799.0&8802.0&8786.0-8788.5&8810.0-8812.0&\cite{kup2000}\\
8805.0\tablefootmark{s}&Fe\,{\sc{i}}&8803.3&8805.6&8786.0-8788.5&8810.0-8812.0&\cite{kir1991}\\
8807.0\tablefootmark{s}&Mg\,{\sc{i}}&8805.6&8808.7&8786.0-8788.5&8810.0-8812.0&\cite{kir1991}\\
8824\tablefootmark{s}&Fe\,{\sc{i}}&8823.2&8825.5&8810.0-8812.0&8828.5-8830.5&\cite{kir1991}\\
8835.9&Y\,{\sc{ii}} ?&8834.3&8837.5&8828.5-8830.5&8850.0-8854.0&\cite{kup2000}\\
8838\tablefootmark{s}&Fe\,{\sc{i}}&8837.5&8840.0&8828.5-8830.5&8850.0-8854.0&\cite{kir1991}\\
\noalign{\smallskip}
\hline
\end{tabular}
\tablefoot{The EWs were integrated over the measurement range indicated in the third and fourth columns, using pseudocontinua calculated through a linear regression of the data from the "pseudocontinuum ranges" listed in the fifth and sixth columns. For details about this methods, see Sect.~\ref{bands}. The last column indicates the reference giving the identification of the chemical species producing each line. In those cases marked as "?", the identification is likely, but not certain.\\
\tablefoottext{s}{used in the shortened input list}\\
\tablefoottext{g}{used in the \textit{Gaia} input list}
}
\end{table*}

\begin{table*}[th!]
\caption{Molecular bandheads measured.}
\label{molecular_ranges}
\centering
\begin{tabular}{c c | c c | c c c c}
\hline\hline
\noalign{\smallskip}
\multicolumn{2}{c|}{Molecular Band}&\multicolumn{2}{c|}{Measurement Ranges (\AA{})}&&\\
Bandhead centre (\AA{})&Chemical species&Pseudo-Continuum (\AA{})&Bandhead Bottom(\AA{})&Reference\\
\noalign{\smallskip}
\hline
\noalign{\smallskip}
8504.5\tablefootmark{s}\tablefootmark{g}&TiO&8500.0-8503.5&8504.5-8505.75&\cite{sol1978}\\
8569.2\tablefootmark{s}\tablefootmark{g}&TiO&8562.5-8568.5&8569.20-8570.45&\cite{car2007}\\
8624.25&VO&8605.0-8620.0&8624.25-8625.5&\cite{sol1978}\\
8859.0\tablefootmark{s}&TiO&8850-8858.5&8859.0-8860.5&\cite{val1998}\\
\noalign{\smallskip}
\hline
\end{tabular}
\tablefoot{For the details about the measurement method see Sect.~\ref{bands}.\\
\tablefoottext{s}{used in the shortened input list}\\
\tablefoottext{g}{used in the \textit{Gaia} input list}
}
\end{table*}

\section{PCA coefficients calculated}

\begin{table*}[th!]
\caption{Coefficients for the PCs 1 to 5 calculated from the shortened data input.}
\label{PC_reduc_a}
\centering
\begin{tabular}{c | c c | c c | c c | c c | c c}
\hline\hline
\noalign{\smallskip}
Line or Bandhead&\multicolumn{2}{| c}{PC1}&\multicolumn{2}{| c}{PC2}&\multicolumn{2}{| c}{PC3}&\multicolumn{2}{| c}{PC4}&\multicolumn{2}{| c}{PC5}\\
(\AA{})&coef. (\AA{}$^{-1}$)&$\pm\sigma$&coef. (\AA{}$^{-1}$)&$\pm\sigma$&coef. (\AA{}$^{-1}$)&$\pm\sigma$&coef. (\AA{}$^{-1}$)&$\pm\sigma$&coef. (\AA{}$^{-1}$)&$\pm\sigma$\\
\noalign{\smallskip}
\hline
\noalign{\smallskip}
EW 8468&-0.38&0.01&-0.37&0.03&0.13&0.05&-0.34&0.08&-0.2&0.1\\
EW 8498&-0.304&0.007&0.168&0.008&-0.16&0.02&-0.08&0.07&-0.19&0.07\\
EW 8542&-0.144&0.003&0.099&0.005&-0.11&0.01&-0.04&0.04&-0.10&0.05\\
EW 8582&-2.00&0.04&0.77&0.06&-1.1&0.1&0.7&0.2&-0.0&0.3\\
EW 8611&-2.06&0.04&0.29&0.06&-2.2&0.2&-0.8&0.3&0.8&0.4\\
EW 8662&-0.188&0.005&0.045&0.005&-0.14&0.02&-0.13&0.03&-0.09&0.03\\
EW 8675&-1.01&0.02&-0.47&0.03&-0.68&0.05&0.08&0.08&-0.1&0.1\\
EW 8710.2&-3.50&0.1&2.0&0.2&3.5&0.4&-1.3&0.6&-0.0&2.0\\
EW 8712.8&-2.37&0.05&0.69&0.06&1.0&0.2&1.1&0.3&-0.8&0.4\\
EW 8730.5&-4.1&0.1&-0.8&0.2&1.4&0.4&2.0&1.0&-5.0&3.0\\
EW 8807.0&-1.39&0.05&0.44&0.05&0.8&0.3&1.7&0.5&1.0&1.0\\
EW 8757&-1.82&0.04&-0.54&0.06&-1.0&0.2&-0.8&0.6&2.4&1.0\\
EW 8764&-2.2&0.1&-1.4&0.2&3.5&0.5&0.0&2.0&10.0&3.0\\
EW 8793.2&-2.83&0.1&2.4&0.1&-0.9&0.3&-0.6&1.0&2.0&2.0\\
EW 8824&-1.09&0.02&-0.56&0.04&-0.6&0.1&-0.6&0.4&1.4&0.5\\
EW 8838&-1.16&0.02&-0.75&0.04&-1.4&0.1&-0.8&0.2&0.5&0.3\\
EW 8514.1&-0.93&0.02&0.12&0.02&-0.55&0.04&0.06&0.07&0.08&0.07\\
EW 8518.1&-1.36&0.03&-1.65&0.08&1.1&0.2&2.1&0.4&-0.8&0.6\\
EW 8679.4&-1.87&0.04&-0.57&0.09&1.7&0.2&-1.6&0.5&-1.7&0.6\\
EW 8683.0&-1.18&0.02&-1.26&0.05&0.17&0.09&0.5&0.3&-1.0&0.4\\
EW 8688.5&-0.94&0.02&-0.27&0.02&-0.42&0.05&0.1&0.1&0.3&0.2\\
EW 8692.0&-1.66&0.04&-2.50&0.09&-0.4&0.2&2.1&0.5&-1.5&0.9\\
EW 8734.5&-1.69&0.05&-3.3&0.1&0.6&0.2&1.4&0.4&-0.9&0.6\\
EW 8736.0&-2.4&0.2&6.4&0.2&12.0&1.0&6.0&1.0&4.0&2.0\\
EW 8805.0&-2.08&0.05&-0.85&0.06&-0.8&0.1&-0.1&0.4&0.8&0.7\\
EW 8699.1&-3.12&0.07&0.97&0.09&-1.1&0.3&1.9&0.4&0.3&1.0\\
EW 8729.0&-2.70&0.06&2.8&0.1&2.4&0.2&-1.5&0.4&-0.6&0.7\\
EW 8740.7&-4.0&0.1&-2.3&0.3&12.0&0.9&5.0&2.0&-3.0&2.0\\
EW 8742.2&-3.2&0.2&9.9&0.5&7.5&1.0&-12.0&2.0&4.0&4.0\\
EW 8751.7&-1.5&0.1&2.0&0.1&2.5&0.4&-5.0&0.8&-3.0&1.0\\
Bandhead 8504.5\tablefootmark{a}&1.08&0.08&-5.7&0.3&2.0&0.4&-3.7&0.7&1.0&1.0\\
Bandhead 8569.2\tablefootmark{a}&1.9&0.2&-9.6&0.5&3.8&0.7&-6.0&1.0&2.0&3.0\\
Bandhead 8859\tablefootmark{a}&0.14&0.07&-2.8&0.1&1.1&0.2&-1.7&0.2&-0.1&0.6\\
\noalign{\smallskip}
\hline
\end{tabular}
\tablefoot{
\tablefoottext{a}{The coefficients for this variable are dimensionless.}
}
\end{table*}

\begin{table*}[th!]
\caption{Coefficients for the PCs 6 to 10 calculated from the shortened data input.}
\label{PC_reduc_b}
\centering
\begin{tabular}{c | c c | c c | c c | c c | c c}
\hline\hline
\noalign{\smallskip}
Line or Bandhead&\multicolumn{2}{| c}{PC6}&\multicolumn{2}{| c}{PC7}&\multicolumn{2}{| c}{PC8}&\multicolumn{2}{| c}{PC9}&\multicolumn{2}{| c}{PC10}\\
(\AA{})&coef. (\AA{}$^{-1}$)&$\pm\sigma$&coef. (\AA{}$^{-1}$)&$\pm\sigma$&coef. (\AA{}$^{-1}$)&$\pm\sigma$&coef. (\AA{}$^{-1}$)&$\pm\sigma$&coef. (\AA{}$^{-1}$)&$\pm\sigma$\\
\noalign{\smallskip}
\hline
\noalign{\smallskip}
EW 8468&-0.15&0.1&-0.7&0.3&-0.1&0.1&-0.10&0.09&-0.2&0.1\\
EW 8498&0.0&0.1&0.1&0.2&0.2&0.3&0.0&0.1&-0.02&0.08\\
EW 8542&0.07&0.1&-0.00&0.09&0.1&0.2&0.01&0.07&-0.00&0.04\\
EW 8582&-0.1&0.4&-2.4&0.8&-0.5&0.9&0.0&1.0&0.0&1.0\\
EW 8611&-0.0&0.4&-0.6&0.5&-0.3&0.6&0.0&1.0&1.0&1.0\\
EW 8662&-0.02&0.04&0.10&0.1&0.1&0.2&0.0&0.1&-0.03&0.07\\
EW 8675&-0.1&0.2&0.5&0.3&0.3&0.3&0.1&0.2&0.1&0.4\\
EW 8710.2&-4.0&4.0&-1.0&4.0&-3.0&6.0&-5.0&9.0&-0.0&4.0\\
EW 8712.8&0.1&0.4&0.0&1.0&-1.0&2.0&-1.0&2.0&0.3&0.9\\
EW 8730.5&6.0&7.0&-4.0&6.0&-4.0&9.0&0.0&4.0&-0.0&3.0\\
EW 8807.0&-2.0&2.0&-0.4&0.6&0.1&0.7&-0.0&1.0&-1.0&1.0\\
EW 8757&2.0&2.0&0.5&0.4&0.3&0.6&0.2&0.7&-0.1&0.7\\
EW 8764&6.0&6.0&-1.0&2.0&-1.0&3.0&-3.0&3.0&-3.0&2.0\\
EW 8793.2&-6.0&6.0&-2.0&2.0&-0.0&1.0&2.0&3.0&1.0&3.0\\
EW 8824&-0.1&0.6&0.3&0.3&0.1&0.5&-0.1&0.5&-0.5&0.8\\
EW 8838&0.4&0.4&0.2&0.5&0.2&0.9&0.5&0.7&0.6&0.6\\
EW 8514.1&0.1&0.2&-0.4&0.2&-0.1&0.2&0.0&0.2&0.1&0.2\\
EW 8518.1&-0.0&0.6&-1.9&0.8&-0.3&0.6&0.1&0.5&0.1&0.6\\
EW 8679.4&-0.6&0.9&2.0&1.0&1.0&2.0&-1.0&2.0&-2.0&3.0\\
EW 8683.0&0.1&0.4&1.2&0.4&0.2&0.4&-0.2&0.6&-0.5&0.6\\
EW 8688.5&-0.3&0.4&0.6&0.2&0.1&0.2&-0.1&0.3&-0.3&0.4\\
EW 8692.0&0.3&0.8&2.8&0.9&0.0&1.0&-0.0&0.9&-0.0&1.0\\
EW 8734.5&0.2&0.5&2.5&0.8&1.0&1.0&1.0&1.0&0.0&1.0\\
EW 8736.0&-3.0&5.0&9.0&4.0&4.0&7.0&0.0&5.0&-1.0&3.0\\
EW 8805.0&-2.0&2.0&-1.4&0.8&-0.5&0.8&0.3&0.8&0.0&1.0\\
EW 8699.1&-1.0&2.0&3.0&3.0&0.0&2.0&2.0&3.0&5.0&4.0\\
EW 8729.0&0.4&0.9&1.0&2.0&-2.0&4.0&-0.0&3.0&2.0&3.0\\
EW 8740.7&2.0&4.0&-8.0&5.0&1.0&9.0&6.0&8.0&7.0&8.0\\
EW 8742.2&3.0&4.0&3.0&3.0&2.0&5.0&8.0&10.0&12.0&10.0\\
EW 8751.7&-0.0&1.0&-0.0&1.0&0.0&1.0&-1.0&3.0&-3.0&3.0\\
Bandhead 8504.5\tablefootmark{a}&-1.0&1.0&0.2&0.6&0.0&0.9&1.0&2.0&2.0&2.0\\
Bandhead 8569.2\tablefootmark{a}&-5.0&6.0&2.0&2.0&1.0&2.0&2.0&2.0&3.0&3.0\\
Bandhead 8859\tablefootmark{a}&-0.9&0.9&-0.2&0.5&-0.5&0.8&0.0&0.5&0.1&0.3\\
\noalign{\smallskip}
\hline
\end{tabular}
\tablefoot{
\tablefoottext{a}{The coefficients for this variable are dimensionless.}
}
\end{table*}

\begin{table*}[th!]
\caption{Coefficients for the PCs 11 to 15 calculated from the shortened data input.}
\label{PC_reduc_c}
\centering
\begin{tabular}{c | c c | c c | c c | c c | c c}
\hline\hline
\noalign{\smallskip}
Line or Bandhead&\multicolumn{2}{| c}{PC11}&\multicolumn{2}{| c}{PC12}&\multicolumn{2}{| c}{PC13}&\multicolumn{2}{| c}{PC14}&\multicolumn{2}{| c}{PC15}\\
(\AA{})&coef. (\AA{}$^{-1}$)&$\pm\sigma$&coef. (\AA{}$^{-1}$)&$\pm\sigma$&coef. (\AA{}$^{-1}$)&$\pm\sigma$&coef. (\AA{}$^{-1}$)&$\pm\sigma$&coef. (\AA{}$^{-1}$)&$\pm\sigma$\\
\noalign{\smallskip}
\hline
\noalign{\smallskip}
EW 8468&-0.4&0.3&-0.5&0.4&-1.1&0.4&-0.5&0.4&-0.3&0.3\\
EW 8498&0.0&0.1&0.0&0.2&-0.1&0.2&-0.0&0.2&0.1&0.2\\
EW 8542&-0.02&0.08&0.1&0.2&-0.1&0.1&0.0&0.1&0.02&0.07\\
EW 8582&-2.0&1.0&0.0&1.0&1.0&1.0&0.0&1.0&-1.0&2.0\\
EW 8611&-1.0&1.0&0.0&1.0&-0.0&1.0&-0.0&1.0&-0.1&0.9\\
EW 8662&-0.03&0.07&0.01&0.09&-0.0&0.1&0.0&0.2&0.1&0.1\\
EW 8675&-0.4&0.3&-0.1&0.2&-0.0&0.3&-0.1&0.4&-0.1&0.4\\
EW 8710.2&7.0&4.0&2.0&4.0&-1.0&3.0&-0.0&2.0&-0.0&3.0\\
EW 8712.8&1.1&0.9&1.0&1.0&1.2&1.0&1.0&1.0&-0.0&2.0\\
EW 8730.5&-0.0&3.0&1.0&3.0&1.0&3.0&1.0&4.0&2.0&5.0\\
EW 8807.0&-0.3&0.9&0.0&1.0&0.0&1.0&-0.0&1.0&0.0&0.9\\
EW 8757&1.2&0.7&0.8&0.8&0.6&0.7&0.4&0.9&-0.0&1.0\\
EW 8764&-1.0&1.0&-1.0&2.0&-0.0&2.0&1.0&3.0&-0.0&1.0\\
EW 8793.2&-0.0&3.0&-2.0&5.0&3.0&5.0&3.0&6.0&-1.0&4.0\\
EW 8824&0.8&0.7&-0.0&0.7&-0.0&1.0&-1.0&2.0&1.0&2.0\\
EW 8838&0.9&0.7&-0.1&0.5&-0.1&0.5&0.0&0.6&-0.1&0.5\\
EW 8514.1&-0.3&0.4&0.4&0.6&0.1&0.3&0.1&0.3&-0.0&0.2\\
EW 8518.1&-1.0&1.0&0.0&1.0&1.0&1.0&0.0&1.0&-1.0&2.0\\
EW 8679.4&-3.0&3.0&-0.0&3.0&2.0&2.0&1.0&2.0&1.0&2.0\\
EW 8683.0&-0.1&0.4&-0.1&0.6&0.4&0.4&0.1&0.4&-0.2&0.5\\
EW 8688.5&0.1&0.4&0.1&0.4&0.1&0.4&0.1&0.6&0.2&0.8\\
EW 8692.0&0.0&1.0&-0.3&1.0&-0.0&0.7&-0.0&0.7&-1.0&1.0\\
EW 8734.5&2.0&2.0&-1.0&2.0&-1.0&1.0&-1.0&1.0&-0.0&2.0\\
EW 8736.0&-3.0&3.0&-1.0&5.0&-5.0&4.0&-2.0&3.0&-1.0&3.0\\
EW 8805.0&1.0&1.0&-1.0&2.0&0.7&1.0&0.3&0.8&0.5&0.9\\
EW 8699.1&1.0&2.0&-1.0&2.0&-1.0&2.0&-1.0&2.0&-1.0&3.0\\
EW 8729.0&-3.0&2.0&-1.0&2.0&0.0&2.0&0.0&2.0&2.0&5.0\\
EW 8740.7&5.0&4.0&0.0&3.0&2.0&3.0&1.0&3.0&5.0&6.0\\
EW 8742.2&-1.0&5.0&-1.0&4.0&-1.0&5.0&-3.0&8.0&-2.0&7.0\\
EW 8751.7&2.0&2.0&1.0&1.0&1.0&1.0&0.0&1.0&-1.0&3.0\\
Bandhead 8504.5\tablefootmark{a}&-1.0&2.0&1.0&2.0&1.0&1.0&1.0&2.0&0.0&2.0\\
Bandhead 8569.2\tablefootmark{a}&-1.0&6.0&10.0&10.0&2.0&7.0&4.0&7.0&1.0&4.0\\
Bandhead 8859\tablefootmark{a}&-0.1&0.4&-0.5&0.6&-0.4&0.5&-0.2&0.6&0.2&0.8\\
\noalign{\smallskip}
\hline
\end{tabular}
\tablefoot{
\tablefoottext{a}{The coefficients for this variable are dimensionless.}
}
\end{table*}

\begin{table*}[th!]
\caption{Coefficients for the PCs 1 to 5 calculated from the \textit{Gaia} data input.}
\label{PC_gaia_1}
\centering
\begin{tabular}{c | c c | c c | c c | c c | c c}
\hline\hline
\noalign{\smallskip}
Line or Bandhead&\multicolumn{2}{| c}{PC1}&\multicolumn{2}{| c}{PC2}&\multicolumn{2}{| c}{PC3}&\multicolumn{2}{| c}{PC4}&\multicolumn{2}{| c}{PC5}\\
(\AA{})&coef. (\AA{}$^{-1}$)&$\pm\sigma$&coef. (\AA{}$^{-1}$)&$\pm\sigma$&coef. (\AA{}$^{-1}$)&$\pm\sigma$&coef. (\AA{}$^{-1}$)&$\pm\sigma$&coef. (\AA{}$^{-1}$)&$\pm\sigma$\\
\noalign{\smallskip}
\hline
\noalign{\smallskip}
EW 8498&-0.384&0.009&0.198&0.009&-0.18&0.07&-0.21&0.07&-0.3&0.1\\
EW 8542&-0.183&0.004&0.125&0.006&-0.14&0.04&-0.08&0.05&-0.17&0.06\\
EW 8582&-2.50&0.05&0.88&0.07&-0.8&0.4&0.5&0.4&1.0&2.0\\
EW 8611&-2.56&0.04&0.26&0.07&-2.4&0.6&-2.2&0.6&1.0&2.0\\
EW 8662&-0.234&0.006&0.039&0.006&-0.15&0.07&-0.27&0.05&-0.24&0.06\\
EW 8675&-1.24&0.02&-0.76&0.03&-1.0&0.2&-0.5&0.2&0.2&0.1\\
EW 8710.2&-4.3&0.1&2.1&0.2&10.0&2.0&-4.0&3.0&7.0&2.0\\
EW 8712.8&-2.94&0.06&0.51&0.08&2.8&0.6&1.8&0.7&1.3&0.5\\
EW 8730.5&-5.1&0.1&-1.8&0.2&3.0&2.0&11.0&2.0&-8.0&5.0\\
EW 8514.1&-1.16&0.02&0.07&0.02&-0.6&0.1&-0.1&0.2&0.2&0.5\\
EW 8518.1&-1.62&0.04&-2.5&0.1&0.7&0.8&3.0&0.5&1.6&0.7\\
EW 8679.4&-2.28&0.05&-1.3&0.1&3.3&1.0&-1.9&1.0&-3.0&2.0\\
EW 8683.0&-1.42&0.02&-1.96&0.07&-0.1&0.2&0.8&0.2&-0.3&0.8\\
EW 8688.5&-1.14&0.02&-0.51&0.03&-0.6&0.2&-0.6&0.2&0.5&0.2\\
EW 8692.0&-1.99&0.04&-3.7&0.1&-1.9&0.8&2.2&0.7&1.0&2.0\\
EW 8699.1&-3.90&0.08&0.9&0.1&-0.7&0.7&-2.0&1.0&7.0&3.0\\
EW 8729.0&-3.39&0.07&3.3&0.2&7.0&1.0&0.0&2.0&-3.0&2.0\\
Bandhead 8504.5\tablefootmark{a}&1.60&0.09&-7.9&0.5&2.0&1.0&-3.5&0.9&-2.0&2.0\\
Bandhead 8569.2\tablefootmark{a}&2.7&0.2&-13.6&0.8&4.0&3.0&-10.0&2.0&-1.0&3.0\\
\noalign{\smallskip}
\hline
\end{tabular}
\tablefoot{
\tablefoottext{a}{The coefficients for this variable are dimensionless.}
}
\end{table*}

\begin{table*}[th!]
\caption{Coefficients for the PCs 6 to 9 calculated from the \textit{Gaia} data input.}
\label{PC_gaia_2}
\centering
\begin{tabular}{c | c c | c c | c c | c c}
\hline\hline
\noalign{\smallskip}
Line or Bandhead&\multicolumn{2}{| c}{PC6}&\multicolumn{2}{| c}{PC7}&\multicolumn{2}{| c}{PC8}&\multicolumn{2}{| c}{PC9}\\
(\AA{})&coef. (\AA{}$^{-1}$)&$\pm\sigma$&coef. (\AA{}$^{-1}$)&$\pm\sigma$&coef. (\AA{}$^{-1}$)&$\pm\sigma$&coef. (\AA{}$^{-1}$)&$\pm\sigma$\\
\noalign{\smallskip}
\hline
\noalign{\smallskip}
EW 8498&-0.2&0.1&-0.33&0.09&-0.3&0.1&-0.4&0.1\\
EW 8542&-0.07&0.07&-0.20&0.07&-0.16&0.08&-0.17&0.07\\
EW 8582&3.0&2.0&-1.0&1.0&2.0&2.0&1.0&2.0\\
EW 8611&4.0&2.0&1.2&0.7&1.0&1.0&1.9&0.8\\
EW 8662&-0.13&0.08&-0.08&0.04&-0.10&0.1&-0.07&0.08\\
EW 8675&-0.0&0.2&0.4&0.2&0.5&0.3&0.7&0.3\\
EW 8710.2&2.0&3.0&-6.0&3.0&-6.0&7.0&3.0&4.0\\
EW 8712.8&0.6&0.7&0.9&0.7&-1.0&1.0&-1.0&1.0\\
EW 8730.5&5.0&7.0&3.0&4.0&-5.0&6.0&1.0&4.0\\
EW 8514.1&0.9&0.6&-0.2&0.2&0.1&0.3&0.1&0.3\\
EW 8518.1&0.0&1.0&-4.0&1.0&1.0&3.0&-2.0&2.0\\
EW 8679.4&-4.0&2.0&-1.0&3.0&3.0&2.0&3.0&2.0\\
EW 8683.0&-1.3&0.9&0.5&0.2&0.2&0.5&0.7&0.4\\
EW 8688.5&0.3&0.3&0.8&0.4&0.2&0.9&1.2&0.6\\
EW 8692.0&-3.0&3.0&1.2&0.7&-1.0&2.0&0.0&1.0\\
EW 8699.1&-1.0&4.0&6.0&2.0&-0.0&4.0&-6.0&2.0\\
EW 8729.0&1.0&3.0&7.0&2.0&2.0&4.0&-4.0&2.0\\
Bandhead 8504.5\tablefootmark{a}&4.0&3.0&1.0&1.0&0.0&2.0&-1.0&2.0\\
Bandhead 8569.2\tablefootmark{a}&5.0&4.0&0.0&3.0&-5.0&4.0&-8.0&4.0\\
\noalign{\smallskip}
\hline
\end{tabular}
\tablefoot{
\tablefoottext{a}{The coefficients for this variable are dimensionless.}
}
\end{table*}

\section{SVM coefficients calculated}

\begin{table*}[th!]
\caption{Coefficients that define the split between the early and late subsamples in the multidimensional (15) space of the PCs.}
\label{SVM_M_red}
\centering
\begin{tabular}{c | c c | c c | c c | c c | c c}
\hline\hline
\noalign{\smallskip}
&\multicolumn{10}{| c}{Boundary between early and late subsamples}\\
&\multicolumn{10}{| c}{ putative boundary at}\\
&\multicolumn{2}{| c}{K5}&\multicolumn{2}{| c}{M0}&\multicolumn{2}{| c}{M1}&\multicolumn{2}{| c}{M2}&\multicolumn{2}{| c}{M3}\\
PC&coef.&$\pm\sigma$&coef.&$\pm\sigma$&coef.&$\pm\sigma$&coef.&$\pm\sigma$&coef.&$\pm\sigma$\\
\noalign{\smallskip}
\hline
\noalign{\smallskip}
0th order&0.3&0.4&-0.3&0.5&-1.1&0.5&-1.7&0.5&-1.2&0.3\\
PC1&-0.06&0.06&0.06&0.05&-0.00&0.07&-0.01&0.06&0.09&0.04\\
PC2&-2.6&0.3&-2.7&0.2&-2.4&0.3&-1.6&0.2&-1.0&0.1\\
PC3&0.8&0.2&1.0&0.2&0.8&0.2&0.6&0.1&0.3&0.1\\
PC4&-0.2&0.3&-0.4&0.3&-0.1&0.3&-0.1&0.3&-0.5&0.2\\
PC5&0.5&0.2&0.1&0.2&0.2&0.1&-0.1&0.1&-0.2&0.1\\
PC6&-1.0&0.3&-1.1&0.2&-0.6&0.2&-0.5&0.2&-0.4&0.2\\
PC7&0.3&0.6&0.4&0.4&0.4&0.3&1.1&0.3&0.9&0.3\\
PC8&-0.7&0.4&-0.4&0.4&-0.8&0.4&-0.7&0.4&-0.7&0.3\\
PC9&0.6&0.4&0.7&0.3&0.3&0.3&0.4&0.3&0.3&0.3\\
PC10&-0.2&0.4&-0.3&0.3&0.1&0.3&-0.3&0.3&0.2&0.3\\
PC11&1.2&0.4&0.8&0.4&0.8&0.4&0.6&0.3&-0.3&0.2\\
PC12&-0.7&0.4&-1.4&0.4&-1.2&0.4&-0.6&0.4&-0.7&0.4\\
PC13&-0.3&0.3&-0.1&0.3&0.6&0.4&0.2&0.3&0.2&0.3\\
PC14&-0.5&0.3&-0.8&0.3&-0.4&0.3&-0.2&0.3&-0.5&0.3\\
PC15&-0.2&0.4&0.1&0.4&-0.1&0.4&0.5&0.4&-0.2&0.4\\
\noalign{\smallskip}
\hline
\end{tabular}
\tablefoot{These coefficients were calculated from the shortened input list. The coefficients are given for the multiple putative boundaries used. Each one of these splits generates two subsamples: the early subsample will be formed by those objects considered earlier than the given  putative boundary by the split, and the late subsample will be formed by those objects considered equal or later than the given  putative boundary.}
\end{table*}

\begin{table*}[th!]
\caption{Coefficients that define the separation between the SGs and non-SGs in the multidimensional (15) space of the PCs, for the early subsample.}
\label{SVM_SGearly_red}
\centering
\begin{tabular}{c | c c | c c | c c | c c | c c}
\hline\hline
\noalign{\smallskip}
&\multicolumn{10}{| c}{Boundary between SG and non-SGs for early subsample}\\
&\multicolumn{10}{| c}{ putative boundary at}\\
&\multicolumn{2}{| c}{K5}&\multicolumn{2}{| c}{M0}&\multicolumn{2}{| c}{M1}&\multicolumn{2}{| c}{M2}&\multicolumn{2}{| c}{M3}\\
PC&coef.&$\pm\sigma$&coef.&$\pm\sigma$&coef.&$\pm\sigma$&coef.&$\pm\sigma$&coef.&$\pm\sigma$\\
\noalign{\smallskip}
\hline
\noalign{\smallskip}
0th order&1.4&0.4&1.5&0.4&1.4&0.4&1.2&0.4&1.1&0.4\\
PC1&-0.44&0.07&-0.45&0.07&-0.47&0.06&-0.48&0.07&-0.49&0.07\\
PC2&-0.3&0.2&-0.3&0.2&-0.2&0.2&-0.1&0.2&-0.1&0.2\\
PC3&-1.0&0.1&-1.0&0.2&-1.1&0.2&-1.2&0.2&-1.2&0.2\\
PC4&-1.1&0.2&-1.1&0.2&-1.1&0.2&-1.2&0.3&-1.1&0.3\\
PC5&-0.3&0.2&-0.3&0.2&-0.4&0.2&-0.5&0.2&-0.4&0.2\\
PC6&1.1&0.2&1.1&0.2&1.1&0.3&0.9&0.3&0.7&0.3\\
PC7&-0.6&0.3&-0.5&0.3&-0.3&0.3&-0.2&0.3&-0.1&0.3\\
PC8&0.3&0.3&0.3&0.3&0.2&0.3&0.3&0.3&0.5&0.4\\
PC9&0.8&0.3&0.7&0.3&0.6&0.3&0.5&0.3&0.6&0.3\\
PC10&0.0&0.3&-0.0&0.3&0.0&0.3&-0.1&0.3&-0.1&0.3\\
PC11&0.5&0.3&0.4&0.3&0.4&0.3&0.3&0.3&0.2&0.3\\
PC12&-0.3&0.2&-0.3&0.2&-0.3&0.2&-0.1&0.2&-0.3&0.3\\
PC13&-0.1&0.3&-0.0&0.3&-0.0&0.3&-0.1&0.3&-0.1&0.3\\
PC14&0.3&0.2&0.3&0.2&0.3&0.2&0.2&0.2&0.2&0.2\\
PC15&0.2&0.3&0.2&0.3&0.2&0.3&0.4&0.4&0.3&0.4\\
\noalign{\smallskip}
\hline
\end{tabular}
\tablefoot{These coefficients were calculated from the shortened input list. The coefficients are given for the multiple  putative boundaries previously used to define the early subsample.}
\end{table*}

\begin{table*}[th!]
\caption{Coefficients that define the split between the SGs and non-SGs in the multidimensional (15) space of the PCs, for the late subsample.}
\label{SVM_SGlate_red}
\centering
\begin{tabular}{c | c c | c c | c c | c c | c c}
\hline\hline
\noalign{\smallskip}
&\multicolumn{10}{| c}{Boundary between SG and non-SGs for late subsample}\\
&\multicolumn{10}{| c}{ putative boundary at}\\
&\multicolumn{2}{| c}{K5}&\multicolumn{2}{| c}{M0}&\multicolumn{2}{| c}{M1}&\multicolumn{2}{| c}{M2}&\multicolumn{2}{| c}{M3}\\
PC&coef.&$\pm\sigma$&coef.&$\pm\sigma$&coef.&$\pm\sigma$&coef.&$\pm\sigma$&coef.&$\pm\sigma$\\
\noalign{\smallskip}
\hline
\noalign{\smallskip}
0th order&1.5&0.8&1.4&0.9&1.8&1.0&2.0&1.0&1.0&1.0\\
PC1&-0.5&0.1&-0.5&0.1&-0.5&0.1&-0.5&0.1&-0.5&0.1\\
PC2&0.1&0.2&0.1&0.2&0.2&0.2&0.2&0.2&0.0&0.3\\
PC3&-1.0&0.3&-1.1&0.3&-1.0&0.3&-1.0&0.3&-1.1&0.2\\
PC4&-1.2&0.4&-1.2&0.4&-1.3&0.4&-1.3&0.4&-1.2&0.4\\
PC5&-0.1&0.3&-0.1&0.3&0.0&0.3&0.2&0.2&0.2&0.2\\
PC6&-0.2&0.3&-0.2&0.3&-0.3&0.3&-0.3&0.3&-0.4&0.2\\
PC7&0.5&0.5&0.5&0.5&0.4&0.5&0.5&0.5&0.6&0.4\\
PC8&-0.2&0.4&-0.3&0.4&-0.3&0.4&-0.5&0.4&-0.7&0.3\\
PC9&0.4&0.4&0.4&0.4&0.5&0.4&0.5&0.4&0.4&0.4\\
PC10&-0.0&0.4&-0.0&0.4&-0.1&0.4&-0.1&0.4&-0.0&0.4\\
PC11&0.6&0.5&0.7&0.5&0.6&0.5&0.8&0.5&1.2&0.4\\
PC12&-0.5&0.4&-0.7&0.4&-0.7&0.4&-0.7&0.4&-0.6&0.3\\
PC13&-0.1&0.4&-0.1&0.4&-0.1&0.4&-0.2&0.4&-0.4&0.3\\
PC14&0.1&0.3&0.1&0.3&0.1&0.3&0.0&0.3&-0.0&0.2\\
PC15&-0.1&0.4&0.0&0.4&-0.0&0.4&0.0&0.4&0.2&0.4\\
\noalign{\smallskip}
\hline
\end{tabular}
\tablefoot{These coefficients were calculated from the shortened input list. The coefficients are given for the multiple  putative boundaries previously used to define the late subsample.}
\end{table*}

\begin{table*}[th!]
\caption{Coefficients that define the split between the early and late subsamples in the multidimensional (9) space of the PCs.}
\label{SVM_M_gaia}
\centering
\begin{tabular}{c | c c | c c | c c | c c | c c}
\hline\hline
\noalign{\smallskip}
&\multicolumn{10}{| c}{Boundary between early and late subsamples}\\
&\multicolumn{10}{| c}{ putative boundary at}\\
&\multicolumn{2}{| c}{K5}&\multicolumn{2}{| c}{M0}&\multicolumn{2}{| c}{M1}&\multicolumn{2}{| c}{M2}&\multicolumn{2}{| c}{M3}\\
PC&coef.&$\pm\sigma$&coef.&$\pm\sigma$&coef.&$\pm\sigma$&coef.&$\pm\sigma$&coef.&$\pm\sigma$\\
\noalign{\smallskip}
\hline
\noalign{\smallskip}
0th order&-0.4&0.2&-0.6&0.3&-1.3&0.3&-1.3&0.4&-1.3&0.3\\
PC1&-0.14&0.06&-0.07&0.07&-0.06&0.06&-0.02&0.07&0.11&0.04\\
PC2&-2.7&0.3&-3.0&0.3&-2.7&0.2&-2.2&0.3&-1.3&0.2\\
PC3&0.0&0.2&0.3&0.2&0.4&0.2&0.5&0.2&0.3&0.2\\
PC4&0.5&0.3&0.5&0.3&0.8&0.3&0.3&0.3&0.1&0.2\\
PC5&1.6&0.4&1.4&0.4&1.4&0.3&0.9&0.4&0.5&0.3\\
PC6&-0.3&0.4&-1.0&0.3&-0.9&0.3&-1.0&0.3&-0.4&0.3\\
PC7&0.6&0.3&0.7&0.3&0.8&0.3&1.3&0.3&0.8&0.3\\
PC8&-0.6&0.4&-0.1&0.4&0.2&0.4&-0.3&0.4&0.5&0.3\\
PC9&0.2&0.4&0.4&0.4&0.2&0.3&-0.0&0.3&-0.3&0.3\\
\noalign{\smallskip}
\hline
\end{tabular}
\tablefoot{These coefficients were calculated from the \textit{Gaia} input list. The coefficients are given for the multiple  putative boundaries used. Each one of these splits generates two subsamples: the early subsample will be formed by those objects considered earlier than the given  putative boundary by the split, and the late subsample will be formed by those objects considered equal or later than the given  putative boundary.}
\end{table*}

\begin{table*}[th!]
\caption{Coefficients that define the split between the SGs and non-SGs in the multidimensional (9) space of the PCs, for the early subsample.}
\label{SVM_SGearly_gaia}
\centering
\begin{tabular}{c | c c | c c | c c | c c | c c}
\hline\hline
\noalign{\smallskip}
&\multicolumn{10}{| c}{Boundary between SG and non-SGs for early subsample}\\
&\multicolumn{10}{| c}{ putative boundary at}\\
&\multicolumn{2}{| c}{K5}&\multicolumn{2}{| c}{M0}&\multicolumn{2}{| c}{M1}&\multicolumn{2}{| c}{M2}&\multicolumn{2}{| c}{M3}\\
PC&coef.&$\pm\sigma$&coef.&$\pm\sigma$&coef.&$\pm\sigma$&coef.&$\pm\sigma$&coef.&$\pm\sigma$\\
\noalign{\smallskip}
\hline
\noalign{\smallskip}
0th order&0.0&0.2&0.2&0.2&0.3&0.2&0.4&0.2&0.6&0.2\\
PC1&-0.51&0.05&-0.52&0.06&-0.53&0.06&-0.54&0.06&-0.53&0.07\\
PC2&0.9&0.3&0.8&0.3&0.7&0.3&0.6&0.2&0.4&0.2\\
PC3&-1.5&0.2&-1.5&0.2&-1.5&0.2&-1.4&0.2&-1.4&0.2\\
PC4&-0.6&0.2&-0.7&0.2&-0.6&0.2&-0.6&0.2&-0.7&0.2\\
PC5&-1.3&0.2&-1.4&0.2&-1.4&0.2&-1.4&0.2&-1.3&0.3\\
PC6&0.4&0.3&0.4&0.3&0.3&0.3&0.4&0.3&0.3&0.3\\
PC7&0.3&0.3&0.3&0.3&0.4&0.3&0.5&0.3&0.6&0.3\\
PC8&-1.5&0.3&-1.5&0.3&-1.6&0.4&-1.5&0.4&-1.5&0.4\\
PC9&0.2&0.3&0.3&0.3&0.4&0.3&0.6&0.3&0.5&0.3\\
\noalign{\smallskip}
\hline
\end{tabular}
\tablefoot{These coefficients were calculated from the \textit{Gaia} input list. The coefficients are given for the multiple  putative boundaries previously used to define the early subsample.}
\end{table*}

\begin{table*}[th!]
\caption{Coefficients that define the split between the SGs and non-SGs in the multidimensional (9) space of the PCs, for the late subsample.}
\label{SVM_SGlate_gaia}
\centering
\begin{tabular}{c | c c | c c | c c | c c | c c}
\hline\hline
\noalign{\smallskip}
&\multicolumn{10}{| c}{Boundary between SG and non-SGs for late subsample}\\
&\multicolumn{10}{| c}{ putative boundary at}\\
&\multicolumn{2}{| c}{K5}&\multicolumn{2}{| c}{M0}&\multicolumn{2}{| c}{M1}&\multicolumn{2}{| c}{M2}&\multicolumn{2}{| c}{M3}\\
PC&coef.&$\pm\sigma$&coef.&$\pm\sigma$&coef.&$\pm\sigma$&coef.&$\pm\sigma$&coef.&$\pm\sigma$\\
\noalign{\smallskip}
\hline
\noalign{\smallskip}
0th order&1.3&0.6&1.5&0.7&1.5&0.8&1.5&0.7&1.0&0.6\\
PC1&-0.5&0.2&-0.5&0.2&-0.5&0.2&-0.5&0.2&-0.32&0.09\\
PC2&0.2&0.3&0.2&0.3&0.3&0.3&0.3&0.3&0.3&0.2\\
PC3&-0.4&0.4&-0.2&0.4&-0.3&0.4&-0.3&0.4&0.0&0.3\\
PC4&-0.8&0.3&-0.8&0.3&-0.8&0.3&-0.9&0.3&-1.1&0.3\\
PC5&-0.4&0.5&-0.4&0.5&-0.4&0.5&-0.1&0.5&-0.1&0.4\\
PC6&0.3&0.5&0.4&0.5&0.5&0.5&0.4&0.5&0.6&0.3\\
PC7&1.3&0.5&1.4&0.5&1.4&0.4&1.7&0.4&1.8&0.3\\
PC8&-0.0&0.4&0.1&0.4&0.0&0.4&-0.1&0.4&-0.1&0.3\\
PC9&0.7&0.4&0.7&0.4&0.8&0.4&0.7&0.4&0.7&0.3\\
\noalign{\smallskip}
\hline
\end{tabular}
\tablefoot{These coefficients were calculated from the \textit{Gaia} input list. The coefficients are given for the multiple  putative boundaries previously used to define the late subsample.}
\end{table*}

\end{document}